\documentclass{ws-ijmpa}

\usepackage{psfrag}

\begin{document}

\markboth{M.-C. Chen and K.T. Mahanthappa}
{Fermion Masses and Mixing and CP-Violation in $SO(10)$}

%
\catchline{}{}{}{}{}
%

\title{Fermion Masses and Mixing and CP-Violation in $SO(10)$ Models with 
Family Symmetries
}

\author{\footnotesize Mu-Chun Chen}

\address{High Energy Theory Group, Department of Physics\\
Brookhaven National Laboratory,
Upton, NY 11973-5000, U.S.A.\\
chen@quark.phy.bnl.gov
}

\author{K.T. Mahanthappa}

\address{Department of Physics, University of Colorado\\
Boulder, CO 80309-0390, U.S.A.\\
ktm@verb.colorado.edu 
}

\maketitle

\pub{Received (Day Month Year)}{Revised (Day Month Year)}

\begin{abstract}
Several ideas for solving the problem of fermion mass hierarchy and mixing  
and specific supersymmetric models that realize it are reviewed. In particular, 
we discuss many models based on $SO(10)$ in four dimensions 
combined with a family symmetry to accommodate fermion 
mass hierarchy and mixing, including the case of neutrinos. These models 
are compared and various tests that can be used to distinguish these 
models are suggested. We also include a discussion of a few $SO(10)$ 
models in higher space-time dimensions.

\end{abstract}

\section{Introduction}\label{intro}

The origin of fermion masses and mixing and CP violation is the least
understood aspect of the Standard Model (SM) of particle physics. 
In the SM, the fermion masses and mixing angles are completely arbitrary. 
In order to accommodate their diverse values
(see Table~\ref{exp}),\cite{Hagiwara:fs,Hocker:2001xe,Fusaoka:1998vc} the Yukawa 
couplings must range over five orders 
of magnitude.
\begin{table}[b]
\tbl{\label{exp}Current status of fermions masses and CKM matrix elements at 
$M_{z}$.}
{\begin{tabular}{@{}lll@{}}\toprule
$m_{u} = 1.88-2.75 \; MeV, \quad$ &
$m_{c} = 616-733 \; MeV, \quad$ &
$m_{t} = 168 - 194 \; MeV$\\
$m_{s}/m_{d} = 17 - 25, \quad$ &
$m_{s} = 80.4 - 105 \; MeV, \quad$ &
$m_{b} = 2.89 - 3.11 \; GeV$\\
$m_{e} = 0.487 \; MeV, \quad$ &
$m_{\mu} = 103 \; MeV, \quad$ &
$m_{\tau} = 1.75 \; GeV$ \\ 
& & \\
& & \\
\end{tabular}}
{\begin{tabular}{@{}c@{}}
\hspace{2cm}$\left|V_{CKM, exp}\right|=
\left(\begin{array}{ccc}
0.9745-0.9757 & 0.219-0.224 & 0.002-0.005 \\
0.218-0.224 & 0.9736-0.9750 & 0.036-0.046 \\
0.004-0.014 & 0.034-0.046 & 0.9989-0.9993
\end{array} \right)$\\
\end{tabular}}
\end{table}
Another relevant aspect is that neutrinos are massless in the framework 
of SM, but recent experiments strongly indicate that they do have 
small but non-vanishing masses. Incorporating these into theory 
leads to an increase in the number of parameters. The {\it flavor problem} 
thus consists the following aspects: how to reduce the number of 
parameters in the Yukawa sector, how to obtain an explanation of 
the mass hierarchy, how to obtain small neutrino masses and large 
leptonic mixing angles.

As we extend the SM to the Minimal Supersymmetric Standard Model (MSSM), 
the particle spectrum is doubled and 
many more parameters are introduced into the model.
One thus expects large flavor changing neutral current 
(FCNC) due to the presence of squarks. The strongest constraints 
come from the lighter two generations due to the type of diagram 
given in Fig.\ref{sflavor}:
\begin{figure}
\centerline{
\psfig{file=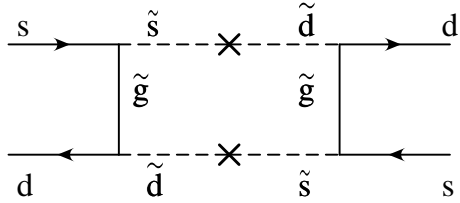,width=6cm}}
\caption{\label{sflavor}Diagram of $\Delta S = 2$ transitions.}
\end{figure}
\begin{eqnarray}
\frac{m_{\tilde{d}}^{2}-m_{\tilde{s}}^{2}}{m_{\tilde{d}}^{2}}
& \lesssim & 6 \times 10^{-3} 
(\frac{m_{\tilde{s}}}{TeV}) 
(\sin\tilde{\theta})^{2}
\\
\frac{m_{\tilde{e}}^{2}-m_{\tilde{\mu}}^{2}}{m_{\tilde{e}}^{2}}
& < & 10^{-1}  
(\frac{m_{\tilde{e}}}{TeV}) 
(\sin\tilde{\theta^{'}})^{2}
\end{eqnarray}
where $\tilde{\theta}$ and $\tilde{\theta^{'}}$ refer to the 
mixing angles in the squark and slepton sectors  
in the bases where the mass matrices of ($d$ $s$) and ($e$ $\mu$) 
are diagonal.\cite{Gabbiani:1996hi,Rosner:fg} 
Thus, in order to suppress the supersymmetric (SUSY) FCNC, 
a near degeneracy between the first and the second generations 
of squarks and sleptons is required. This is sometimes called 
{\it SUSY flavor problem}.

There have been many SUSY models proposed to accommodate the observed 
masses and mixing angles. These models can be classified according to 
the family symmetry implemented in the model. We also discuss other 
mechanisms that have been proposed to solve the problem of the 
fermion mass hierarchy and mixing.

In the following two sections, we introduce various tools that have 
been used to solve the flavor problem in quark sector and 
lepton sector, respectively; in Sec. \ref{so10}, $SO(10)$ is reviewed. This is then 
followed by a review in Sec. \ref{modelt} and Sec. \ref{models} on various models 
based on $SO(10)$ in $4D$, combined with mass texture ansatz and 
family symmetry, respectively;  
a comparison of these models is given at the end of Sec. \ref{models}; 
Sec. \ref{modeled} is devoted to models based 
on $SO(10)$ in higher space-time dimensions. Sec. \ref{conclude} concludes this review.

\section{Quark Masses and Mixing}\label{qmass}

\begin{figure}
\psfrag{a}[][]{$\alpha$}
\psfrag{b}[][]{$\beta$}
\psfrag{c}[][]{$\gamma$}
\psfrag{v1}[][]{$V_{ud}V_{ub}^{\ast}$}
\psfrag{v2}[][]{$V_{td}V_{tb}^{\ast}$}
\psfrag{v3}[][]{$V_{cd}V_{cb}^{\ast}$}
\centerline{
\psfig{file=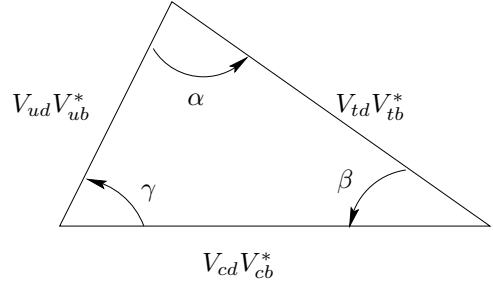,width=6cm}}
\caption{\label{triangle}The Unitarity triangle.}
\end{figure}

Because the strong interaction eigenstates (same as the mass eigenstates) and the
weak interaction eigenstates do not match, flavor mixing arises. 
However, currently we do not have any fundamental understanding for 
such a mismatch between the strong eigenstates and the weak eigenstates. 
In the weak basis, the quark mass terms are 
\begin{equation}  
\mathcal{L}_{mass} =
-Y_{u} \overline{U}_{R} Q_{L} H_{u}  - Y_{d} \overline{D}_{R} Q_{L} H_{d} +
h.c.   
\end{equation}
where $Q$ stands for the $SU(2)$ doublet quark; $U$ and $D$ are up- and down-type 
fermion $SU(2)$ singlet; $H_{u}$ and $H_{d}$ are Higgs fields giving masses 
to up- and down-type quarks. The charged current interaction is given by
\begin{equation} \mathcal{L}_{cc} = \frac{g}{\sqrt{2}} (W_{\mu}^{+}
\overline{U}_{L} \gamma_{\mu} D_{L}) + h.c..  
\end{equation}
The Yukawa couplings $Y_{i}, \; (i=u,d,e,\nu_{LR})$ are in general 
non-diagonal. They are diagonalized by the 
bi-unitary transformations
\begin{eqnarray}
Y_{u}^{diag} = V_{u_{R}} Y_{u} V_{u_{L}}^{\dagger} 
= diag(y_{u},y_{c},y_{t})
\\
Y_{d}^{diag} = V_{d_{R}} Y_{d} V_{d_{L}}^{\dagger} 
= diag(y_{d},y_{s},y_{b})
\end{eqnarray}
where $V_{R}$ and $V_{L}$ are the right-handed and left-handed rotations
respectively, and all the eigenvalues $y_{i}$'s are real and non-negative, 
and are obtained by diagonalizing the hermitian quantity 
$Y^{\dagger}Y$ and $Y Y^{\dagger}$. The Cabbibo-Kobayashi-Maskawa 
(CKM) matrix is then given by\cite{Kobayashi:fv} 
\begin{equation}
V_{CKM} = V_{u_{L}} V_{d_{L}}^{\dagger}
= \left(
\begin{array}{ccc}
V_{ud} & V_{us} & V_{ub}\\
V_{cd} & V_{cs} & V_{cb}\\
V_{td} & V_{ts} & V_{tb}
\end{array}
\right).
\end{equation}
The unitary matrix $V_{CKM}$ has in general $6$ phases. By phase
redefinition of various quark fields, one can remove $5$ of the $6$ phases. 
The remaining one phase is one of the sources for CP violation 
in the quark sector. There are many ways to parameterize 
the CKM matrix, for example,
\begin{equation}
V_{CKM} = 
\left(
\begin{array}{ccc}
c_{12}^{q} c_{13}^{q} & s_{12}^{q} c_{13}^{q} & s_{13}^{q} e^{-i\delta_{q}}\\
-s_{12}^{q} c_{23}^{q}-c_{12}^{q}s_{23}^{q}s_{13}^{q}e^{i\delta_{q}} &
c_{12}^{q}c_{23}^{q}-s_{12}^{q}s_{23}^{q}s_{13}^{q}e^{i\delta_{q}} &
s_{23}^{q}c_{13}^{q}\\
s_{12}^{q}s_{23}^{q}-c_{12}^{q}c_{23}^{q}s_{13}^{q}e^{i\delta_{q}} &
-c_{12}^{q}s_{23}^{q}-s_{12}^{q}c_{23}^{q}s_{13}^{q}e^{i\delta_{q}} &
c_{23}^{q}c_{13}^{q}
\end{array}
\right).
\end{equation}
Defining $\lambda=s_{12}^{q}, \; A\lambda^{2} = s_{23}^{q}$ and 
$A\lambda^{3}(\rho-i\eta)=s_{13}^{q}e^{-i\delta_{q}}$, we obtain
an alternative parameterization, the Wolfenstein 
parameterization,\cite{Wolfenstein:1983yz}
\begin{equation}
V_{CKM} = 
\left(
\begin{array}{ccc}
1-\frac{1}{2}\lambda^{2} & \lambda & A\lambda^{3}(\rho-i\eta)\\
-\lambda & 1-\frac{1}{2}\lambda^{2} & A\lambda^{2}\\
A\lambda^{3}(1-\rho-i\eta) & -A\lambda^{2} & 1
\end{array}
\right).
\end{equation}
Here the parameters $A, \; \rho, \; \eta$ are of 
order $1$, and $\lambda$ is the sine of the Cabbibo angle which is 
about $0.22$, and thus a good choice as an expansion parameter.  
A parameterization independent 
measure for the CP violation is the Jarlskog invariant,\cite{Jarlskog:1985cw} 
defined as  
\begin{equation}
J_{CP}^{q} \equiv Im \{ V_{ud}V_{us}^{\ast}V_{cd}^{\ast}V_{cs} \}.
\end{equation}
Unitarity of $V_{CKM}$ applied to the first and the third columns 
leads to the following condition\footnote{There are
five other similar conditions one can write down.}
\begin{equation}
V_{td}V_{tb}^{\ast} + V_{ud}V_{ub}^{\ast} + V_{cd}V_{cb}^{\ast} = 0.
\end{equation}
A geometrical representation of this equation on the complex plane gives rise
to the CKM ``unitarity triangle'' shown in Fig.\ref{triangle}.
The three angles of the CKM unitarity triangle are 
\begin{equation}
\alpha \equiv 
Arg(-\frac{V_{td}V_{tb}^{\ast}}{V_{ud}V_{ub}^{\ast}}), \quad
\beta \equiv
Arg(-\frac{V_{cd}V_{cb}^{\ast}}{V_{td}V_{tb}^{\ast}}), \quad
\gamma \equiv
Arg(-\frac{V_{ud}V_{ub}^{\ast}}{V_{cd}V_{cb}^{\ast}}).
\end{equation}
The Jarlskog invariant $J_{CP}^{q}$ is proportional to the area of the unitarity
triangle. A non-vanishing value for $J_{CP}^{q}$ thus implies 
non-vanishing values for $(\alpha,\beta,\gamma)$ which in turn indicates CP
violation. 

\subsection{Textures of Mass Matrices}

There have been many mass textures, with different elements having zeros 
in the mass matrices, proposed in order to accommodate the observed 
fermion mass hierarchy and mixing pattern. Imposing texture ansatz 
on mass matrices reduces the number of parameters in the Yukawa sector; 
as a consequence, masses and mixing angles may be related in 
some simple ways. This is illustrated in a simplified two family example in 
the up- and down-quark sectors, in which the mass matrices of the 
up- and down-type quarks are assumed to be symmetric and each contains 
one zero entry,
\begin{equation}
M_{U}=
\left(\begin{array}{cc}
0 & a\\
a & b
\end{array}
\right), \qquad 
M_{D}=
\left(\begin{array}{cc}
0 & c\\
c & d
\end{array}
\right).
\end{equation}
In this case, there is a very simple relation between the Cabbibo angle 
and the quark masses,
\begin{equation}
|V_{us}| = |\sqrt{\frac{m_{d}}{m_{s}}} - e^{i\alpha}\sqrt{\frac{m_{u}}{m_{c}}}|
\end{equation}
where the CP violating phase $\alpha$ arises as the relative phase that enters 
when combining the up- and down-quark rotaion matrices. 
This relation is in good agreement with experiment. 
Note that because $m_{u}/m_{c}$ is quite small compared to 
$m_{d}/m_{s}$, the value of $|V_{us}|$ is not very sensitive to the 
complex phase $\alpha$. One can then generalize this to consider the 
three family case. The ultimate goal of studying 
texture zero ansatz is that it may help us to understand the 
underlying theory of flavor, if the zeros are protected by some family symmetry. 
In the quark sector, assuming symmetric mass matrices, 
the total number of texture zeros is at most six, because there are 
six different quark masses. Nevertheless, Ramond, Robert and Ross\cite{Ramond:1993kv}
found that the observed masses and mixing angles cannot be 
accommodated with six texture zeros. They found five combinations 
of five-zero texture for up- and down-type quark mass 
matrices which give rise to predictions that are consistent with current 
observations for fermion masses and mixing angles. One should note that 
in the context of a grand unified theory (GUT), the texture ansatz is valid 
only at the GUT scale. The vanishing entries in the mass matrices at the GUT scale 
will be filled in by radiactive corrections at lower energy scales. 
These five solutions are summarized in Table.\ref{5zero}.
\begin{table}
\tbl{\label{5zero} Mass Texture combinations for up- and down-type quarks 
with five zeros proposed by Ramond, Roberts and Ross.}
{\begin{tabular}{@{}cccccc@{}}\toprule
 & I & II & III & IV & V\\
\hline
\\
$M_{u}$ 
&
$\left(\begin{array}{ccc}
0 & A_{u} & 0\\
A_{u} & B_{u} & 0\\
0 & 0 & D_{u}
\end{array}\right)$
& 
$\left(\begin{array}{ccc}
0 & A_{u} & 0\\
A_{u} & 0 & C_{u}\\
0 & C_{u} & D_{u}
\end{array}\right)$
&
$\left(\begin{array}{ccc}
0 & 0 & E_{u}\\
0 & B_{u} & 0\\
E_{u} & 0 & D_{u}
\end{array}\right)$
&
$\left(\begin{array}{ccc}
0 & A_{u} & 0\\
A_{u} & B_{u} & C_{u}\\
0 & C_{u} & D_{u}
\end{array}\right)$
&
$\left(\begin{array}{ccc}
0 & 0 & E_{u}\\
0 & B_{u} & C_{u}\\
E_{u} & C_{u} & D_{u}
\end{array}\right)$
\\
&&&&&
\\
$M_{d}$
& 
$\left(\begin{array}{ccc}
0 & A_{d} & 0\\
A_{d} & B_{d} & C_{d}\\
0 & C_{d} & D_{d}
\end{array}\right)$
&
$\left(\begin{array}{ccc}
0 & A_{d} & 0\\
A_{d} & B_{d} & C_{d}\\
0 & C_{d} & D_{d}
\end{array}\right)$
&
$\left(\begin{array}{ccc}
0 & A_{d} & 0\\
A_{d} & B_{d} & C_{d}\\
0 & C_{d} & D_{d}
\end{array}\right)$
&
$\left(\begin{array}{ccc}
0 & A_{d} & 0\\
A_{d} & B_{d} & 0\\
0 & 0 & D_{d}
\end{array}\right)$
&
$\left(\begin{array}{ccc}
0 & A_{d} & 0\\
A_{d} & B_{d} & 0\\
0 & 0 & D_{d}
\end{array}\right)$
\\
\botrule
\end{tabular}}
\end{table}

Lop-sided textures have also been considered in model building. In section $V$, 
we classify various $SO(10)$ models according to whether the mass textures in 
the models are symmetric or lop-sided. Symmetric textures arise if $SO(10)$ 
breaks down to the SM group with the left-right 
symmetric group as the intermediate symmetry, while lop-sided textures 
arise if the intermediate symmetry is $SU(5)$.

\subsection{Froggatt-Nielsen Mechanism and Family Symmetry}

\begin{figure}
\centerline{
\psfig{file=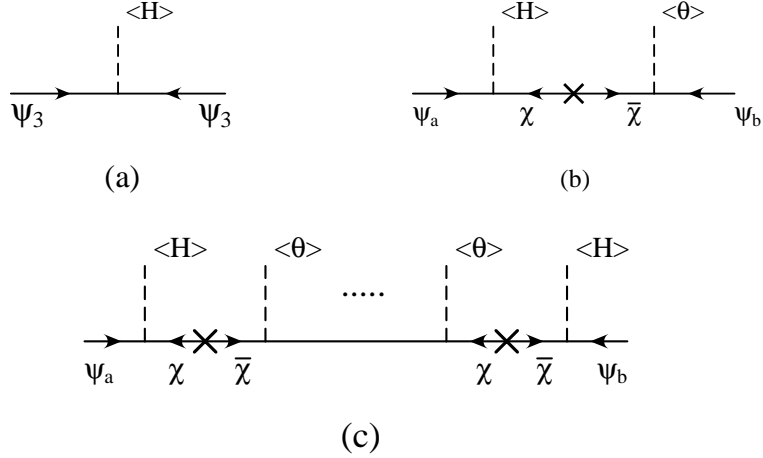,width=10cm}}
\caption{\label{fn}Schematic diagrams for Froggatt-Nielsen mechanism. Here $a$ and $b$
are the family indices. $(\chi,\overline{\chi})$ are the vector-like
Froggatt-Nielsen fields. Figure (a): The tree level diagram generating the
mass  of the third family is given; (b): The mass of the lighter
matter fields generated by this diagram is $\sim O((\frac{<\theta>}{M})^{2})$;
(c): Higher order diagrams generate mass $\sim
O((\frac{<\theta>}{M})^{n})$.}
\end{figure}

A prototype scenario which produces hierarchy in the fermion 
mass matrices is the Froggatt-Nielsen mechanism.\cite{Froggatt:1978nt} 
The idea is that the heaviest matter fields acquire their masses through
tree level interactions with the Higgs fields while masses of lighter matter
fields are produced by higher dimensional interactions involving, in addition
to the regular Higgs fields, exotic vector-like pairs of matter fields and the
so-called flavons (flavor Higgs fields). Schematic diagrams for these
interactions are shown in Fig.\ref{fn}.
After integrating out superheavy vector-like matter fields of mass $M$, 
the mass terms of the light matter fields get suppressed by a factor of
$\frac{<\theta>}{M}$, where $<\theta>$ is the VEVs of the flavons and $M$ is
the UV-cutoff of the effective theory above which the flavor symmetry is
exact. When the family symmetry is exact, only the $(33)$ entry is non-zero.
When the family symmetry is spontaneously broken, the zero entries will be
filled in at some order $O(\frac{<\theta>}{M})$. Suppose the family symmetry
allows only the $(23)$ and $(32)$ elements at order  $O(\frac{<\theta>}{M})$, 
\begin{equation} \left(\begin{array}{ccc} 0 & 0 & 0\\
0 & 0 & 0\\
0 & 0 & 1\\
\end{array}\right)
\qquad
{\mbox{SSB} \atop \longrightarrow} \qquad
\left(\begin{array}{ccc}
0 & 0 & 0\\
0 & 0 & \frac{<\theta>}{M}\\
0 & \frac{<\theta>}{M} & 1\\
\end{array}\right).
\end{equation}
Then a second fermion mass is generated at order $O((\frac{<\theta>}{M})^{2})$
after the family symmetry is spontaneous broken. The fermion mass 
hierarchy thus arises. 

To illustrate how the Froggatt-Nielsen mechanism works, suppose
there is a vector-like pair of matter fields $(\chi \oplus \overline{\chi})$
with mass $M$ and carrying the same quantum numbers as $\psi_{R}$ under the
vertical gauge group (e.g. SM or $SO(10)$), but different quantum numbers
under the family symmetry. It is therefore possible to have a Yukawa coupling
$y\overline{\chi}\psi_{L} H$ where $H$ is the SM doublet Higgs if the family
symmetry permits such a coupling. In addition, there is a gauge singlet
$\theta$ which transforms non-trivially under the family symmetry. Suppose the
coupling $y^{'}\overline{\psi}_{R}\chi\theta$ is allowed by the family
symmetry, we then obtain the following seesaw mass matrix, upon $H$ and
$\theta$ acquiring VEV's 
\begin{equation}
\left(\overline{\psi}_{R} \; \overline{\chi}\right)
\left(\begin{array}{cc}
0 & y^{'}<\theta>\\
y<H> & M
\end{array}
\right)
\left(\begin{array}{c}
\psi_{L}\\
\chi
\end{array}\right).
\end{equation}
Diagonalizing this matrix gives the following mass term for $\psi$
\begin{equation}
m_{\psi} \simeq \frac{y y^{'}<H><\theta>}{M}.
\end{equation}
So the suppression factor $\frac{<\theta>}{M}$ is due to the mixture between
the light states and the heavy states. This is very similar to how the light
neutrino masses are generated in the seesaw 
mechanism.\cite{Gell-Mann:fmass1979a,Yanagida:1979,Mohapatra:1979ia} 

So what are the possible family symmetries that can be incorporated with 
the Froggatt-Nielsen mechanism? The kinetic terms and gauge interactions 
of the SM have a very large family symmetry, 
$[U(3)]^{5}$, where the $U(3)$ factors act on the right- and
left-handed multiplets of quarks and leptons, respectively. If right-handed
neutrinos are included, the family symmetry becomes $[U(3)]^{6}$. Therefore,
any family symmetry group proposed to incorporate the Froggatt-Nielsen
mechanism must be contained in $[U(3)]^{6}$. If all the particles in 
each family are unified into one single multiplet, as in the case of 
$SO(10)$, the maximal possible family symmetry group is reduced 
to $U(3)$. The family symmetry can either be global or gauged. However, in the 
case of a global symmetry, there are problems associated with the massless 
Goldstone bosons when the symmetry is broken, and with possibly large 
gravitational quantum corrections. These problems do not arise in the case 
of gauged symmetries. In what follows, we discuss separately the abelian 
group and non-abelian group as a family symmetry. 

\subsubsection{Abelian Case}

In compactified string theories, one usually obtains aboundant 
Abelian symmetries below the compactification scale, 
in additional to the SM gauge group (or a GUT gauge group). 
Suppose that the flavon field $\theta$ and $\overline{\theta}$ carry 
$+1$ and $-1$ charges under $U(1)$, and the ratios $\frac{<\theta>}{M}$ and 
$\frac{<\overline{\theta}>}{M}$ 
have approximately the value of the Cabbibo angle $0.22$
(or equivalently, the parameter $\lambda$ in the 
Wolfenstein parameterization). By assigning different $U(1)$ 
charges to different family, one in general, obtains a mass 
matrix of the form
\begin{equation}
\left(\begin{array}{ccc}
\lambda^{n_{11}} & \lambda^{n_{12}} & \lambda^{n_{13}}\\
\lambda^{n_{21}} & \lambda^{n_{22}} & \lambda^{n_{23}}\\
\lambda^{n_{31}} & \lambda^{n_{32}} & \lambda^{n_{33}}\\
\end{array}
\right).
\end{equation}
The exponent $n_{ij}$ is given by $Q_{L,i}+Q_{R,j}$, where 
$Q_{L,i} \; (Q_{R,i})$ is 
the $U(1)$ charge of the left-handed (right-handed) field of the $i$-th family.
It is usually not difficult to find solutions for the charge assignments 
that are consistent with experiments, and the solutions are not unique. 
An interesting model for both quarks and leptons based on anomalous  
$U(1)_{H}$ was proposed by Irges, Lavigna and Ramond which makes 
use of the Green-Schwartz anomaly cancellation condition to constrain  
the $U(1)$ charge assignments.\cite{Irges:1998ax}  
Due to the Abelian nature, models of this type have limited predictive power, 
because only the powers of the small expansion parameter, $\lambda$, are 
determined while the relative strengths between different entries are left 
un-determined. As a consequence, every entry in the above mass matrix has 
an un-known $\mathcal{O}(1)$ coefficient associated with it. 

Non-anomalous $U(1)_{H}$ has also been considered in a model constructed 
by Mira, Nardi and Restrepo.\cite{Mira:1999fx} In this model, anomaly 
cancellation condition leads to a massless up-quark, thus solving the strong 
CP problem.

\subsubsection{Non-Abelian Case Based on $SU(2)$}

Models with non-Abelian family symmetry generally have more predictive power 
because the relative strengths between different matrix elements can be 
determined. The original motivation of using non-Abelian group as the family 
symmetry is to solve the SUSY flavor problem. $SU(2)$ was proposed by 
Barbieri {\it et al}~\cite{Barbieri:1996ww} as a family symmetry.  
It has two attractive features: (i) As we have seen previously, 
the constraints from the SUSY FCNC  requires that
\begin{equation} \frac{m_{2}^{2}-m_{1}^{2}}{m^{2}} \le
\frac{10^{-3}}{\sin\phi} (\frac{m}{300 \mbox{GeV}}) 
\end{equation}
where $(m_{3}^{2}-m_{1,2}^{2}) \sim m^{2}$ is the average scalar
mass squared and $\phi$ is some relevant CP phase. $SU(2)$ gives rise to the
degeneracies between 1-2 families needed to suppress the supersymmetric FCNC
in the squark sector; (ii) A multi-step breaking of $SU(2)$ gives rise to the 
observed inter-family hierarchy naturally. Unlike models based on the 
$U(1)$ family symmetry, in which one has the freedom in choosing $U(1)$
charges for various matter fields, a $SU(2)$ family symmetry appears to be a
much more constrained framework for constructing realistic models. 

The heaviness of the top-quark and suppression of 
the SUSY FCNC together suggest that the three families 
of matter fields transform under a $SU(2)$ family symmetry as 
\begin{equation}  
\psi_{a} \oplus \psi_{3} = 2 \oplus 1
\end{equation}
where $a =1,2$ and the subscripts refer to family indices. In the symmetric
limit, only the third family of matter fields have non-vanishing masses. 
This can be understood easily since the third family of matter
fields have much higher masses compared to the other two families of matter
fields. $SU(2)$ breaks down in two steps:
\begin{equation}
\label{eq:steps} SU(2) \stackrel{\epsilon M}{\longrightarrow} 
U(1) \stackrel{\epsilon' M}{\longrightarrow}
nothing
\end{equation}
with $\epsilon' \ll \epsilon \ll 1$ and $M$ is the UV cut-off of the 
effective theory mentioned before. These small
parameters $\epsilon$ and $\epsilon'$ are the ratios of the vacuum expectation
values of the flavon fields to the cut-off scale. Note that because 
\begin{equation} 
\psi_{3}\psi_{3} \sim 1_{S},\qquad \psi_{3}\psi_{a} \sim 2, \qquad  
\psi_{a}\psi_{b} \sim 2 \otimes 2 = 1_{A} \oplus 3 
\end{equation}
the only relevant flavon fields are in the $1_{A}, 2$ and $3$
dimensional representations of $SU(2)$, namely, 
\begin{equation}
A^{ab} \sim 1_{A}, \qquad \phi^{a} \sim 2, \qquad S^{ab} \sim 3.
\end{equation}
So a generic mass matrix constrained by $SU(2)$ family symmetry 
is of the following form:
\begin{equation}
\left(\begin{tabular}{c|c}
&\\
$<S>$, $<1_{A}>$  & $<\phi>$\\
&\\
\hline
$<\phi>^{T}$ & 1
\end{tabular}
\right).
\end{equation}

To see how the vacuum alignment in the flavon sector is achieved, let us first
consider the supersymmetric limit with only one conjugate pair of doublets 
$(\phi \oplus \overline{\phi})$, anti-symmetric singlets 
$(A \oplus \overline{A})$ and triplets $(S \oplus
\overline{S})$ of $SU(2)$. The most general renormalizable 
superpotential is then given by\cite{Barbieri:1998em}
\begin{equation}
W_{\mbox{flavon}}(\phi,\overline{\phi},S,\overline{S}) =
\phi \overline{S}\phi + \overline{\phi}S\overline{\phi} 
+X_{\phi} \phi\overline{\phi} + X_{s}S\overline{S} 
+ X_{A} A \overline{A} 
\end{equation}
where $X_{s}$, $X_{\phi}$ and $X_{A}$ are dimensionful parameters in the
superpotential. Note that the anti-symmetric singlet fields $A$ and $\overline{A}$ 
are decoupled from other fields. From the F-flat conditions, one obtains the 
following solutions,
\begin{eqnarray}
<S_{ab}> & = & -<\phi_{a}><\phi_{b}>/X_{s}
\\
\sum_{a} \; <\phi_{a}\overline{\phi}_{a}> & = & \frac{1}{2}(X_{s}X_{\phi})
\\
X_{A} A^{ab} & = & 0.
\end{eqnarray}
Thus the relative strengths of $<S>$ and $<\phi>$ are determined. For $X_{A} \ne 0$, 
the F-flat conditions imply $<A> = <\overline{A}> = 0$.
Non-vanishing $<A>$ and $<\overline{A}>$ can be obtained if non-renormalizable 
operators are introduced.\cite{Barbieri:1998em} 

If all the $16$ observed matter fields in one family form a single representation 
as in the case of $SO(10)$, the most general effective superpotential, 
after integrating out all the heavy
Froggatt-Nielsen fields, that generates fermion masses for a $SO(10) \times
SU(2)$ model has the following very simple form 
\begin{equation} 
W = H(\psi_{3}
\psi_{3} + \psi_{3} \frac{\phi^{a}}{M} \psi_{a} 
+ \psi_{a}\frac{S^{ab}}{M} \psi_{b} + \psi_{a}\frac{A^{ab}}{M} \psi_{b}). 
\end{equation}
In a specific $SU(2)$ basis, 
\begin{equation}
\label{genU2}
\frac{ \left< \phi \right> }{M} \sim O \left( \begin{array}{c}
\epsilon'\epsilon\\
\epsilon
\end{array} \right),
\qquad
\frac{ \left< A^{ab} \right> }{M} \sim O \left( \begin{array}{cc}
0 & -\epsilon' \\
\epsilon' & 0
\end{array} \right),
\qquad
\frac{ \left< S^{ab} \right> }{M} \sim O \left( \begin{array}{cc}
\epsilon'^{2} & \epsilon'\epsilon\\
\epsilon'\epsilon & \epsilon
\end{array} \right).
\end{equation}
Here we have indicated the VEVs \textit{all} the flavon fields could acquire
for symmetry breaking in Eq.(\ref{eq:steps}). 
The mass matrix $\widetilde{M}$ takes the following form
\begin{equation}
\widetilde{M} \sim \mathcal{O} \left( \begin{array}{ccc}
\epsilon'^{2} & \epsilon' & \epsilon'\epsilon \\
-\epsilon' & \epsilon & \epsilon \\
\epsilon'\epsilon & \epsilon & 1 
\end{array} \right).
\end{equation}
The hierarchy is thus built into this mass matrix.  

\subsubsection{Non-Abelian Case Based on $SU(3)$}

Ultimately, the $SU(2)$ family symmetry can be embedded into $SU(3)$, under 
which the three families form a triplet. A model based on $SU(3)$ 
is presented in Ref. 18. 
The three families form a $SU(3)$ triplet 
before the symmetry is broken. The symmetry breaking takes place at two steps
\begin{displaymath}
SU(3) \longrightarrow SU(2) \longrightarrow {\mbox nothing}.
\end{displaymath}
The $SU(3)$ anti-triplet flavon fields, $\overline{\phi}_{3}$ and 
$\overline{\phi}_{23}$, acquire VEV's along the following directions, 
breaking the $SU(3)$ symmetry
\begin{equation}\label{vev1}
<\overline{\phi}_{3}> = \left(\begin{array}{c}
0 \\ 0 \\ a_{3}
\end{array}\right), \qquad
<\overline{\phi}_{23}> = \left(\begin{array}{c}
0 \\ b \\ b
\end{array}\right)
\end{equation}
where $<\overline{\phi}_{3}>$ triggers the first stage of breaking, and 
$<\overline{\phi}_{23}>$ triggers the second stage of breaking. 
The leading order Yukawa couplings of the matter fields are
\begin{equation}\label{leading}
H(\frac{1}{M_{3}^{2}} \psi_{i}\overline{\phi}_{3}^{i}\psi_{j}^{c}
\overline{\phi}_{3}^{j}
+ \frac{1}{M_{23}^{2}} \psi_{i}\overline{\phi}_{23}^{i}\psi_{j}^{c}
\overline{\phi}_{23}^{j}).
\end{equation}
This gives rise to a mass matrix of the following form
\begin{equation}
\left(\begin{array}{ccc}
0 & 0 & 0\\
0 & \frac{b^{2}}{M_{23}^{2}} & \frac{b^{2}}{M_{23}^{2}}\\
0 & \frac{b^{2}}{M_{23}^{2}} & \frac{a_{3}^{2}}{M_{3}^{2}} 
+ \frac{b^{2}}{M_{23}^{2}}
\end{array}
\right) \sim 
\left(\begin{array}{ccc}
0 & 0 & 0\\
0 & \epsilon^{2} & \epsilon^{2}\\
0 & \epsilon^{2} & 1
\end{array}\right).
\end{equation}
Assuming $M_{3} \simeq a_{3} \gg M_{23} \gg b$ and $\epsilon=b/M_{23}$,
one thus obtains a hierarchical structure in the $(2,3)$ block. 
There are also operators that mix $\overline{\phi}_{3}$ with $\overline{\phi}_{23}$,
\begin{equation}\label{mix}
\frac{\epsilon^{2}}{M_{23}M_{3}} \; H
(\psi_{i}\overline{\phi}_{23}^{i}\psi_{j}^{c}\overline{\phi}_{3}^{j}
+\psi_{i}\overline{\phi}_{3}^{i}\psi_{j}^{c}\overline{\phi}_{23}^{j}).
\end{equation}
These operators give rise to contributions to the $(23)$ and $(32)$ matrix 
elements of order $\mathcal{O}(\epsilon)$, which is larger than the $(22)$ 
element of order $\mathcal{O}(\epsilon^{2})$, leading to a wrong prediction 
for $V_{cb}$. One way to suppress  these operators is to impose a discrete 
$Z_{2}$ symmetry, under which $\overline{\phi}_{3}$ and $\overline{\phi}_{23}$ 
have opposite parity. Thus the operators given in Eq.(\ref{leading}) are allowed 
by the $Z_{2}$ symmetry while the operators given in Eq.(\ref{mix}) only arise 
at higher order with a suppression factor $\epsilon^{2}$. To fill in the first 
row and column, one has to introduce additional $SU(3)$ triplet flavon fields, 
$\phi_{3}$ and $\phi_{23}$, which acquire VEV's along the following directions
\begin{equation}\label{vev2}
<\phi_{3}> = \left(\begin{array}{c}
0 \\ 0 \\ a_{3}
\end{array}\right), \qquad
<\phi_{23}> = \left(\begin{array}{c}
0 \\ b \\ -b
\end{array}\right)
\end{equation}
and consider the following operators
\begin{eqnarray}
& \frac{\epsilon^{2}}{M_{23}} (\epsilon^{ijk} \psi_{i}
\phi_{23,j} \psi_{k}^{c}) H \hspace{3.3cm}
\label{13op}\\
& \frac{\epsilon^{6}}{M_{3}^{2}M_{23}^{2}}
(\epsilon^{ijk} \psi_{i}\phi_{3,j}\phi_{23,k})
(\epsilon^{lmn}\psi_{l}^{c}\phi_{3,m}\phi_{23,n}) H.
\label{11op}
\end{eqnarray}
The operators given in Eq.(\ref{13op}) generate the entries $(12) = (13) = -(21) = -(31)$ of 
order $\mathcal{O}(\epsilon^{3})$. These entries are anti-symmetric due to the presence of 
the anti-symmetric tensor $\epsilon^{ijk}$ in the couplings. The operators given in 
Eq.(\ref{11op}) generate the $(11)$ matrix element of order $\mathcal{O}(\epsilon^{8})$. 
The suppression in this operator is due to the presence of  
the $Z_{2}$ symmetry discussed above and an additional $R$-symmetry; 
these symmetries also forbid all the operators which 
could lead to un-realistic predictions. The operators given in Eq.(\ref{leading}), 
(\ref{mix}), (\ref{13op}) and (\ref{11op}) together give rise to 
a Yukawa matrix of the form, in the leading order of the expansion parameter, 
$\epsilon$,
\begin{equation}
\left(\begin{array}{ccc}
\mathcal{O}(\epsilon^{8}) & \lambda \epsilon^{3} & \lambda \epsilon^{3}\\
-\lambda \epsilon^{3} & \epsilon^{2} & \epsilon^{2}\\
-\lambda \epsilon^{3} & \epsilon^{2} & 1
\end{array}\right)
\end{equation}
which can accommodate realistic fermion masses and mixing angles.
The vacuum alignment leading to Eq.(\ref{vev1}) and (\ref{vev2}) 
is discussed in detail in Ref.18.

We comment that the absolute mass scale of the family symmetry, $M$, 
cannot be determined in the Froggatt-Nielsen type of scenario, 
because it is the ratio, $(\epsilon, \epsilon^{'})$, rather than the
absolute mass scale, $M$, that is phenomenologically relevant. Some attempts
have been made by having the SUSY breaking messenger fields play also the
role of Froggatt-Nielsen fields such that the family symmetry scale is linked
to the SUSY breaking scale. Though attractive, these models have difficulties
getting low SUSY breaking scale.\cite{Arkani-Hamed:1996xm}  

\subsection{Ideas from Extra Dimensions}

\subsubsection{Split Fermion Scenario in Factorizable Geometry}

It has been proposed that the existence of extra compact spatial 
dimensions could account for the large hierarchy between 
the Planck scale and the electroweak scale. Arkani-Hamed, Dimopoulos 
and Dvali\cite{Arkani-Hamed:1998rs} showed that extra dimensions of 
size $\sim 1/TeV$ may provides a solution to the gauge hierarchy problem.

Based on this framework, Arkani-Hamed and Schmaltz\cite{Arkani-Hamed:1999dc} 
proposed a non-supersymmetric model in which the mass hierarchy is 
generated by localizing zero modes 
of the weak doublet and singlet fermions at different locations. 
Note that this mechanism works despite of the size of the extra dimensions. 
Consider a chiral fermion 
$\Psi$ in $5D$. The action for $\Psi$ coupled to a background 
scalar field $\Phi$ is given by
\begin{equation}\label{split1}
S = \int d^{4} dy \overline{\Psi} \left[ 
i \gamma^{\mu} \partial_{\mu} + i\gamma^{5} \partial_{5} + \Phi(y) \right] \Psi.
\end{equation}
The chiral fermion $\Psi(x,y)$ can be expanded in the product basis
\begin{equation}
\Psi(x,y) = \sum_{n} <y|L,n> P_{L} \psi_{n}(x) + <y|R,n> P_{R} \psi_{n}(x).
\end{equation}
Here $|L,n>$ and $|R,n>$ satisfy the following equations
\begin{eqnarray}
aa^{\dagger} |L,n> & = & (-\partial_{5}^{2} + \Phi^{2} + \dot{\Phi})|L,n> 
= \mu^{2}_{n} |L,n>
\\
a^{\dagger}a |R,n> & = & (-\partial_{5}^{2} + \Phi^{2} - \dot{\Phi})|R,n> 
= \mu_{n}^{2} |R,n>,
\end{eqnarray}
where $\dot{\Phi} \equiv \partial_{5}\Phi$ and 
\begin{equation}
a = \partial_{5} + \Phi(y), \qquad a^{\dagger} = -\partial_{5} + \Phi(y).
\end{equation}
For a special choice of linearized background field, $\Phi(y) = 2\mu^{2}y$ where 
$\mu$ is a constant of mass dimension one, the operator $a$ and $a^{\dagger}$ 
become the usual creation and annihilation operators of a simple 
harmonic oscillator. In what follows, we will concentrate on 
this special case. Expanding the $5D$ action in terms of $|L,n>$ and 
$|R,n>$, and then integrating out $y$, we obtain the $4D$ effective action
\begin{equation}
S = \int d^{4}[\overline{\psi}_{L} i \gamma^{\mu} \partial_{\mu} P_{L} \psi_{L} 
+ \overline{\psi}_{R} i \gamma^{\mu} \partial_{\mu} P_{R} \psi_{R}
+ \sum_{n=1}^{\infty} \overline{\psi}_{n} (i\gamma^{\mu}\partial_{\mu} + \mu_{n}) 
\psi_{n} ],
\end{equation}
where the first two terms are kinetic terms of the chiral zero modes  
whose wave functions are Gaussian centered at $y=0$,
\begin{equation}
<y|L,0> = \frac{\mu^{1/2}}{(\pi/2)^{1/4}}e^{-\mu^{2}y^{2}}.
\end{equation}
In general, there is a bulk mass term, $m\overline{\Psi}\Psi$, 
in the $5D$ action, Eq.(\ref{split1}), because mass terms for $5$-dimensional 
field are allowed by all symmetries. In this case, instead of being centered 
at $y=0$, the Gaussian wave function is centered at $y=m/2\mu$. 
Thus by having different bulk mass term, $m$, different bulk fields 
can be localized at different $4D$ slices in the $5D$ bulk.

For simplicity, the wave function of the Higgs doublet is assumed to have  
constant spread along the fifth dimension. In this case, the Yukawa coupling is 
\begin{equation}
S_{yuk} = y_{ij} \int d^{4} x H(x) f_{i}(x) \tilde{f}_{j}^{c}(x) 
\int dy \phi_{f_{i}}(y) \phi_{\tilde{f}^{c}_{j}}(y),
\end{equation} 
where $\phi_{f_{i}}(y)$ and $\phi_{\tilde{f}^{c}_{j}}(y)$ are the zero 
mode wave functions of the $SU(2)$ doublet, $f_{i}$, 
and singlet, $\tilde{f}_{j}^{c}$, respectively. 
Integrating out the $y$ coordinate, the effective Yukawa coupling is then given by
\begin{equation}
y_{ij} \int dy \phi_{f_{i}} (y) \phi_{\tilde{f_{j}}^{c}} (y) = y_{ij}
\frac{\sqrt{2}\mu}{\sqrt{\pi}} 
\int dy e^{-\mu^{2}(y-r_{f_{i}})^{2}} e^{-\mu^{2}(y-r_{\tilde{f}^{c}_{j}})^{2}}
= e^{-\mu^{2}r_{ij}^{2}/2},
\end{equation}
where $r_{f_{i}}=m_{f_{i}}/2\mu$ and $r_{\tilde{f}^{c}_{j}}
=m_{\tilde{f}^{c}_{j}}/2\mu$ 
are the locations at which the Gaussian wave functions 
$\phi_{f_{i}}$ and $\phi_{\tilde{f}^{c}_{j}} (y)$ are centered, and 
$r_{ij} = |r_{\tilde{f}^{c}_{j}}-r_{f_{i}}|$ is the ``distance'' between the peaks 
of these two wave functions. Thus the large hierarchy among fermion masses can be 
generated by having different $m_{i}$. 

Variations based on this mechanism have been investigated in 
Ref. 22, 23, 24. 

\subsubsection{Non-Factorizable Geometry}

Randall and Sundrum\cite{Randall:1999ee}  
proposed an alternative based on non-factorizable 
geometry from which the gauge hierarchy can be derived. In their 
original proposal, only gravity propagates in the bulk. If the SM Higgs 
doublet is confined to the TeV brane, which is required if the gauge hierarchy 
is assumed to arise from the warped geometry, while all other SM particles 
are allowed to propagate in the bulk, it is possible to understand the fermion 
mass hierarchy in this setup. The equations of motion for various bulk 
fields are given in the following compact 
form\cite{Grossman:1999ra,Gherghetta:2000qt}
\begin{equation}\label{geneom}
(e^{2\sigma}\eta^{\mu\nu}\partial_{\mu}\partial_{\nu}
+ e^{s\sigma}\partial_{5}(e^{-s\sigma}\partial_{5}) - M_{\Phi}^{2})
\Phi(x^{\mu},y) = 0,
\end{equation}
where for $\Phi=(\phi,e^{-2\sigma}\Psi_{L,R})$ we have 
$M_{\Phi}^{2}=(ak^{2}+b\sigma^{''}(y),C(C \pm 1)k^{2} 
\pm C\sigma^{''}(y))$ and $s=(4,1)$, where $a$ and $C$ are bulk mass terms 
of the scalar and fermionic fields, and $b$ is a boundary mass term for the scalar 
field. The field $\Phi(x^{\mu},y)$ is decomposed into an infinite sum of
Kaluza-Kline (KK) modes as follows,
\begin{equation}\label{KKdecompose}
\Phi(x^{\mu},y)=\frac{1}{\sqrt{2\pi R}} \sum_{n} 
\Phi_{(n)} (x^{\mu}) f_{n}(y).
\end{equation}
The profile of the n-th mode, $f_{n}(y)$, satisfies
\begin{equation}\label{eomfn}
(-e^{s\sigma}\partial_{5}(e^{-s\sigma} \partial_{5})+\hat{M}_{\Phi}^{2})
f_{n}(y) = e^{2\sigma} m_{n}^{2} f_{n}(y),
\end{equation}
where $\hat{M}_{\Phi}^{2} = (ak^{2}, C(C \pm 1)k^{2})$. 
$m_{n}$ is the mass of the $n$-th KK
mode. The solution for the zero modes of the bulk spin-$1/2$ fields are found to be
\cite{Grossman:1999ra,Gherghetta:2000qt}  
\begin{equation}\label{LH}
f_{0}(y) = \frac{1}{N_{0}} e^{-c\sigma},
\end{equation}
where $c=C$ for left-handed fermions and $c=-C$ for right-handed fermions. 
The normalization constant $1/N_{0}$ is given by
\begin{equation}
\frac{1}{N_{0}^{2}} = \frac{(1-2c) \pi k R}{e^{(1-2c)\pi k R} -1}.
\end{equation}
Thus the bulk fermion can be decomposed into
\begin{equation}
\Psi_{L}(x^{\mu},y) = e^{2\sigma} \Phi(x^{\mu},y) 
= \frac{1}{\sqrt{2\pi R}} \frac{1}{N_{0}} e^{(2-c)\sigma} 
\Phi_{L(0)}(x^{\mu}) + \cdots.
\end{equation}
The Yukawa coupling for the charged fermions to the Higgs doublet reads
\begin{equation}\label{yuk}
\frac{Y_{ij}}{M_{pl}} \int d^{4}x \int dy \sqrt{-g}
\overline{\Psi}_{R,i}(x,y) \Psi_{L,j}(x,y) H(x) \delta(y-\pi R),
\end{equation}
where $Y_{ij}$ are dimensionless $\mathcal{O}(1)$ coefficients. 
The effective Yukawa coupling in four dimensions is obtained after
integrating out the fifth coordinate, $y$:
\begin{eqnarray}
\tilde{Y}_{ij} & = & \frac{Y_{ij}}{M_{pl}}
(\frac{1}{2\pi R})
\left[
\frac{(1-2c_{R,i})\pi kR}{e^{(1-2c_{R,i})\pi kR} -1 }
\right]^{1/2}
\left[
\frac{(1-2c_{L,j})\pi kR}{e^{(1-2c_{L,j})\pi kR} -1 }
\right]^{1/2}
\nonumber\\
& & \qquad \quad \cdot 
\int_{-\pi R}^{\pi R} dy \sqrt{-g} \;
e^{(2-c_{R,i})\sigma} e^{(2-c_{L,j})\sigma}
e^{\sigma} \delta(y-\pi R)
\nonumber\\
& = & \frac{Y_{ij}}{2} \frac{k}{M_{pl}} 
\frac{\sqrt{(1-2c_{R,i})(1-2c_{L,j})}}
{\sqrt{(e^{(1-2c_{R,i})\pi kR}-1)(e^{(1-2c_{L,i})\pi kR}-1)}}
e^{(1-c_{R,i} - c_{L,j})\pi k R}.
\end{eqnarray}
Thus by choosing $c_{L,i}$ and $c_{R,i}$ all of $\mathcal{O}(1)$, we can reproduce 
the observed mass hierarchy and mixing. We note that this mechanism is not predictive  
in the sense that it does not reduce the number of parameters; 
the virtue of this mechanism is that the large hierarchy observed in fermion 
masses arises from parameters all of $\mathcal{O}(1)$. A configuration that 
reproduce the observed mass hierarchy has been found by Huber 
and Shafi.\cite{Huber:2000ie}

\section{Lepton Masses and Mixing}\label{lmass}

If neutrinos are massive, the mixing arises in the leptonic 
charged current interaction, 
\begin{equation} 
\mathcal{L}_{cc} = \frac{g}{\sqrt{2}} U_{LM}^{\dagger} (W_{\mu}^{+} 
\overline{\nu}_{L} \gamma_{\mu} E_{L}) + h.c..  
\end{equation}
The leptonic mixing (LM) matrix,\cite{pontecorvo:1967,Maki:mu,Lee:1977qz,Lee:1977ti}
$U_{LM}$,\footnote{The LM matrix is sometimes called Pontecorvo-Maki-Nakagawa-Sakata 
(PMNS) or Maki-Nakagawa-Sakata (MNS) matrix. It was first discussed, 
in a two flavor case, by Pontecorvo\cite{pontecorvo:1967} and by Maki, 
Nakagawa and Sakata.\cite{Maki:mu} The mixing matrix with $3$ flavors  
was first discussed by Lee, Pakvasa, Shrock and Sugawara.\cite{Lee:1977qz,Lee:1977ti}} 
is obtained by diagonalizing the Yukawa matrix of 
the charged leptons and the effective neutrino mass 
matrix, assuming neutrinos are Majorana 
particles,\footnote{If neutrinos are Dirac particles, 
Eq.(\ref{nudiag}) becomes 
\begin{displaymath}
M_{\nu}^{diag} =  V_{e_{L}} M_{\nu}^{eff} V_{\nu_{R}}^{\dagger}
= diag(m_{\nu_{1}},m_{\nu_{2}},m_{\nu_{3}}).\end{displaymath}}
\begin{eqnarray}
Y_{e}^{diag} & = & V_{e_{R}} Y_{e} V_{e_{L}}^{\dagger} 
= diag(y_{e},y_{\mu},y_{\tau})\\
M_{\nu_{LL}}^{diag} & = & V_{\nu_{LL}} M_{\nu_{LL}}^{eff} V_{\nu_{LL}}^{T}
= diag(m_{\nu_{1}},m_{\nu_{2}},m_{\nu_{3}}),\label{nudiag}
\end{eqnarray}
where $V_{\nu_{LL}}$ is an orthogonal matrix, 
and it can be parameterized as a product of a CKM-like mixing matrix, 
which has three mixing angles and one CP violating phase, with 
a diagonal phase matrix,  
\begin{eqnarray}
& U_{LM} \equiv V_{e_{L}}V_{\nu_{LL}}^{\dagger} \hspace{10cm}\nonumber\\
& \simeq 
\left(
\begin{array}{ccc}
c_{12}^{l} c_{13}^{l} & s_{12}^{l} c_{13}^{l} & s_{13}^{l} e^{-i\delta_{l}}\\
-s_{12}^{l} c_{23}^{l}-c_{12}^{l}s_{23}^{l}s_{13}^{l}e^{i\delta_{l}} &
c_{12}^{l}c_{23}^{l}-s_{12}^{l}s_{23}^{l}s_{13}^{l}e^{i\delta_{l}} &
s_{23}^{l}c_{13}^{l}\\
s_{12}^{l}s_{23}^{l}-c_{12}^{l}c_{23}^{l}s_{13}^{l}e^{i\delta_{l}} &
-c_{12}^{l}s_{23}^{l}-s_{12}^{l}c_{23}^{l}s_{13}^{l}e^{i\delta_{l}} &
c_{23}^{l}c_{13}^{l}
\end{array}
\right)
\cdot \left(
\begin{array}{ccc}
1 & &\\
& e^{i \frac{\alpha_{21}}{2}} & \\
& & e^{i\frac{\alpha_{31}}{2}}
\end{array}
\right),\nonumber\\
\end{eqnarray}
which relates the neutrino mass eigen states to the flavor eigenstates by
\begin{eqnarray}
|\nu_{e}> & = & 
U_{e\nu_{1}} |\nu_{1}> + U_{e\nu_{2}} |\nu_{2}> + U_{e\nu_{3}} |\nu_{3}>
\\
|\nu_{\mu}> & = &
U_{\mu\nu_{1}} |\nu_{1}> + U_{\mu\nu_{2}} |\nu_{2}> + U_{\mu\nu_{3}} |\nu_{3}>
\\
|\nu_{\tau}> & = & 
U_{\tau\nu_{1}} |\nu_{1}> + U_{\tau\nu_{2}} |\nu_{2}> + U_{\tau\nu_{3}} |\nu_{3}>.
\end{eqnarray}
Note that the Majorana condition,
\begin{equation}
C(\overline{\nu}_{j})^{T} = \nu_{j},
\end{equation}
where $C$ is the charge conjugate operator, forbids the rephasing of the
Majorana fields. Therefore, we can only remove $3$ of the $6$ phases
present in the unitary matrix $U_{LM}$ by redefining the 
charged lepton fields and are left with three CP violating phases in the leptonic 
sector, if neutrinos are Majorana 
particles.\cite{Bilenky:1980cx,Schechter:1980gr,Doi:1980yb}
Thus $U_{LM}$ can be parameterized as a product of a unitary matrix, analogous 
to the CKM matrix which has one phase (the so-called universal phase), $\delta_{l}$,
and a diagonal phase matrix which contains two phases 
(the so-called Majorana phases), $\alpha_{21}$ and $\alpha_{31}$.
The leptonic analog of the Jarlskog invariant, which measures the
CP violation due to the universal phase, is given by
\begin{equation}
J_{CP}^{l} \equiv Im\{ U_{\mu \nu_{2}} U_{e\nu_{3}} 
U_{\mu \nu_{3}}^{\ast} U_{e\nu_{2}}^{\ast} \}.
\end{equation}
For the Majorana phases, the rephasing invariant CP violation measures are
\cite{Nieves:1987pp}
\begin{equation}
S_{1} \equiv Im\{ U_{e\nu_{1}} U_{e\nu_{3}}^{\ast} \}, \qquad
S_{2} \equiv Im\{ U_{e\nu_{2}} U_{e\nu_{3}}^{\ast} \}.
\end{equation}
From $S_{1}$ and $S_{2}$, one can then determine the Majorana phases
\begin{eqnarray}
\cos \alpha_{31} = 1 - 2 \frac{S_{1}^{2}}{\vert U_{e\nu_{1}} \vert^{2}
\vert U_{e\nu_{3}} \vert^{2}}
\\
\cos (\alpha_{31} - \alpha_{21})
= 1 - 2 \frac{S_{2}^{2}}{\vert U_{e\nu_{2}} \vert^{2}
\vert U_{e\nu_{3}} \vert^{2}}.
\end{eqnarray}

The recently reported measurements from KamLAND reactor 
experiment\cite{Eguchi:2002dm} confirm the large mixing 
angle (LMA) solution to be the unique oscillation 
solution to the solar neutrino problem 
at $4.7 \; \sigma$ level.\cite{Maltoni:2002aw,Bahcall:2002ij,Fogli:2002au} 
The global analysis including Solar + KamLAND + CHOOZ\cite{Apollonio:1999ae}
indicate the following allowed region at 
$3\sigma$,\cite{Maltoni:2002aw} 
\begin{eqnarray}
5.1 \times 10^{-5} < & \Delta m_{21}^2 & < 9.7 \times 10^{-5} eV^{2}
\\
0.29 \le & \tan^{2}\theta_{12} & \le 0.86.
\end{eqnarray}
The allowed regions at $3\sigma$ level based on a global fit including 
SK\cite{Fukuda:2000np} 
+ Solar + CHOOZ for the atmospheric parameters 
and the CHOOZ angle are\cite{Gonzalez-Garcia:2002dz}
\begin{eqnarray}
1.4 \times 10^{-3} < & \Delta m_{32}^2 & < 6.0 \times 10^{-3} eV^{2}
\\
0.4 \le & tan^{2} \theta_{23} & \le 3.0\\
& \sin^{2} \theta_{13} & < 0.06.
\end{eqnarray}
And the magnitudes of $U_{LM}$ elements at $1 \sigma$ ($3 \sigma$) are given 
by\cite{Gonzalez-Garcia:2003qf}
\begin{equation}
|U_{LM}| = \left(
\begin{array}{ccc}
(0.73) 0.79 - 0.86 (0.88) & (0.47) 0.50 - 0.61 (0.67) & 0 - 0.16 (0.23)\\
(0.17) 0.24 - 0.52 (0.57) & (0.37) 0.44 - 0.69 (0.73) & (0.56) 0.63 - 0.79 (0.84)\\
(0.20) 0.26 - 0.52 (0.58) & (0.40) 0.47 - 0.71 (0.75) & (0.54) 0.60 - 0.77 (0.82)
\end{array}
\right)
\end{equation} 

To see what a bi-large mixing means, let us assume, to a good approximation, 
that $\theta_{13}=\eta \ll 1$. In this case, the LM matrix reads,
\begin{equation}
U_{LM} = \left(\begin{array}{ccc}
c_{12} & s_{12} & \eta \\
-\frac{(s_{12}+\eta c_{12})}{\sqrt{2}} 
& \frac{(c_{12}-\eta s_{12})}{\sqrt{2}} & 1/\sqrt{2}\\
\frac{(s_{12}-\eta c_{12})}{\sqrt{2}} & -\frac{(c_{12}+\eta s_{12})}{\sqrt{2}} 
& 1/\sqrt{2}
\end{array}
\right),
\end{equation}
which in turn implies the neutrino mass eigenstates given in terms of 
flavor eigenstates as
\begin{eqnarray}
|\nu_{1}> & = & 
c_{12} |\nu_{e}> - \frac{1}{\sqrt{2}}s_{12} 
( \; |\nu_{\mu}> \; - \; |\nu_{\tau}> \; ) 
- \frac{\eta}{\sqrt{2}} c_{12} ( \; |\nu_{\mu}> \; + \; |\nu_{\tau}> \; )
\\
|\nu_{2}> & = & 
s_{12} |\nu_{e}> + \frac{1}{\sqrt{2}} c_{12} 
( \; |\nu_{\mu}> \; - \; |\nu_{\tau}> \; )
-\frac{\eta}{\sqrt{2}} s_{12} ( \; |\nu_{\mu}> \; + \; |\nu_{\tau}> \; )
\\
|\nu_{3}> & = & \frac{1}{\sqrt{2}} ( \; |\nu_{\mu}> \; + \; |\nu_{\tau}> \; ) 
+ \eta |\nu_{e}>.
\end{eqnarray}

\subsection{Generation of Small Neutrino Masses}

\subsubsection{Small Neutrino Masses from See-saw Mechanism}

The observation of neutrino oscillations provides the first indication 
of beyond the Standard Model physics. It has two implications: neutrinos 
have non-zero masses, and lepton family numbers are violated. 
In the SM, neutrinos are massless because there are no $SU(2)$ singlet 
neutrinos nor are there $SU(2)$ triplet Higgss. Adding one $SU(2)$ 
singlet neutrino for each family is the simplest way to introduce neutrino 
masses. Because the right-handed neutrinos are SM singlets, the Majorana 
mass terms for the $SU(2)$ singlet neutrinos are not forbidden by the symmetry. 
In the Lagrangian, there are Dirac mass term for the neutrinos and the 
right-handed Majorana mass 
terms,\cite{Gell-Mann:fmass1979a,Yanagida:1979,Mohapatra:1979ia} 
\begin{eqnarray}
\mathcal{L}_{seesaw} & = &  
-M_{\nu_{LR}}\overline{\nu}_{R}\nu_{L}
- \frac{1}{2}M_{\nu_{RR}}\overline{\nu}_{R}^{T}\overline{\nu}_{R} + h.c. 
\nonumber\\
& = & -\frac{1}{2}
\left(\begin{array}{cc}\nu_{L} & \overline{\nu}_{R}
\end{array}\right)
\left(\begin{array}{cc}
0 & M_{LR}^{T}\\
M_{LR} & M_{RR}
\end{array}\right)
\left(\begin{array}{c}
\nu_{L} \\ \overline{\nu}_{R}
\end{array}\right).
\end{eqnarray}
(In general, there could be a non-vanishing mass term for 
$(\nu_{L}\nu_{L})$; this is the Type II see-saw 
mechanism.\cite{Mohapatra:1979ia} Such a mass term
can be obtained in $SO(10)$ from the Yukawa coupling given in
Eq.(\ref{typeII})). 
After integrating out the heavy right-handed neutrinos, 
\begin{eqnarray}
-\frac{\partial\mathcal{L}}{\partial \overline{\nu}_{R}}
& = & M_{LR}\nu_{L} + M_{RR} \overline{\nu}_{R}^{T}
\nonumber\\
\overline{\nu}_{R} & = & -\nu^{T}M_{LR}^{T}M_{RR}^{-1},
\end{eqnarray}
one then obtains the effective light neutrino Majorana mass matrix 
\begin{equation} 
\label{seesaw}
M_{LL} = M_{\nu_{LR}}^{T} M_{RR}^{-1} M_{\nu_{LR}}.
\end{equation}
For $M_{LR}$ of the order of the weak scale as the mass scale of other 
charged fermions, the Majorana mass term $M_{RR}$ must be around $10^{12-14} GeV$ 
to give rise to neutrino masses of the order of $0.1 \; eV$. As we will see later, 
most grand unified theories predict the existence of the right-handed 
neutrinos. Furthermore, the GUT scale provides an understanding 
why $M_{RR}$ is large.

\subsubsection{Small Neutrino Masses from Large Extra Dimension}

An interesting way to generate small {\it Dirac} neutrino masses arises in models with 
large extra dimensions of size $\sim (1/TeV)$.\cite{Dienes:1998sb,Dienes:2000ph}  
Consider the case in which only gravity can propagate in the bulk, while 
the SM particles and interactions are confined to the brane. Because the 
right-handed neutrinos are SM singlets, they are the only particles that 
can propagate in the bulk. Its Yukawa coupling to the charged lepton $L(x)$, and 
the Higgs doublet, $H(x)$, which are confined to the brane, is given by
\begin{equation}
S = y \int d^{4}x \int dy L(x) H(x) \nu_{R}(x,y) \delta(y).
\end{equation}
Compactified on a circle $S^{1}$, $\nu_{R}(x,y)$ can be decomposed into
\begin{equation}
\nu_{R}(x,y) = \frac{1}{\sqrt{2\pi R M_{pl}^{5D}}} \sum_{-\infty}^{\infty}
\nu_{R}^{(n)} (x) e^{i n y /R}.
\end{equation}
Below the compactification scale, we thus obtain a Dirac neutrino mass
\begin{equation}
m_{\nu} = \frac{y <H>}{\sqrt{2 \pi R M_{pl}^{5D}}} 
= \frac{y <H> M_{pl}^{5D}}{M_{pl}^{4D}},
\end{equation}
where tha last steps follows from the relation between the $4D$ Planck scale and $5D$ 
Planck scale, $2 \pi R M_{pl}^{5D} = (M_{pl}^{4D}/M_{pl}^{5D})^{2}$. 
With $M_{pl}^{5D} \sim (1-10) TeV$ which 
could avoid the gauge hierarchy problem, one obtains a highly 
suppressed $m_{\nu}$ which is 
consistent with the experimental observations.

\subsubsection{Small Neutrino Masses from Warped Geometry}

Small neutrino masses of the Dirac type are possible if right-handed 
neutrinos are localized toward the Planck brane while the lepton doublets 
are localized toward the TeV brane. This results in a small overlap between 
the zero mode profiles of the lepton doublets and the right-handed neutrinos 
based on the formulation given in Sec. II. Models of this type have been 
constructed in Ref. 26, 28, 43. 

\subsection{Bi-Large Neutrino Mixing Angles}

To obtain the bi-large mixing pattern for the neutrinos, in additional to 
having the hierarchical mass pattern, let us consider, 
for example, the following mass texture,\cite{Chen:2002pa} 
\begin{equation}\label{texture}
\left(\begin{array}{ccc}
0 & 0 & t\\
0 & 1 & 1+t^{n}\\
t & 1+t^{n} & 1
\end{array}
\right),
\end{equation}
with $t < 1$ which is a special case of the following texture 
\begin{equation}\label{cmtexture}
\left(\begin{array}{ccc}
0 & 0 & \ast\\
0 & \ast & \ast\\
\ast & \ast & \ast
\end{array}
\right)
\end{equation}
first proposed in Ref. 49 
in which the elements in $(2,3)$ block 
are taken to have equal strengths to accommodate near bi-maximal mixing. 
The modification of adding the term $t^{n}$ in the $(23)$ and $(32)$ entries 
is needed in order to accommodate a large, but non-maximal 
solar angle in the so-called ``light side'' region 
($0 < \theta < \pi/4$).\cite{deGouvea:2000cq}   
It is possible to obtain the LMA solution at $3\sigma$ level   
with $n$ ranging from $1$ to $2$. 

An interesting alternative in which a $3 \times 2$ neutrino Dirac mass 
matrix is considered was proposed recently by Kuchimanchi and 
Mohapatra\cite{Kuchimanchi:2002yu,Kuchimanchi:2002fi}. 
A $3 \times 2$ neutrino Dirac mass matrix arises if there are only two right-handed 
neurtinos, instead of three. The existence of two right-handed neutrinos is required 
by the cancellation of Witten anomaly, if a global leptonic $SU(2)$ family symmetry is 
imposed\cite{Kuchimanchi:2002yu,Kuchimanchi:2002fi}. 
Along this line, Frampton, Glashow and Yanagida proposed a model, which has 
the following Lagrangian,\cite{Frampton:2002qc} 
\begin{equation}\label{FGY}
\mathcal{L} = \frac{1}{2} (N_{1} N_{2}) 
\left(\begin{array}{ccc}
M_{1} & 0\\
0 & M_{2}
\end{array}\right)
\left(\begin{array}{c} N_{1} \\ N_{2} \end{array}\right)
+ (N_{1} N_{2}) \left(\begin{array}{ccc}
a & a^{'} & 0\\
0 & b & b^{'}
\end{array}\right)
\left(\begin{array}{c} l_{1} \\ l_{2} \\ l_{3}
\end{array}\right) H + h.c..
\end{equation} 
The effective neutrino mass matrix due to this Lagrangian is obtained, 
using the see-saw formula,
\begin{equation}
\left(\begin{array}{ccc}
\frac{a^{2}}{M_{1}} & \frac{aa^{'}}{M_{1}} & 0
\\
\frac{aa^{'}}{M_{1}} & \frac{a^{'2}}{M_{1}} + \frac{b^{2}}{M_{2}} & \frac{bb^{'}}{M_{2}}
\\
0 & \frac{bb^{'}}{M_{2}} & \frac{b^{'2}}{M_{2}}
\end{array}\right),
\end{equation}
where $a, b, b^{'}$ are real and $a^{'} = |a^{'}| e^{i\delta}$. 
By takinging all of them to be real, with the choice $a^{'} = \sqrt{2} a$ 
and $b=b^{'}$, and assuming $a^{2}/M_{1} \ll b^{2}/M_{2}$, 
the effective neutrino masses and mixing matrix are obtained
\begin{equation}
m_{\nu_{1}} = 0, \quad m_{\nu_{2}} = \frac{2a^{2}}{M_{1}}, \quad 
m_{\nu_{3}} = \frac{2b^{2}}{M_{2}}
\end{equation}
\begin{equation}
U = \left(\begin{array}{ccc}
1/\sqrt{2} & 1/\sqrt{2} & 0\\
-1/2 & 1/2 & 1/\sqrt{2}\\
1/2 & -1/2 & 1/\sqrt{2}
\end{array}\right) \times
\left(\begin{array}{ccc}
1 & 0 & 0\\
0 & \cos\theta & \sin\theta \\
0 & -sin\theta & \cos\theta
\end{array}\right),
\end{equation}
where $\theta \simeq m_{\nu_{2}}/\sqrt{2}m_{\nu_{3}}$, and the 
observed bi-large mixing angles and 
$\Delta m_{atm}^{2}$ and $\Delta m_{\odot}^{2}$ can be accommodated. 
An interesting feature of this model 
is that the sign of the baryon number asymmetry ($B \propto \xi_{B} = 
Y^{2} a^{2} b^{2} \sin2\delta$) 
is related to the sign of the 
CP violation in neutrino oscillation ($\xi_{osc}$) in the following way 
\begin{equation}
\xi_{osc} = -\frac{a^{4}b^{4}}{M_{1}^{3}M_{2}^{3}} (2 + Y^{2}) \xi_{B} \propto - B
\end{equation}
assuming the baryon number asymmetry is resulting from leptogenesis 
due to the decay of the 
lighter one of the two heavy neutrinos, $N_{1}$, 
whose mass is of $\mathcal{O}(10^{10} \; GeV)$. A $SO(10)$ model which gives rise 
to the neutrino mass ansatz, Eq.(\ref{FGY}), has been 
constructed.\cite{Raby:2003ay} 
A more detailed discussion on this model is given in Sec. 6.

Other phenomenologically viable textures for neutrino mass matrix are analyzed in 
Ref. 55. 

\subsubsection{$SO(10)$ GUT realization}

In $SO(10)$ models in $4D$, the bi-large mixing in the leptonic 
sector arises in two ways (A detailed classification according 
to how the maximal $\nu_{\mu}-\nu_{\tau}$ 
mixing arises is given by Barr and Dorsner\cite{Barr:2000ka}):\\

{\noindent (i) Symmetric mass textures for the charged fermions:}

{\noindent This scenario is realized in symmetric textures arising from left-right 
symmetric breaking chain of $SO(10)$. In this case, both the large solar mixing 
angle and the maximal atmospheric mixing angle come from the diagonalization of 
the effective neutrino mass matrix. A characteristic of this class of models is that 
the predicted value for $|U_{e\nu_{3}}|$ element tends to be larger than the value 
predicted by models in class (ii) below.}\\

{\noindent (ii) Lop-sided mass textures for charged fermions:}

{\noindent In this scenario, the large atmospheric mixing angle comes from 
charged lepton mixing matrix. This scenario is realized in models with 
$SU(5)$ as the intermediate symmetry which gives rise to the 
so-called ``lop-sided'' mass textures, due to the $SU(5)$ relation,
\begin{equation}
M_{e} = M_{d}^{T}.
\end{equation}
Due to the lop-sided nature of $M_{e}$ and $M_{d}$, 
the large atmospheric neutrino mixing is 
related to the large mixing in the $(23)$ sector of 
the RH charged lepton diagonalization matrix, 
instead of $V_{cb}$.   
It thus provides an explanation why the small value of $V_{cb}$ and the large 
value of $U_{\mu\nu_{3}}$ exist simultaneously. 
The large solar mixing angles comes from the diagonalization matrix 
for the neutrino mass matrix. Because the two large mixing angles come 
from different sources, the constraint 
on $U_{e\nu_{3}}$ is not as strong as in class (i). In fact, the prediction for 
$U_{e\nu_{3}}$ in this class of models tend to be quite small. On the other hand, 
this mechanism also predicts an enhanced decay rate for the flavor-violating 
process, $\mu \rightarrow e \; \gamma$, which is close to current experimental limit.}

We will discuss these two classes of models in detail in Sec.\ref{models}.

\subsubsection{Large Mixing from Renormalization Group Evolution}

It has been shown that it is possible to obtain large neutrino mixing angles 
through the renormalization group 
evolution.\cite{Balaji:2000gd,Balaji:2000au,Balaji:2000ma,Antusch:2002hy,Antusch:2002fr} 
Recently, Mohapatra, Parida and Rajasekaran observed in 
Ref. 62 
that bi-large mixing angles can be driven by the renormalization group evolution, 
assuming that the CKM matrix and the LM matrix are identical at the 
GUT scale, which is a natural consequence of quark-lepton unification. 
The only requirement for this mechanism to work is that the masses of the 
three neutrinos are nearly degenerate of the form $m_{3} \gtrsim m_{2} \gtrsim 
m_{1}$ and have same CP parity. 
The one-loop RGE of the effective left-handed Majorana neutrino  
mass operator is given 
by\cite{Chankowski:1993tx,Babu:qv,Antusch:2001ck,Antusch:2001vn}
\footnote{Note that some of the earlier results were not entirely correct; 
re-derivation of these results has been done in 
Ref. 65, 66.} 
\begin{equation}
\label{rgem}
\frac{d m_{\nu}}{dt}=-\{\kappa_{u}m_{\nu}+m_{\nu}P+P^{T}m_{\nu}\},
\end{equation}
where $t \equiv \ln \mu$.
In the MSSM, $P$ and $\kappa_{u}$ are given by,
\begin{eqnarray}
P & = & -\frac{1}{32\pi^{2}} \frac{Y_{e}^{\dagger}Y_{e}}{\cos^{2} \beta} 
\simeq -\frac{1}{32\pi^{2}} \frac{h_{\tau}^{2}}{\cos^{2}\beta} diag(0,0,1)
\equiv diag(0,0,P_{\tau})\\ 
\kappa_{u} & = &
\frac{1}{16\pi^{2}}[\frac{6}{5}g_{1}^{2} + 6g_{2}^{2} 
- 6 \frac{Tr(Y_{u}^{\dagger}Y_{u})}{\sin^{2}\beta}] 
\simeq  
\frac{1}{16\pi^{2}}[\frac{6}{5}g_{1}^{2} + 6g_{2}^{2} 
- 6 \frac{h_{t}^{2}}{\sin^{2}\beta}], 
\end{eqnarray}
where $g_{1}^{2}=\frac{5}{3}g_{Y}^{2}$ is the $U(1)$ gauge coupling
constant, $Y_{u}$ and $Y_{e}$ are the $3 \times 3$ Yukawa coupling matrices for
the up-quarks and charged leptons respectively, and $h_{t}$ and $h_{\tau}$ are
the $t$- and $\tau$-Yukawa couplings. One can then follow the ``diagonal-and-run'' 
procedure, and obtain the RGE's at scales between $M_{R} \ge \mu \ge M_{SUSY}$ 
for the mass eigenvalues and the three mixing angles,
assuming CP violating phases vanish,
\begin{eqnarray}
\frac{d \; m_{i}}{d t} & = & 
-4 P_{\tau} m_{i} U_{\tau \nu_{i}}^{2} - m_{i} \kappa_{u}, \quad (i=1,2,3)
\label{nurge1}\\
\frac{d \; s_{23}}{d t} & = & -2P_{\tau} c_{23}^{2} 
(-s_{12}U_{\tau\nu_{1}}\nabla_{31} + c_{12}U_{\tau\nu_{2}}\nabla_{32})
\\
\frac{d \; s_{13}}{dt} & = & -2P_{\tau}c_{23}c_{13}^{2}
(c_{12}U_{\tau\nu_{1}}\nabla_{31}+s_{12}U_{\tau\nu_{2}}\nabla_{32})
\\
\frac{d \; s_{12}}{dt} & = & -2P_{\tau}c_{12}
(c_{23}s_{13}s_{12}U_{\tau\nu_{1}}\nabla_{31}
-c_{23}s_{13}c_{12}U_{\tau\nu_{2}}\nabla_{32}
+U_{\tau\nu_{1}}U_{\tau\nu_{2}}\nabla_{21}),
\label{nurge4}
\end{eqnarray}
where $\nabla_{ij} \equiv (m_{i}+m_{j})/(m_{i}-m_{j})$. 
Because the LM matrix is identical to the CKM matrix, we have, at the GUT scale, 
the following initial conditions, $s_{12}^{0} \simeq \lambda, \quad 
s_{23}^{0} \simeq \mathcal{O}(\lambda^{2})$ and 
$s_{13}^{0} \simeq \mathcal{O}(\lambda^{3})$, where $\lambda$ is the Wolfenstein 
parameter. (Note that the RG evolution has negligible effect on the 
Wolfenstein parameter, 
see Eq.(\ref{rgeV})). When the masses $m_{i}$ and $m_{j}$ are nearly degenerate, 
$\nabla_{ij}$ approaches infinity. Thus it drives the mixing angles to become large. 
Starting with the values of $(m_{1}^{0},m_{2}^{0},m_{3}^{0}) 
= (0.2983,0.2997,0.3383) \; eV$ at the GUT scale, the solutions at the weak 
scale for the masses are $(m_{1},m_{2},m_{3}) 
= (0.2410,0.2411,0.2435) \; eV$, which correspond to $\Delta m_{atm}^{2} = 
1.1 \times 10^{-3} eV^{2}$ and $\Delta m_{\odot}^{2} = 4.8 \times 10^{-5} eV^{2}$. 
The mixing angles predicted at the weak scale are $\sin\theta_{23} = 0.68$, 
$\sin\theta_{12} = 0.568$ and $\sin\theta_{13} = 0.08$. Because the masses are 
larger than $0.1 \; eV$, they are testable at the present searches for 
the neutrinoless double beta decay. 

Models based on GUT and horizontal symmetry often suffer from fine-tunning or  
the difficulty of constructing a viable scalar potential that gives rise to the 
required vacua. Along the line discussed in the above paragraph, some attempts 
have been made to show that the maximal mixing 
angle\cite{Chankowski:1999xc,Casas:1999tp,Casas:1999ac} 
and nearly degenerate neutrino 
masses\cite{Ellis:1999my,Ma:1999xq,Haba:1999xz,Chankowski:2000fp,Chen:2001gk,Miura:2002nz,Bhattacharyya:2002aq,Frigerio:2002in}
are manifestations of infrared fixed points (IRFP) of the RGEs given above 
in Eq.(\ref{nurge1})-(\ref{nurge4}), under certain assumptions. 

\subsubsection{Bi-large Mixing and $b-\tau$ Unification}

In the minimal $SO(10)$ model utilizing Type II see-saw mechanism 
with one $10$ and one $\overline{126}$, 
we have the following relations:
\begin{eqnarray}
M_{u} & = & f <10> + h <\overline{126}^{+}>
\\
M_{d} & = & f <10> + h <\overline{126}^{-}>
\\
M_{e} & = & f <10> -3 h <\overline{126}^{-}>
\\
M_{\nu_{LR}} & = & f <10> -3 h <\overline{126}^{+}>
\end{eqnarray}
where $f$ and $h$ are Yukawa matrices; 
the mass terms $M_{\nu,LL}$ and $M_{\nu,RR}$ are both due to the coupling to 
$\overline{126}$,
\begin{eqnarray}
M_{\nu,LL} & = & h <\overline{126}^{'+}>
\\
M_{\nu,RR} & = & h <\overline{126}^{'0}>
\end{eqnarray}
where the superscripts $+/-/0$ refer to the sign of the hypercharge $Y$   
(see Table 3). The small neutrino masses are explained by the Type II see-saw 
mechanism with the {\it assumption} that the LH Majorana mass 
term dominates over the usual Type I see-saw term, thus it is proportional 
to the Yukawa matrix $h$, which can be determined by calculating the difference 
between $M_{d}$ and $M_{e}$. Using down-type quark masses, charged lepton masses, 
and CKM matrix elements, which have roughly the form
\begin{equation}
M_{b,\tau} \sim
\left(
\begin{array}{ccc}
\lambda^{3} & \lambda^{3} & \lambda^{3}\\
\lambda^{3} & \lambda^{2} & \lambda^{2}\\
\lambda^{3} & \lambda^{2} & 1
\end{array}
\right) m_{b,\tau} \quad,
\end{equation}
predictions for neutrino masses and LM matrix elements 
have been made.\cite{Bajc:2002iw,Goh:2003sy} 
The large atmospheric mixing results from a small deviation of 
$\mathcal{O}(\lambda^{2})$ from $b-\tau$ unification.  

Predictions from a detailed numerical study made in Ref. 79 
are $\sin^{2}2\theta_{23} < 0.9$ and $\sin^{2}2\theta_{12} > 0.9$, 
which are experimentally allowed only at $3 \sigma$ level; 
these unique predictions can thus be used to test this type of models. 
Note that the best fit value of $\sin^{2}2\theta_{23}$ {\em cannot} 
be accommodated in these models.
The prediction for $U_{e3}$ is about $0.16$, very close to the sensitivity 
of current experiments. We also note that as this type of models do not address 
the origin of the flavor structure, they are not as predictive as $SO(10)$ 
models combined with family symmetry, in which as non-minimal Higgs content 
is usually present (see Sec. 6).

\subsection{CP violation in Leptonic Sector}

As we mentioned previously, if neutrinos are Dirac particles, there is 
only one phase in the LM matrix. On the other hand, if neutrinos are 
Majorana particles, which is the case if see-saw mechanism is implemented 
to give small neutrino masses, there are two additional phases. 
These two types of phases have very different impacts on phenomenology. 
The universal phase, $\delta_{l}$, affects the transition probability 
in the neutrino oscillation
\begin{eqnarray}
P(\nu_{\alpha}\rightarrow \nu_{\beta}) & = &
\delta_{\alpha\beta} - 4 \sum_{i>j} 
Re\{ U_{\alpha i} U_{\beta j} U_{\alpha j}^{\ast}
U_{\beta i}^{\ast} \} \sin^{2} (\Delta m_{ij}^{2} \frac{L}{4E})\nonumber\\
& & 
+ 2 \sum_{i>j} 
Im\{ U_{\alpha i} U_{\beta j} U_{\alpha j}^{\ast}
U_{\beta i}^{\ast} \} \sin^{2} (\Delta m_{ij}^{2} \frac{L}{2E}).
\end{eqnarray}
Note that the only chance that one might observe CP violation 
in neutrino oscillation is to have LMA solution in the solar sector, and 
to have large value for $\theta_{13}$.
The Majorana phases affect the matrix element for the neutrinoless double beta  
$(\beta\beta_{0\nu})$ decay, 
$\vert < m > \vert$, 
given in terms of the rephasing invariant quantities by
\begin{eqnarray}
\vert < m > \vert ^{2} & = &
m_{1}^{2} \vert U_{e1} \vert^{4} +  m_{2}^{2} \vert U_{e2} \vert^{4}
+ m_{3}^{2} \vert U_{e3} \vert^{4} 
\nonumber\\
& & + 
2m_{1}m_{2} \vert U_{e1} \vert^{2} \vert U_{e2} \vert^{2} \cos\alpha_{21} 
\nonumber\\
& & + 
2m_{1}m_{3} \vert U_{e1} \vert^{2} \vert U_{e3} \vert^{2} \cos\alpha_{31} 
\nonumber\\
& & +  
2m_{2}m_{3} \vert U_{e2} \vert^{2} \vert U_{e3} \vert^{2} 
\cos(\alpha_{31}-\alpha_{21}).
\end{eqnarray}
The current bound on $|<m>|$ from Heidelberg-Moscow experiment is $0.11-0.56 \; eV$ 
at $95\%$ confidence level.\cite{Klapdor-Kleingrothaus:2001ke}


\section{Grand Unified Theories Based on $SO(10)$}\label{so10}

The smallest GUT group $SU(5)$ in its minimal form is very strongly constrained 
due to the non-observation of proton 
decay.\cite{Murayama:2001ur,Bajc:2002bv,Emmanuel-Costa:2003pu} 
The next candidate is the rank-$5$ $SO(10)$, which is a very attractive 
candidate as a GUT group for many reasons: First of all, all of its 
irreducible representations are free of anomaly, 
unlike in the case of $SU(5)$ where the representations of the matter fields,
$\overline{5} \oplus 10$, are carefully chosen to cancel the anomaly. It
unifies all the $15$ known fermions with the right-handed neutrino for each
family into one $16$-dimensional spinor representation. The seesaw mechanism
then arises very naturally, and the small but non-zero neutrino masses can thus
be explained, as evidenced by recent atmospheric neutrino oscillation data
from Super-Kamiokande indicating small non-zero neutrino masses. Because a
complete quark-lepton symmetry is achieved, it has the promise for explaining
the pattern of fermion masses and mixing. In some $SO(10)$ models,
R-parity is conserved automatically at all energy scales. This is to be
contrasted with MSSM and SUSY $SU(5)$ where R-parity must be imposed by hand.
Because $(B-L)$ is a gauge symmetry contained in $SO(10)$, it has the
promises of baryogenesis. In what follows, we briefly review the structure 
of $SO(10)$ models. A detail discussion can be found in Ref. 84.

\subsection{The Algebra of $SO(2n)$}
\label{susygut-algebra}
It is convenient to discuss the $SO(2n)$ algebra in the $SU(n)$ 
basis.\cite{Mohapatra:1979nn} 
Consider a set of $n$ operators $\; \xi_{i} \; (i = 1, ..., n)$, and
their hermitian conjugates, $\; \xi_{i}^{\dagger}$, satisfying
\begin{equation}
\{\xi_{i},\xi_{j}^{\dagger}\} = \delta_{ij},
\qquad
\{\xi_{i},\xi_{j}\} = 0,
\end{equation}
where $\{ \; , \; \}$ denotes an anti-commutator and 
$[ \; , \; ]$ denotes a commutator. 
The operators $K^{i}_{j}$ defined as
\begin{equation}
K^{i}_{j} \equiv \xi_{i}^{\dagger}\xi_{j}
\end{equation}
satisfy the algebra of the $U(n)$ group 
\begin{equation}
[K^{i}_{j},K^{m}_{n}] = \delta^{m}_{j} K^{i}_{n} 
- \delta^{i}_{n} K^{m}_{j}.
\end{equation}
We can then define the following $2n$ operators, $\Gamma_{\mu}\;
(\mu=1,...,2n)$ \begin{eqnarray}
\Gamma_{2j-1} & = & -i \; ( \; \xi_{j} - \xi_{j}^{\dagger} \; )
\nonumber\\
\Gamma_{2j} & = & ( \; \xi_{j} + \xi_{j}^{\dagger}\; ), \qquad j = 1,...,n.
\end{eqnarray}
The $\Gamma_{\mu}$ form the Clifford algebra of rank $2n$
\begin{equation}
\{ \Gamma_{\mu}, \Gamma_{\nu} \} = 2 \delta_{\mu\nu}
\end{equation}
and they can then be used to construct the generators
of $SO(2n)$ as follows:
\begin{equation}
\Sigma_{\mu\nu} = \frac{1}{2i} \; [\Gamma_{\mu}, \Gamma_{\nu}].
\end{equation}
The dimensionality of the spinor representation of $SO(2n)$ is $2^{n}$. In
terms of the $SU(n)$ basis, the spinor representation of $SO(2n)$ can then be
constructed by,
\begin{eqnarray}
|0> & \sim & 1\\
\xi_{i}^{\dagger} \; |0> & \sim & n\\
\xi_{i}^{\dagger} \; \xi_{j}^{\dagger} \; |0> & \sim & \frac{n(n-1)}{2}\\
\xi_{i}^{\dagger} \; \xi_{j}^{\dagger} \; \xi_{l}^{\dagger} \; |0>
& \sim & \frac{n(n-1)(n-2)}{6}\\
.....\\
\xi_{1}^{\dagger} \; ... \; \xi_{n}^{\dagger} \; |0> & \sim & n
\end{eqnarray}
where $|0>$ is the $SU(n)$ invariant vacuum state. The spinor representation
can then be split into two $2^{n-1}$-dimensional representations by a chiral
projection operator. Let us define
\begin{equation}
\Gamma_{0} \equiv i^{n} \Gamma_{1} \; \Gamma_{2} \; ... \; \Gamma_{2n}
\end{equation}
and the number operator
\begin{equation}
N_{i} \equiv \xi_{i}^{\dagger} \xi_{i}.
\end{equation}
$\Gamma_{0}$ then can be written as
\begin{eqnarray}
\Gamma_{0} & = & [\xi_{1},\xi_{1}^{\dagger}] \; [\xi_{2},\xi_{2}^{\dagger}] \;
... \; [\xi_{n},\xi_{n}^{\dagger}]
\nonumber\\
& = & \prod_{i-1}^{n} \; (1-2n_{i})
\nonumber\\
& = & (-1)^{n}.
\end{eqnarray}
To arrive at the last step, we have used the property of the number operator
$n_{i}^{2} \; = \; n_{i}$ to get $1-2n_{i} = (-1)^{n_{i}}$ and $n = \sum_{i}
\; n_{i}$. One can then check that
\begin{equation}
[\Sigma_{\mu\nu},\Gamma_{0}]=0.
\end{equation}
The chirality projection operator is therefore defined by 
\begin{equation}
\frac{1}{2} (1 \pm \Gamma_{0}).
\end{equation}
Consider the case $n=5$ and define a column vector $|\psi>$:
\begin{eqnarray}
|\psi> & = & |0> \psi_{0} 
+ \xi_{j}^{\dagger} |0> \psi_{j} 
+ \frac{1}{2}\xi_{j}^{\dagger}\xi_{k}^{\dagger} |0> \psi_{jk} 
+ \frac{1}{12}
\epsilon^{ijklm} \xi_{k}^{\dagger}
\xi_{l}^{\dagger}\xi_{m}^{\dagger}|0>\overline{\psi}_{ji} \nonumber\\
& & + \frac{1}{24}
\epsilon^{jklmn} \xi_{k}^{\dagger}
\xi_{l}^{\dagger}\xi_{m}^{\dagger}\xi_{n}^{\dagger}|0>\overline{\psi}_{j}
+ \xi_{1}^{\dagger}
\xi_{2}^{\dagger}\xi_{3}^{\dagger}\xi_{4}^{\dagger}\xi_{5}^{\dagger}|0>
\overline{\psi}_{0}
\end{eqnarray}
where $\overline{\psi}$ is not the complex conjugate of $\psi$ but an
independent vector. This can be generalized to any $n$ if we write 
\begin{equation}
\psi = \left(
\begin{array}{cccccc}
\psi_{0} &
\psi_{i} &
\psi_{ij} &
\overline{\psi}_{ij} &
\overline{\psi}_{i} &
\overline{\psi}_{0}
\end{array}
\right)^{T}.
\end{equation}
The spinor representation is then split under the chirality projection
operator as
\begin{equation}
\psi = \left(
\begin{array}{c}
\psi_{+}\\
\psi_{-}
\end{array}
\right)
\end{equation}
where
\begin{equation}
\psi_{\pm} = \frac{1}{2} (1 \pm \Gamma_{0}) \; \psi 
\end{equation}
and
\begin{equation}
\psi_{+} =
\left(
\begin{array}{c}
\psi_{0}\\
\psi_{ij}\\
\overline{\psi}_{j}
\end{array}
\right), \qquad
 \psi_{-} =
\left(
\begin{array}{c}
\overline{\psi}_{0}\\
\overline{\psi}_{ij}\\
\psi_{j}
\end{array}
\right).
\end{equation}
In the case of $n=5$, $\overline{\psi}_{i}$ and $\psi_{ij}$ are $\overline{5}$
and $10$-dimensional representations of $SU(5)$ and $\psi_{0}$ is the singlet. 
All the SM fermions are assigned to $\psi_{+}$. The electric charge formula
for $SO(10)$ is given by
\begin{equation}
Q = \frac{1}{2}\Sigma_{78} - \frac{1}{6}(\Sigma_{12} + \Sigma_{34} +
\Sigma_{56}).
\end{equation}

The dimensionality of the adjoint representation of $SO(2n)$ is
$\frac{(2n)(2n-1)}{2}$. For $SO(10)$, it is $45$-dimensional. 
Under the decomposition with respect to $SU(3)
\times SU(2)_{L} \times SU(2)_{R}$ these $45$ gauge bosons are:
\begin{eqnarray}
45 & = & (8,1,1) + (1,3,1) + (1,1,1)\nonumber\\
& & + (1,1,3) + (\overline{3},2,2) + (3,2,2) + (3,1,1) + (\overline{3},1,1).
\end{eqnarray}
In this basis, the $12$ Standard Model gauge fields are in the $(8,1,1), \; 
(1,3,1)$ and $(1,1,1)$ multiplets. The rest are $33$ new gauge bosons 
which could potentially mediate proton decay. 

\subsection{Symmetry Breaking}

Because $SO(10)$ is a rank-$5$ group while SM is a rank-$4$ group, there exist
several intermediate symmetries through which $SO(10)$ can descend to
$SU(3)\times SU(2)_{L} \times U(1)_{Y}$. There are four maximal subgroups of
$SO(10)$:  $SU(4)\times SU(2)_{L} \times SU(2)_{R}$, $SU(5)\times U(1)$,
$SO(9)$, and $SO(7)\times SU(2)$. Only through the breaking chains of
$SU(4)\times SU(2)_{L} \times SU(2)_{R}$ and $SU(5)\times U(1)$ can one obtain
the correct quantum numbers for the SM particle content. Due to the
presence of these intermediate scales, the predictions for proton lifetime and
$\sin^{2}\theta_{w}$ are much less certain, compared to the case of
$SU(5)$. Details of breaking chains giving the SM are as follows:

\noindent{(i) The left-right symmetry breaking chain is}
\begin{eqnarray}\label{sblr}
SO(10) 
& {<54> \atop \longrightarrow} 
& SU(4) \times SU(2)_{L} \times SU(2)_{R} \nonumber\\
& {<45> \atop \longrightarrow}
& SU(3) \times SU(2)_{L} \times SU(2)_{R} \times U(1)_{B-L}\nonumber\\  
& {<126 \oplus \overline{126}> \atop \longrightarrow} 
& SU(3) \times SU(2)_{L} \times U(1)_{Y} \nonumber\\
& {<10> \atop \longrightarrow} 
& SU(3) \times U(1)_{EM}.
\end{eqnarray}
In this case, the hypercharge $Y$ is given by $Y=2T_{3R}+(B-L)$. The first
step of breaking to the left-right symmetry group is achieved by a symmetric
two-index tensor, $<54>$. $SU(4)$ in the left-right symmetry group is then
broken to $SU(3)\times U(1)_{B-L}$ by the adjoint $<45>$. The subsequent
breaking to the SM gauge group is achieved by the anti-symmetric $5$-index
tensor, $<126>$ and $<\overline{126}>$; the electroweak symmetry breaking 
is then achieved by $<10>$. In realistic models, the two Higgs doublets 
in MSSM are linear combinations of the $SU(2)$ doublet components from different 
$SO(10)$ representations of Higgses, for example, $10$ and $\overline{126}$. 
Thus all fields in the linear combinations contribute to electroweak 
symmetry breaking.  

\noindent{(ii)} For the $SU(5)$ breaking chain,
\begin{eqnarray}
SO(10) 
& \rightarrow 
& SU(5) \times U(1)_{x}
\nonumber\\ 
& \rightarrow 
& SU(3) \times SU(2) \times U(1)_{z} \times U(1)_{x}
\nonumber\\
& \rightarrow
& SU(3) \times SU(2) \times U(1)_{Y}
\nonumber\\
& \rightarrow
& SU(3) \times U(1)_{EM}
\end{eqnarray}
the hypercharge $Y$ is given by $\frac{1}{2} Y = \alpha z + \beta x$ where
$z$ and $x$ are the charges under $U(1)_{z}$ and $U(1)_{x}$ respectively. There
are two possible ways to embed the SM under this route: $(\alpha,\beta) =
(1/6,0)$ or $(-1/15,-2/5)$. In the case of $(\alpha,\beta) = (1/6,0)$, we
obtain $Y = \frac{1}{6}(2z)$. This corresponds to the $SU(5)$ breaking chain
\begin{eqnarray}
SO(10) 
& {<16 \oplus \overline{16}> \atop \rightarrow} 
& SU(5)
\nonumber\\ 
& {<45> \atop \rightarrow}
& SU(3) \times SU(2) \times U(1)_{Y}
\nonumber\\
& {<10> \atop \rightarrow}
& SU(3) \times U(1)_{EM}.
\end{eqnarray}
In this case, the spinors 
$<16 \oplus \overline{16}>$ break the symmetry down to $SU(5)$; $<45>$ then
breaks $SU(5)$ down to the SM. 
In the case of $(\alpha,\beta) = (-1/15,-2/5)$, we have 
$Y = \frac{-1}{15}(z+6x)$. This corresponds to flipped $SU(5)$ (that is,
$SU(5) \times U(1)$) breaking chain 
\begin{eqnarray}
SO(10) 
& {<45> \atop \rightarrow} 
& SU(5) \times U(1)_{x}
\nonumber\\ 
& {<16 \oplus \overline{16}> \atop \rightarrow}
& SU(3) \times SU(2) \times U(1)_{Y}
\nonumber\\
& {<10> \atop \rightarrow}
& SU(3) \times U(1)_{EM}.
\end{eqnarray}
For this breaking to occur, again we need
$<16 \oplus \overline{16}>$ and $<45>$. Each breaking chain has its 
characteristic set of Higgs fields and symmetry breaking superpotential. 
The electroweak symmetry breaking is then achieved by $<10>$. In realistic models, 
the two Higgs doublets in MSSM are linear combinations of the $SU(2)$ doublet 
components from different $SO(10)$ representations of Higgses, for example, 
$10$ and $16$. As we will see below, different symmetry breaking 
chains give rise to different mass relations between various sectors.

\subsection{Renormalization Group Equation and the 
Georgi-Jarlskog (GJ) relations}

Before describing the Yukawa sector of the $SO(10)$ models, we discuss how to 
relate the weak scale observables to the GUT scale parameters. 
We use the expressions derived from 1-loop RGEs given
by:\cite{Arason:1991ic,Castano:1993ri,Barger:1992ac,Berezhiani:1998vn} 
\begin{eqnarray}
m_{u} = Y_{u}^{0}R_{u}\eta_{u}B_{t}^{3}v_{u}, \quad & 
m_{c} = Y_{c}^{0}R_{u}\eta_{c}B_{t}^{3}v_{u}, \quad &
m_{t} = Y_{c}^{0}R_{u}B_{t}^{6}v_{u} \nonumber\\
m_{d} =  Y_{d}^{0}R_{d}\eta_{d}v_{d}, \quad &
m_{s} = Y_{s}^{0}R_{d}\eta_{s}v_{d}, \quad &
m_{b} =  Y_{b}^{0}R_{d}\eta_{b}B_{t}v_{d} \nonumber\\
m_{e} =  Y_{e}^{0}R_{e}v_{d}, \quad &
m_{\mu} = Y_{\mu}^{0}R_{e}v_{d}, \quad &
m_{\tau} = Y_{\tau}^{0}R_{e}v_{d}  
\end{eqnarray}
and
\begin{equation}\label{rgeV}
V_{ij} = \bigg\{
\begin{array}{lll}
V_{ij}^{0}, & & ij = ud, us, cd, cs, tb \\
V_{ij}^{0} B_{t}^{-1}, & & ij = ub, cb, td, ts
\end{array}
\end{equation}
where $V_{ij}$ are CKM matrix elements; quantities with superscript $0$ are
evaluated at GUT scale, and all the $m_{f}$ and $V_{ij}$ are the experimental
values at $M_{Z}$;\cite{Hagiwara:fs,Hocker:2001xe,Abe:2001xe,Aubert:2001nu}  
$Y_{f}^{0}$ are Yukawa couplings, and $v_{u}$ and $v_{d}$ are VEV of the Higgs fields 
$H_{u}$ and $H_{d}$.   
The SM Higgs VEV is $v=\sqrt{v_{u}^{2}+v_{d}^{2}}=246/\sqrt{2} \; GeV.$
The running factor $\eta_{f}$ includes QCD + QED contributions:  
For $f=b,c$, $\eta_{f}$ is for the range $m_{f}$ to $m_{t}$, and for
$f=u,d,s$, $\eta_{f}$ is for the range $1 GeV$ to $m_{t}$;
\begin{eqnarray}
\eta_{u}=\eta_{d}=\eta_{s}=2.38_{-0.19}^{+0.24}\nonumber\\
\eta_{c}=2.05_{-0.11}^{+0.13}\nonumber\\
\eta_{b}=1.53_{-0.04}^{+0.03}.
\end{eqnarray}
$R_{u,d,e}$ are contributions of the gauge-coupling constants
running from weak scale $M_{z}$ to the SUSY breaking scale, taken to be
$m_{t}$, with the SM spectrum, and from $m_{t}$ to the GUT scale with MSSM
spectrum;
\begin{equation} 
R_{u}=3.53_{-0.07}^{+0.06}, \qquad 
R_{d}=3.43_{-0.06}^{+0.07}, \qquad 
R_{e}=1.50.
\end{equation}
$B_{t}$ is the running induced by large top-quark Yukawa coupling defined
by
\begin{equation}
B_{t}=\exp\left[\frac{-1}{16 \pi^{2}} \int_{\ln M_{SUSY}}^{\ln M_{GUT}}
Y_{t}^{2}(\mu) d(\ln \mu)\right]  
\end{equation}
which varies from $0.7$ to $0.9$ corresponding to the perturbative
limit $Y_{t}^{0}\approx 3$ and the lower limit $Y_{t}^{0} \approx 0.5$ imposed
by the top-pole mass. 

Naively, one would expect that masses of the down type quarks are identical to 
masses of the charged leptons, because of the quark-lepton unification. This 
turns out to be not true.
Taking the experimentally measured values for masses of the down-type quarks
and those of the charged leptons, evolving these values using the RGE's 
from the weak scale to the GUT scale, however, one finds the
following approximate relations\cite{Georgi:1979df,Georgi:1979ga}
\begin{equation}
m_{d} \simeq 3 m_{e}, \qquad m_{s} \simeq \frac{1}{3}m_{\mu}, \qquad
m_{b} \simeq m_{\tau}.
\end{equation}
These are known as the {\it Georgi-Jarlskog relations}. As it will become
apparent later, one way to satisfy these relations is by having a
relative factor of $-3$ in the $(22)$ entry of the charged lepton mass matrix
with respect to that of the down-type quarks, and all other elements are
identical in these two mass matrices. An example suggested by Georgia and Jarlskog 
is:
\begin{equation}
M_{u} = \left(
\begin{array}{ccc}
0 & A & 0\\
A & 0 & B\\
0 & B & C
\end{array}\right), 
M_{d} = \left(
\begin{array}{ccc}
0 & E & 0\\
E & F & 0\\
0 & 0 & G
\end{array}\right), 
M_{e} = \left(
\begin{array}{ccc}
0 & E & 0\\
E & -3F & 0\\
0 & 0 & G
\end{array}\right).
\end{equation}
This factor arises naturally as a Clebsch-Gordon (CG) coefficient in some 
models of $SO(10)$.

\subsection{Yukawa sector in $SO(10)$}

In $SO(10)$, at the renormalizable level, only three types of Higgs fields can
couple to fermions,
\begin{equation}
16 \otimes 16 = 10_{S} \oplus 120_{A} \oplus 126_{S}
\end{equation}
namely, $10$, $120_{A}$, and $\overline{126}_{S}$, where the subscripts $S$
and $A$ refer to the symmetry property under interchanging two family indices
in the Yukawa couplings $\mathcal{Y}_{ab}$. That is,
\begin{equation}
\mathcal{Y}_{ab}^{10} = \mathcal{Y}_{ba}^{10},\quad
\mathcal{Y}_{ab}^{120} = -\mathcal{Y}_{ba}^{120},\quad
\mathcal{Y}_{ab}^{\overline{126}} = \mathcal{Y}_{ba}^{\overline{126}}.
\end{equation}
The gauge invariant Yukawa couplings are then given by
\begin{eqnarray}
\mathcal{Y}_{ab}^{10} (16)_{a} (16)_{b} (10) & = &
\mathcal{Y}_{ab}^{10} \Psi_{a}^{T} B C^{-1} \Gamma_{\alpha} \Psi_{b}
H_{\alpha}
\\ 
\mathcal{Y}_{ab}^{120} (16)_{a} (16)_{b} (120) & = &
\mathcal{Y}_{ab}^{120} \Psi_{a}^{T}BC^{-1} \Gamma_{\alpha} 
\Gamma_{\beta} \Gamma_{\gamma} \Psi_{b} \Lambda_{\alpha\beta\gamma}
\\
\mathcal{Y}_{ab}^{126} (16)_{a} (16)_{b} (\overline{126}) & = &
\mathcal{Y}_{ab}^{126} \Psi_{a}^{T}BC^{-1} \Gamma_{\alpha} 
\Gamma_{\beta} \Gamma_{\gamma} \Gamma_{\delta} \Gamma_{\xi} \Psi_{b}
\overline{\Delta}_{\alpha\beta\gamma\delta\xi}
\end{eqnarray}
where $\Psi$ denotes the matter fields; $H_{\alpha}, \;
\Lambda_{\alpha\beta\gamma}$ and
$\overline{\Delta}_{\alpha\beta\gamma\delta\xi}$ denote the $10$-, $120$-
and $\overline{126}$-dim Higgs fields, respectively. $C$ is the usual Dirac
charge conjugate operator and $B$ is the charge conjugate operator for
$SO(10)$ defined by
\begin{equation}
B^{-1} \Gamma_{\mu}^{T}B=-\Gamma_{\mu}
\end{equation}
and we can choose
\begin{equation}
B=\prod_{\mu=\mbox{odd}} \Gamma_{\mu}.
\end{equation}
Under this charge conjugation
\begin{equation}
B\left(
\begin{array}{c}
\psi_{0}\\
\psi_{ij}\\
\overline{\psi}_{i}\\
\psi_{i}\\
\overline{\psi}_{ij}\\
\overline{\psi}_{0}
\end{array}\right)
=\left(\begin{array}{c}
\overline{\psi}_{0}\\
-\overline{\psi}_{ij}\\
\psi_{i}\\
-\overline{\psi}_{i}\\
\psi_{ij}\\
\psi_{0}
\end{array}\right).
\end{equation}

Note that $SO(10)$ can break down to SM through many different 
breaking chains. Different breaking chains give rise to 
different mass relations among the up-quark, down-quark,
charged lepton and neutrino sectors. In what follows, we discuss the 
two symmetry breaking separately.\\

{\noindent (i) $SU(5)$ breaking chain:}\\

Under the $SU(5)$ decomposition, we have
\begin{eqnarray}
16 & = & 1 + \overline{5} + 10\\
10 & = & 5 + \overline{5}\\
120 & = & 5 + \overline{5} + 10 + \overline{10} + 45 + \overline{45}\\
126 & = & 1 + \overline{5} + 10 + \overline{15} + 45 + \overline{50}
\end{eqnarray}
where the $SU(5)$ component $5$ and $45$ contain the $Y = +1$ $SU(2)_{L}$ Higgs doublet, 
and the $SU(5)$ component $\overline{5}$ and $\overline{45}$ 
contain the $Y = -1$ $SU(2)_{L}$ Higgs doublet. The $SU(5)$ singlet contained in 
$\overline{126}$ gives masses to the RH neutrinos through the coupling 
$(16_{i})(16_{j})(\overline{126}_{H})$. (As we will discuss later, some models 
utilize non-renormalizable operators $(16_{i})(16_{j})(16_{H})(16_{H})$ 
to generate RH neutrino masses. This can be achieved because $16_{H}$ also 
contains a $SU(5)$ singlet component.) The $\overline{15}$ of $SU(5)$ 
contained in $\overline{126}$ has a $(1,3,2)$ component under $SU(3) \times SU(2)_{L} 
\times U(1)_{Y}$ which couples to two lepton doublets as $(1,2,-1)(1,2,-1)(1,3,2)_{H}$
under $G_{SM}$ and gives the LH neutrino Majorana masses 
in the Type II see-saw mechanism.

As the neutral components in these $SU(2)_{L}$ doublets acquires VEV's 
of the electroweak scale, the following mass matrices are obtained
\begin{eqnarray}\label{su5mass}
& M_{u} =  \mathcal{Y}^{10}_{ab} \left< 5(10) \right> 
+ \mathcal{Y}^{120}_{ab} \left< 45(120) \right>
+ \mathcal{Y}^{\overline{126}}_{ab} \left< 5(\overline{126}) \right>
\equiv Y_{u} v_{u} \hspace{2.9cm}
\\
& M_{d} = \mathcal{Y}^{10}_{ab} \left< \overline{5}(10) \right> 
+ \mathcal{Y}^{120}_{ab} ( \left< \overline{5}(120) \right> 
+ \left< \overline{45}(120) \right> )
+ \mathcal{Y}^{\overline{126}}_{ab} \left< \overline{45}(\overline{126}) \right>
\equiv Y_{d} v_{d} \qquad
\\
& M_{e} = \mathcal{Y}^{10}_{ab} \left< \overline{5}(10) \right>
+ \mathcal{Y}^{120}_{ab} 
(\left< \overline{5}(120) \right> 
-3 \left< \overline{45}(120) \right> )
-3 \mathcal{Y}^{\overline{126}}_{ab} \left< \overline{45}(\overline{126})\right>
\equiv Y_{e} v_{d}
\qquad \\
& M_{\nu_{LR}} =   \mathcal{Y}^{10}_{ab} \left< 5(10) \right> 
+ \mathcal{Y}^{120}_{ab} \left< 5(120) \right>
-3 \mathcal{Y}^{\overline{126}}_{ab} \left< 5(\overline{126}) \right>
\equiv Y_{\nu_{LR}} v_{u} \hspace{2.6cm}
\end{eqnarray}
where we denote the $m$-dim $SU(5)$ component of the $n$-dim 
representation of $SO(10)$ by 
$m(n)$; $v_{u}$ and $v_{d}$ are the vacuum expectation values of the two Higgs
doublets in MSSM. A Clebsch-Gordon coefficient $(-3)$ is generated in the 
lepton sectors when the $SU(5)$ component $\overline{45}$ are involved 
in the Yukawa couplings. This factor of $(-3)$ is very crucial for 
obtaining the Georgi-Jarlskog 
relations as we have seen in the previous section.  
The neutrino Majorana mass matrices are given by
\begin{eqnarray}
M_{\nu,RR} & = & \mathcal{Y}^{\overline{126}}_{ab} \left< 1(\overline{126})
\right>
\\
M_{\nu,LL} & = & \mathcal{Y}^{\overline{126}}_{ab} \left<
15(\overline{126}) \right>.\label{typeIIsu5}
\end{eqnarray}

To see how these CG coefficients $(-3)$ come about, let us
decompose the following representations under $SU(5)$. 
The $120$ and $\overline{126}$-dimensional representation of $SO(10)$ 
both contain a component which transforms as $\overline{45}$ under $SU(5)$. The
Yukawa interactions then can be written as
\begin{equation}
(16)(16)(120), \; (16)(16)(\overline{126}) 
\supset (10)(\overline{5})(\overline{45}) 
= \psi^{\alpha\beta}\psi_{\gamma}<H_{\alpha\beta}^{\gamma}>.
\end{equation} 
We then write out all the terms in the summation
\begin{equation}
\psi^{\alpha\beta}\psi_{\gamma} <H_{\alpha\beta}^{\gamma}>
\supset
(\psi^{45}\psi_{4} <H_{45}^{4}> + \psi^{a5} \psi_{a} <H_{a5}^{a}>)
=(-3 e^{+}e^{-} + d d^{c}) v^{'}.
\end{equation}
Here the index $a = 1, 2, 3$. Note that $H_{\alpha\beta}^{\gamma}$
is anti-symmetric under inter-changing $\alpha \leftrightarrow \beta$ and 
it is traceless
\begin{equation}
H_{\alpha\beta}^{\gamma} = - H_{\beta\alpha}^{\gamma}, \qquad
\sum_{\beta} H_{\beta 5}^{\beta} = 0.
\end{equation}
This implies that
\begin{equation}
3H_{a5}^{a} + H_{45}^{4} = 0, \qquad
<H_{45}^{4}> = - 3<H_{a5}^{a}> \equiv -3 v^{'}.
\end{equation}
Essentially, the CG factor of $(-3)$ is related to the fact that there are
three colors.\\

We note that if $10$ and $\overline{126}$ are the only fields utilized in the 
Yukawa sector, we have the up-quark mass matrix related to the Dirac neutrino 
mass matrix, and the down-quark mass matrix related to the charged lepton mass 
matrix. When $120$ is introduced, the relation between the up-quark sector and the 
Dirac neutrino sector is lost because these two sectors receive contributions 
from different components of $120$.\\

{\noindent (ii) $SU(4) \times SU(2)_{L} \times SU(2)_{R}$ breaking chain:}\\

Under $SU(4) \times SU(2)_{L} \times SU(2)_{R}$, the relevant $SO(10)$ 
representations have the following decomposition
\begin{eqnarray}
16 & = & (4,2,1) + (\overline{4},1,2)\\
10 & = & (6,1,1) + (1,2,2)\\
120 & = & (15,2,2) + (6,3,1) + (6,1,3) + (1,2,2) + (10,1,1) + (\overline{10},1,1)\\
126 & = & (10,1,3) + (\overline{10},3,1) + (15,2,2) + (6,1,1)
\end{eqnarray}
where the components $(15,2,2)$ and $(1,2,2)$ both contain 
a pair of the $Y = \pm 1$ $SU(2)_{L}$ Higgs doublets, whose neutral components 
give masses to the fermions. The component $(10,1,3)$ contained in 
$\overline{126}$ gives masses to the RH neutrinos through the coupling 
$(16_{i})(16_{j})(\overline{126}_{H})$. The LH neutrino Majorana masses 
are generated due to the $(\overline{10},3,1)$ component of $\overline{126}$.
As these $SU(2)_{L}$ doublets acquire VEV's, the following mass matrices 
are generated,
\begin{eqnarray}\label{so10mass}
M_{u} & = & \mathcal{Y}^{10}_{ab} \left< 10^{+} \right> 
+ \mathcal{Y}^{120}_{ab} ( \left< 120^{+} \right> + 
\frac{1}{3} \left< 120^{'+} \right> )
+ \frac{1}{3} \mathcal{Y}^{\overline{126}}_{ab} \left< \overline{126}^{+} \right>
\equiv Y_{u} v_{u}
\\
M_{d} & = & \mathcal{Y}^{10}_{ab} \left< 10^{-} \right> 
+ \mathcal{Y}^{120}_{ab} (- \left< 120^{-} \right>
+\frac{1}{3} \left< 120^{'-} \right> )
-\frac{1}{3} \mathcal{Y}^{\overline{126}}_{ab} \left< \overline{126}^{-} \right>
\equiv Y_{d} v_{d}
\qquad \\
M_{e} & = & \mathcal{Y}^{10}_{ab} \left< 10^{-} \right>
+ \mathcal{Y}^{120}_{ab} (-\left< 120^{-} \right>
-\left< 120^{'-} \right>)
+ \mathcal{Y}^{\overline{126}}_{ab} \left< \overline{126}^{-} \right>
\equiv Y_{e} v_{d}
\qquad \\
M_{\nu_{LR}} & = & \mathcal{Y}^{10}_{ab} \left< 10^{+} \right> 
+ \mathcal{Y}^{120}_{ab} (\left< 120^{+} \right>
-\left< 120^{'+} \right>)
+ \mathcal{Y}^{\overline{126}}_{ab} \left< \overline{126}^{+}\right>
\equiv Y_{\nu_{LR}} v_{u}.
\end{eqnarray}
Note that a Clebsch-Gordon coefficient $(-3)$ is generated in the lepton 
sectors when the $SU(4) \times SU(2)_{L} \times SU(2)_{R}$ components $(15,2,2)$ 
are involved in the Yukawa couplings.   
The neutrino Majorana mass matrices are given by
\begin{eqnarray}
M_{\nu,RR} & = & \mathcal{Y}^{\overline{126}}_{ab} \left< \overline{126}^{'0}
\right>
\\
M_{\nu,LL} & = & \mathcal{Y}^{\overline{126}}_{ab} \left<
\overline{126}^{'+} \right>\label{typeII} 
\end{eqnarray}
where various VEVs are those of the neutral components of $SO(10)$
representations as indicated in Table~\ref{notation}.
\begin{table}[th!]
{\center
\tbl{\label{notation}Standard Model singlet components contained in various $SO(10)$ 
Higgs representations. Here the subscripts refer to the symmetry groups on 
the right-hand side of Eq.(\ref{sblr}); and superscripts $+/0/-$ referring 
to the sign of the hypercharge Y.}
{\begin{tabular}{@{}rlllll@{}}\toprule
$\left< 10^{+} \right>$ : & $(1,0)_{31}$
& $\subset (1,2,1)_{321}$ & $\subset 
(1,2,2,0)_{3221}$ & $\subset (1,2,2)_{422}$ & $\subset 10$
\\ 
$\left< 10^{-} \right>$ : & $(1,0)_{31}$
& $\subset (1,2,-1)_{321}$ & $\subset
(1,2,2,0)_{3221}$ & $\subset (1,2,2)_{422}$ & $\subset 10$
\\
$\left< 120^{+} \right>$ : & $(1,0)_{31}$
& $\subset (1,2,1)_{321}$ & $\subset 
(1,2,2,0)_{3221}$ & $\subset (1,2,2)_{422}$ & $\subset 120$\\
$\left< 120^{-} \right>$ :
& $(1,0)_{31}$
& $\subset (1,2,-1)_{321}$ & $\subset
(1,2,2,0)_{3221}$ & $\subset (1,2,2)_{422}$ & $\subset 120$
\\
$\left< 120^{'+} \right>$ : & $(1,0)_{31}$ 
& $\subset (1,2,1)_{321}$ & $\subset
(1,2,2,0)_{3221}$ & $\subset (15,2,2)_{422}$ & $\subset 120$\\
$\left< 120^{'-} \right>$ : & $(1,0)_{31}$
& $\subset (1,2,-1)_{321}$ & $\subset
(1,2,2,0)_{3221}$ & $\subset (15,2,2)_{422}$ & $\subset 120$\\
$\left< \overline{126}^{+} \right>$ : & $(1,0)_{31}$ 
& $\subset (1,2,1)_{321}$ & $\subset
(1,2,2,0)_{3221}$ & $\subset (15,2,2)_{422}$ & $\subset \overline{126}$\\
$\left< \overline{126}^{-} \right>$ : & $(1,0)_{31}$
& $\subset (1,2,-1)_{321}$ & $\subset
(1,2,2,0)_{3221}$ & $\subset (15,2,2)_{422}$ & $\subset \overline{126}$\\
$\left< \overline{126}^{'0} \right>$ : & $(1,0)_{31}$
& $\subset (1,1,0)_{321}$ & $\subset
(1,1,3,-2)_{3221}$ & $\subset (10,1,3)_{422}$ & $\subset \overline{126}$\\
$\left< \overline{126}^{'+} \right>$ : & $(1,0)_{31}$
& $\subset (1,3,2)_{321}$ & $\subset
(1,3,1,2)_{3221}$ & $\subset (\overline{10},3,1)_{422}$ & $\subset \overline{126}$
\\ &&&&&
\botrule\end{tabular}}}
\end{table}\\

{\noindent \underline{Non-renormalizable Operators:}}\\

\begin{figure}[b!]
\centerline{
\psfig{file=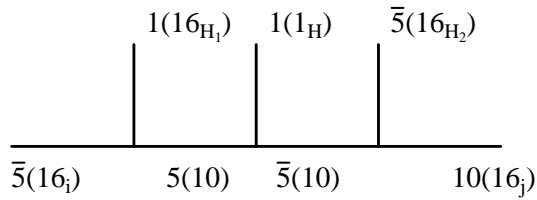,width=7.0cm}}
\caption{\label{lopsideddiagram} Froggatte-Nielsen diagram which generates lop-sided 
texture. Here we denote the field by $m(n)$ where $m$ is the $SU(5)$ representation which 
is a component of the $n$-dim representation of $SO(10)$.}
\end{figure}

If $SO(10)$ breaks down through the $SU(5)$ breaking chain, the Higgs 
fields needed are in the $(16 \oplus 16)$, $45$ and $54$-dim representaions.
A lop-sided texture can arise, when non-renormalizable operators 
involing the $16$'s are utilized to generate fermion masses. 
The $16$-dim Higgs fields are needed in this case to break $SO(10)$ 
down to $SU(5)$. A lop-sided texture 
is generated by the operator\cite{Albright:1998vf}
\begin{equation}\label{lopsided}
\lambda (16_{i}16_{H_{1}})(16_{j}16_{H_{2}}).
\end{equation}
If $16_{H_{1}}$ acquires a VEV along the $SU(5)$ singlet direction which breaks 
$SO(10)$ down to $SU(5)$, and $16_{H_{2}}$ acquires a VEV along the $\overline{5}$ 
direction of $SU(5)$ which breaks the electroweak symmetry $SU(2) \times U(1)$, 
we obtain
\begin{equation}
(\overline{5}_{i})(10_{j})<1_{H_{1}}> <\overline{5}_{H_{2}}>.
\end{equation} 
Inside the first parenthesis in Eq.(\ref{lopsided}), 
the two $16$'s contract to from a $\overline{5}$ 
of $SU(5)$, while inside the second parenthesis, the two $16$'s contract to 
form a $5$ of $SU(5)$. This contraction arises by integrating out a pair of $5$ 
and $\overline{5}$ of $SU(5)$ from the $10$'s of $SO(10)$, 
as shown in Fig.\ref{lopsideddiagram}.
Because the $\overline{5}$ contains the $SU(2)$ lepton doublet and the singlet down-type 
quarks, the resulting mass terms
\begin{equation}
\lambda (d_{L,i}^{c}d_{L,j} + e_{L,i}e_{L,j}^{c}) v_{d}
\end{equation}
are related by $M_{d} = M_{e}^{T}$ and the lop-sided mass texture arises,
\begin{equation}
(\overline{d}_{R,2} \; \overline{d}_{R,3})
\left(\begin{array}{cc}
0 & \lambda\\
0 & 0
\end{array}\right)
\left(\begin{array}{c}
d_{L,2}\\
d_{L,3}
\end{array}\right)v_{d}
+ 
(\overline{e}_{R,2} \; \overline{e}_{R,3})
\left(\begin{array}{cc}
0 & 0\\
\lambda & 0
\end{array}\right)
\left(\begin{array}{c}
e_{L,2}\\
e_{L,3}
\end{array}\right)v_{d},
\end{equation}
if $(i,j)$ is chosen to be $(2,3)$. The $(33)$ entry which is expected to be 
of order $\mathcal{O}(1)$ 
is generated by tree level diagram involving a $10$. The $(23)$ and $(32)$ entries of 
$M_{d}$ and $M_{e}$ also receive contributions from other non-renormalizable operators, 
for example, $16_{i} 16_{j} 45_{H} 10_{H}$. If $\lambda$ is of order 
$\mathcal{O}(1)$ and other contributions to the $(23)$ and $(32)$ entries 
are much smaller than one, a large mixing angle is then generated in the 
right-handed down quark sector, and the left-handed charged lepton sector, 
while the corresponding mixing angle in the left-handed down quark sector 
is small. This thus provides a way to explain the large mixing angle in atmospheric 
neutrinos while the quark mixing $V_{cb}$ is small. 

As we have seen above, when additional matter fields are introduced into the model and 
non-renormalizable operators that generate fermion masses are taken 
into account, as in the case of the Froggatt-Nielsen mechanism, 
other Higgs representations can play a role in the mass generation. 
An interesting case is the $45$-dimensional Higgs representation which 
has Yukawa couplings to a $16$- and a $10$-dim matter fields. 
These non-renormalizable operators can be expressed 
generically by\cite{Anderson:1993fe}
\begin{equation}
\mathcal{O}_{ij} = 16_{i} \cdot \mathcal{O}_{n} \cdot 10 \cdot \mathcal{O}_{m} 
\cdot 16_{j}
\end{equation}
where the operator $\mathcal{O}_{n}$ is given by
\begin{equation}
\mathcal{O}_{n} = 
\frac{M_{G}^{p} \; 45_{p+1} ... 45_{q}}{M_{Pl}^{q} \; 45_{X}^{n-q}}, \qquad 
n, \; p, \; q = {\mbox integer}.
\end{equation}
Here $M_{G}$ and $M_{Pl}$ refer to the GUT scale and the Planck scale respectively. 
The $45$-dimensional representation Higgs can acquire VEV along the following four 
directions: $X, Y, B-L, T_{3R}$, where $X$ and $Y$ are defined as
\begin{eqnarray}
X & = & -(\Sigma_{12} + \Sigma_{34} + \Sigma_{56} + \Sigma_{78} + \Sigma_{9 \; 10})
\\
Y & = & y \otimes \zeta
\nonumber\\
& & {\mbox where} \qquad  
\zeta = \left(\begin{array}{cc}
0 & -i\\
i & 0
\end{array}\right), \qquad
y = diag(1/3,1/3,1/3,-1/2,-1/2).
\end{eqnarray}
A systematical way to search for these effective operators is discussed in 
Ref. 95. 
The coefficient $-3$ is obtained in this case 
when the $45$-dimensional Higgs acquires a VEV along the $B-L$ direction.
We summarize in Table.~\ref{CG} all possible CG coefficients for the four possible 
directions of $<45>$.
\begin{table}[bh!]
\tbl{\label{CG}CG coefficients for the four possible 
directions of $<45>$.}
{\begin{tabular}{ccccccccc}\toprule
 & $u$ & $\overline{u}$ & $d$ & $\overline{d}$ 
 & $e$ & $\overline{e}$ & $\nu$ & $\overline{\nu}$
\\
\hline
B-L & 1 & -1 & 1 & -1 & -3 & 3 & -3 & 3
\\
$T_{3R}$ & 0 & -1/2 & 0 & 1/2 & 0 & 1/2 & 0 & -1/2
\\
X & 1 & 1 & 1 & -3 & -3 & 1 & -3 & 5
\\
Y & 1/3 & -4/3 & 1/3 & 2/3 & -1 & 2 & -1 & 0
\\
\botrule
\end{tabular}}
\end{table}

\subsection{Automatic R-parity conservation and 
$\overline{126}$ v.s. $(16 \times 16)$ for neutrino masses}

One of the salient features of $SO(10)$ is that in some classes of models,
R-parity is conserved automatically. 
Group theoretically, the congruence number is defined as follows:
For an irreducible representation of $SO(2n)$ whose Dynkin index reads
$(a_{1} \; a_{2} \; ... a_{n})$, this representation has congruence number
\begin{eqnarray}
(c_{1},c_{2}) & \equiv & (a_{n-1}+a_{n},2a_{1}+2a_{3} + ... +
2a_{n-2}+(n-2)a_{n-1}+na_{n})
\nonumber\\
&& \mbox{mod} \; (2,4) \qquad \mbox{for} \; n = \; \mbox{odd}.
\end{eqnarray}
There are four possible classes of $(c_{1},c_{2})$: $(0,0), (0,2), (1,1),
(1,3)$. The first two classes are tensor-like while the later two classes are
spinor-like which is troublesome. To see how $(c_{1},c_{2})$
relates to R-parity, alternatively we can define, in $SO(10)$, 
\begin{equation}
c = 3 (B-L) \; \mbox{mod} \; 4
\end{equation}
and $c = \mbox{Max} \; (c_{1},c_{2})$.
It has been shown that if all the Higgs representations that break $SO(10)$
down to the SM are chosen to have congruence number $c = 0 \; \mbox{or} \; 2$,
then R-parity is preserved at all energies.\cite{Mohapatra:su,Martin:1992mq}  
Representations having $c=0$ are: $45,\; 54, \; 210, ...$; those having $c=2$
are: $10, \; 126, \; \overline{126}, ...$. Note that the spinor representations
$16$ and $\overline{16}$ have $c=1$ and $3$ respectively. 

Some models avoid the use of $\overline{126}$-dim Higgses by introducing 
non-renormalizable operators of the form
$\frac{1}{M} \psi_{a}\psi_{b}(\overline{16}_{H})(\overline{16}_{H})$ instead of a
renormalizable $\psi_{a}\psi_{b} \overline{126}_{H}$. Models utilizing the spinor 
representation $16$ to construct the neutrino mass operators generally have R-parity
broken at some high energy scale. Such models also appear to be
less constrained due to the inclusion of non-renormalizable operators. Also, a
discrete symmetry, the R-parity symmetry, must be imposed by hand to avoid
dangerous dim-$4$ baryon number violating operators in the effective
potential at low energies which otherwise could lead to fast proton decay
rate. Utilizing $\overline{126}$-dim representation of Higgses 
has the advantage that R-parity symmetry is automatic. 
The $\overline{126}$ representation has been
used in model building
before.\cite{Lee:1994je,Brahmachari:1997cq,Aulakh:1999cd,Aulakh:2000sn,Chen:2000fp,Chen:2001pr,Chen:2002pa,Nath:2001uw,Nath:2001yj}  
It is to be noted that the contribution of the $\overline{126}$-dimensional
representation to the $\beta$-function makes the model nonperturbative  (with
the onset of the Landau pole) above the unification scale $M_{GUT}$. One could
view these models as an effective theories valid below this scale where coupling
constants are perturbative. One may also argue against the use of
$\overline{126}$ with the fact that it is not possible to obtain such a large
representation from string theory. It has been shown that in heterotic string
theory it is not possible to get $126$ of $SO(10)$ up to Kac-Moody 
level-$5$.\cite{Dienes:1996wx} Nevertheless, no such constraints have been 
found in other types of string theories.\cite{Dienes:comm} 

\subsection{Some Related Issues}

\subsubsection{Proton Decay}

One of the signatures of any GUT model is baryon number violating processes.
These processes include proton decay ($\Delta(B-L)=0$), $N-\overline{N}$
oscillation ($\Delta(B-L)\ne0$), etc. The theoretical prediction for
oscillation time in $N-\overline{N}$ oscillation can naturally satisfy the
experimental lower limit $\tau_{N-\overline{N}} \ge 0.86 \times 10^{8}$
sec,\cite{Baldo-Ceolin:1994jz} if a high $B-L$ breaking scale is assumed.
On the other hand, the non-observation of proton decay has put many GUT models
under siege. There are three kinds of operators leading to proton decays in
SUSY GUT's:\\

\noindent{(i) Dimension-$6$ operators:}

\noindent{As we have mentioned previously, the extra gauge bosons 
in GUT models, the $X$ and $Y$ gauge bosons, can lead to proton decay. 
The terms in the Lagrangian containing the $X$ $Y$ gauge bosons are
\begin{eqnarray}
\mathcal{L}_{X,Y} & = &
\frac{ig_{X}}{\sqrt{2}}X_{\mu,i} 
(\epsilon_{ijk}\overline{u}^{c}_{kL}\gamma_{\mu}u_{jL}
+ \overline{d}_{i}\gamma_{\mu}e^{+}) \nonumber\\
& & + \frac{ig_{Y}}{\sqrt{2}}Y_{\mu,i} 
(\epsilon_{ijk}\overline{u}^{c}_{kL}\gamma_{\mu}d_{jL}
- \overline{u}_{iL}\gamma_{\mu}e^{+}_{L} 
+ \overline{d}_{iR}\gamma_{\mu}\nu^{c}_{R}) + h.c.
\end{eqnarray}
and the following vertices are allowed
\begin{equation}
\overline{X} \rightarrow uu, \; \overline{d}e^{+}, \qquad
\overline{Y} \rightarrow ud, \; \overline{u} e^{+}.
\end{equation}
These can thus lead to proton decay via the dim-$6$ operators. 
Note that these type of operators exist in both non-SUSY and SUSY GUT's. The
dominant decay mode is $p \rightarrow e^{+}\pi^{0}$, and the decay amplitude
associated with this mode is\cite{Mohapatra:1999vv}
\begin{equation}
\mathcal{M}_{p \rightarrow e^{+} \pi^{0}} \simeq 
4\pi \alpha_{GUT} / M_{GUT}^{2}
\end{equation}
leading to a life-time of\cite{Mohapatra:1999vv}
\begin{equation}
\tau_{p} \simeq 
\frac{1}{\mathcal{M}_{p \rightarrow e^{+} \pi^{0}}^{2}m_{p}^{5}}  
\simeq 4.5 \times 10^{29{+\atop-}0.7} (\frac{M_{GUT}}{2.1\times 10^{14}
GeV})^{4} 
\end{equation}
where $m_{p}$ is the mass of the proton. 
For $M_{GUT} \simeq 2 \times 10^{16} GeV$, we get $\tau_{p} \sim 4.5 \times  
10^{37{+\atop-}0.7} \; \mbox{years}$ which is far above the current capability
of SuperKamiokande experiments whose limit is $\sim 10^{34}$ years.}
\\

\noindent{(ii) Dimension-$5$ operators}

\noindent{In SUSY GUT's, a new channel for proton decay is possible via the dim-$5$
operator through the exchange of the color triplet Higgsinos where $QQH$ and
$QL\overline{H}$ via $H\overline{H}$ mixing generate an effective operator
\begin{equation}
QQQL/M_{H}.
\end{equation} 
In order to suppress this operator, the masses of the
color triplet Higgsinos must have superheavy masses. From the point of view of
unification, we would like to have the spectrum of MSSM below $M_{GUT}$. This
requires that all the Higgs fields, including the color triplet Higgsinos, to
be very heavy, with masses of the order of $M_{GUT}$, except the pair of
$SU(2)_{L}$ doublet Higgses remaining light which are then identified as the
pair of Higgs doublets in MSSM. How to achieve such a mass splitting is
referred to as the  {\it doublet-triplet splitting} (DTS) and {\it doublet-doublet 
splitting} (DDS) problems. The dominant decay mode of these operator is 
$\tau(p\rightarrow K^{+}\tilde{\nu}) \sim
m_{\tilde{h}}^{2} \sim M_{GUT}^{2}$. Its decay amplitude is\cite{Mohapatra:1999vv}
\begin{equation}
\mathcal{M}_{p\rightarrow K^{+}\tilde{\nu}} \simeq
\frac{h_{u}h_{d}}{M_{H}}
\frac{m_{gaugino}g_{GUT}^{2}}{16\pi^{2}M_{\tilde{Q}}^{2}}.
\end{equation}
Dimopolous and Wilczek proposed a mechanism\cite{Dimopoulos:zu} 
to achieve such mass splittings using $<45_{H}>$ along the $(B-L)$ direction
\begin{equation}
<45_{H}> = i\tau_{2}\otimes {\mbox diag}(a,a,a,0,0).
\end{equation}
Chacko and Mohapatra\cite{Chacko:1998zn} found that with a complimentary VEV pattern 
to that of the Dimopoulos-Wilczek type, that is,
\begin{equation}
<45_{H}> = i\tau_{2}\otimes {\mbox diag}(0,0,0,b,b)
\end{equation}
the same goals can be achieved. A solution utilizing $126_{H}$ instead of $45_{H}$ 
to achieve the DTS and DDS was proposed by Lee and Mohapatra.\cite{Lee:1994je}
Detailed calculations have shown, nevertheless, that even with these 
mechanisms in place, in order for the prediction of $\tau_{p}$ to be consistent 
with the experimental limit, the effective $M_{H}$ must be larger than 
$M_{GUT}$ by at least 
a factor of $10$. This in turn requires some couplings to be much larger 
than $1$ which is somewhat unnatural.}\\ 

\noindent{(iii) Dimension-$4$ operators:}

\noindent{As we have seen, the dim-$4$ operators are
forbidden if there is R-parity in the model. In SUSY $SU(5)$ one has to impose
R-parity by hand, while in some $SO(10)$ models, R-parity is
automatically conserved if certain type of Higgs fields are chosen to
construct the model, as we have discussed in the previous section.}

\subsubsection{Baryogenesis}

The three Sakharov conditions,\cite{Sakharov:1967dj} 
({\it i}) baryon number ($B$) violating processes, ({\it ii}) $C$ and
$CP$ violation, and ({\it iii}) the deviation from thermal equilibrium, for the
generation of the cosmological matter anti-matter asymmetry can be naturally
satisfied in the $SO(10)$ model. Due to the presence of sphaleron effects,
the only chance for GUT baryogenesis to work is to produce an asymmetry in
$B-L$ at a high scale. To see this, let us first write the baryon number $B$
as  \begin{equation} 
B=\frac{1}{2} (B+L) + \frac{1}{2} (B-L).
\end{equation}
The electroweak sphaleron transitions rapidly erase the asymmetry 
$B+L$ as soon as the temperature drops down to about $10^{12} \; GeV$.
Therefore, to have a non-vanishing baryonic asymmetry requires a
non-vanishing $B-L$ asymmetry. The crucial point to note is that even though
the $B+L$ asymmetry is erased by the sphaleron transitions, the orthogonal
combination $B-L$ is left untouched, and it opens up the possibility that the
baryonic asymmetry can be generated through 
leptogenesis.\cite{Fukugita:1986hr,Luty:1992un,Buchmuller:1996pa} 
The basic idea is that, since $B+L$ must vanish at all
times due to the sphaleron transitions, the asymmetry in the lepton number
will consequently be converted into the asymmetry in the baryon number (with
an opposite sign). The primordial leptonic asymmetry is generated by the
out-of-equilibrium decay of the heavy right-handed Majorana neutrinos and
their scalar partners in the supersymmetric case.  The relevant superpotential
is \begin{equation}
W_{Leptogenesis} =
(Y_{e})_{ij} E_{i}^{c}L_{j}H_{1} 
+ (Y_{\nu_{LR}})_{ij} N_{i}^{c}L_{j}H_{2}
+ \frac{1}{2} (M_{RR})_{ij} N_{i}^{c}N_{j}^{c}. 
\end{equation}
The heavy right-handed neutrinos and their scalar
partners can decay through the following four decay modes: 
\begin{eqnarray}
N_{1} & \longrightarrow & \tilde{l} \quad + \quad \tilde{h^{c}}\\ 
N_{1} & \longrightarrow & l \quad + \quad H_{2}\\ 
\tilde{N_{1}^{c}} & \longrightarrow & \tilde{l} \quad + \quad H_{2}\\
\tilde{N_{1}^{c}} & \longrightarrow & l \quad + \quad \tilde{h^{c}}.
\end{eqnarray} 
The interference between the tree-level and one-loop diagrams   
thus gives rise to the CP asymmetry.

In the basis where both charged lepton Yukawa couplings and 
the right-handed neutrino mass matrix are diagonal, the amount
of CP asymmetry due to the interference between the tree level 
and one-loop diagrams      
for {\it each} decay mode is given by\cite{Buchmuller:1996pa}
\begin{equation}
\epsilon_{1} =
-\frac{1}{8\pi} 
\frac{1}{(h_{\nu}h_{\nu}^{\dagger})_{11}}
\sum_{i=2,3} Im \{ (h_{\nu}h_{\nu}^{\dagger})_{1i}^{2} \}
f(\frac{M_{i}^{2}}{M_{1}^{2}})
\end{equation}
where
\begin{equation}
f(x)=\sqrt{x} \; [ \; \ln(\frac{1+x}{x}) + \frac{2}{x-1} \;],
\end{equation}
and $h_{\nu}$ is the Dirac neutrino Yukawa matrix in the new basis. 
The right-handed Majorana neutrino mass matrix is
diagonalized by 
\begin{equation}
P_{M} O_{R} M_{RR} O_{R}^{T} P_{M} = diag(M_{1},M_{2},M_{3})
\end{equation}
where $M_{i}$'s are real and non-negative, and $P_{M}$ is the
diagonal Majorana phase matrix. In terms of the diagonalization matrices and
the original Dirac neutrino Yukawa coupling, $(h_{\nu}h_{\nu}^{\dagger})$ can
then be rewritten as 
\begin{equation}
h_{\nu}h_{\nu}^{\dagger} 
= P_{M_{RR}} O_{R}Y_{\nu_{LR}}U_{e_{L}}^{\dagger}U_{e_{L}}
Y_{\nu_{LR}}^{\dagger}O_{R}^{\dagger} P_{M_{RR}}^{-1}
=P_{M_{RR}} O_{R}Y_{\nu_{LR}}Y_{\nu_{LR}}^{\dagger}
O_{R}^{\dagger} P_{M_{RR}}^{-1}.
\end{equation}
We see that the phases in the right-handed neutrino mass matrix  $M_{RR}$
and the Dirac neutrino mass matrix $Y_{\nu_{LR}}$ are responsible for the CP
asymmetry needed for the leptogenesis. For a hierarchical heavy right-handed
neutrino mass spectrum,  $M_{3} \gg M_{2} \gg M_{1}$, the argument of the
function $f(x)$, $x \equiv \frac{M_{i}^{2}}{M_{1}^{2}}$, is much greater than
$1$. We can then approximate $f(x)$ as 
\begin{equation}
f(x) = \sqrt{x} \;
[ \; (\frac{1}{x} - \frac{1}{2 x^{2}} + \ldots) +\frac{2}{x-1} \;] 
\simeq \frac{3}{\sqrt{x}}
= 3 \frac{M_{1}}{M_{i}}.
\end{equation}
The asymmetry $\epsilon_{1}$ can thus be rewritten as
\begin{equation}\label{epsilon1}
\epsilon_{1} \simeq -\frac{4}{8\pi} 
\frac{1}{(h_{\nu}h_{\nu}^{\dagger})_{11}}
\sum_{i=2,3} Im \{ (h_{\nu}h_{\nu}^{\dagger})_{1i}^{2} \}
(\frac{3M_{1}}{M_{i}}) 
\end{equation}
where the factor $4$ accounts for the fact that there are four decay modes.
Using the fact that the mixing in $Y_{\nu_{LR}}$ is small and that
$(Y_{\nu_{LR}})_{33}$ dominates other elements, we can further approximate
\begin{eqnarray}
(h_{\nu}h_{\nu}^{\dagger})_{1i} & \simeq & 
(P_{11}P_{ii}^{-1}) \vert y_{\nu_{\tau}} \vert^{2}
(O_{R})_{13}(O_{R}^{\ast})_{i3}
\\
(h_{\nu}h_{\nu}^{\dagger})_{11} & \simeq & 
\vert y_{\nu_{\tau}} \vert^{2}
\vert (O_{R})_{13} \vert^{2}
\end{eqnarray}
and
\begin{equation}
\frac{Im \{ (h_{\nu}h_{\nu}^{\dagger})_{1i}^{2} \} }
{(h_{\nu}h_{\nu}^{\dagger})_{11}}
\simeq  
\vert y_{\nu_{\tau}} \vert^{2}
Im \{ (P_{11}P_{1i}^{-1})^{2} \frac{(O_{R})_{13}}{(O_{R})_{13}^{\ast}} 
(O_{R}^{\ast})_{i3}^{2} \}.
\end{equation}
To have a large amount of CP asymmetry, $\epsilon_{1}$, thus requires that the
hierarchy among the three right-handed neutrino masses cannot be too large
(that is, $\frac{M_{1}}{M_{2,3}}$ cannot be too small), and that the imaginary
part of 
$\{ (P_{11}P_{1i}^{-1})^{2} \frac{(O_{R})_{13}}{(O_{R})_{13}^{\ast}} 
(O_{R}^{\ast})_{i3}^{2} \}$
together with the neutrino Dirac Yukawa couplings 
cannot be too small.

The amount of the lepton asymmetry generated is given by
\begin{equation}
Y_{L} \equiv \frac{n_{L}-\overline{n}_{L}}{s} = 
\kappa \frac{\epsilon_{1}}{g_{\ast}}.
\end{equation}
Here $g_{\ast}$ is the number of relativistic degrees of freedom. For MSSM,
it is $g_{\ast}=228.75$. 
The out-of-equilibrium decay of the heavy Majorana neutrinos requires the
decay width of the lightest neutrino, $\Gamma_{1}$, smaller than the Hubble
constant at the temperature of the decay. That is, 
\begin{equation}
r \equiv \frac{\Gamma_{1}}{H \vert_{T=M_{1}}} 
= \frac{M_{pl}}{(1.7)(32\pi)\sqrt{g_{\ast}}} 
\frac{( h_{\nu} h_{\nu}^{\dagger} )_{11}}{M_{1}}
< 1
\end{equation}
where $M_{Pl}$ is the Planck scale taken to be $1.2 \times 10^{19} \; GeV$. In
general, one can still have a sizable CP asymmetry remains even for 
$1 < r < 10$. 
The wash-out effects due to inverse decays and lepton number violating
scattering processes together with the time evolution of the system is then 
accounted for by the factor $\kappa$. 
It is obtained by solving the Boltzmann equation for the system. 
An approximation is given by\cite{Kolb:vq}
\begin{eqnarray}
10^{6} \le r: & \quad 
\kappa = (0.1 r)^{1/2} e^{-(\frac{4}{3})(0.1r)^{1/4}} \quad & 
( < 10^{-7} )\\
10 \le r \le 10^{6}: & \quad \kappa = \frac{0.3}{r (\ln r)^{0.6}} \quad &
(10^{-2} - 10^{-7})\\
0 \le r \le 10: & \quad \kappa = \frac{1}{2\sqrt{r^{2}+9}} \quad &
(10^{-1} - 10^{-2})   
\end{eqnarray}
where inside the parentheses we give the order of magnitude of $\kappa$ for
each corresponding $r$. We note that in order to have a small 
dilution factor, the lightest right-handed neutrino cannot be too
light.  The electroweak sphaleron effect will convert the lepton asymmetry 
$Y_{L}$ into baryon asymmetry $Y_{B}$, and they are related by 
\begin{equation}
Y_{B} \equiv \frac{n_{B}-\overline{n}_{B}}{s} = c Y_{B-L} = \frac{c}{c-1} Y_{L}
\end{equation}
with
\begin{equation}
c = \frac{8N_{F}+4N_{H}}{22N_{F}+13N_{H}}
\end{equation}
where $N_{F}$ is the number of families and $N_{H}$ is the number of 
$SU(2)$ Higgs doublets. 
For the MSSM spectrum, $(N_{F},N_{H})=(3,2)$, we have the conversion factor 
$(\frac{c}{c-1}) \simeq -0.53$. 

Models with only two RH neutrinos have been constructed by Frampton 
{\it et al}~\cite{Frampton:2002qc} and by Raby\cite{Raby:2003ay} (see Sec. 3 and 6),  
which give rise to bi-large mixing pattern and a correlation between 
the sign of the baryon number asymmetry and the sign of the CP violation 
in neutrino oscillation. General analyses on the consistency between 
constraints from simplest $SO(10)$ models and leptogenesis 
can be found in Ref. 117, 118. 

\subsection{SUSY Breaking}

SUSY breaking can be incorporated into models by including explicitly the
soft SUSY breaking terms. Since the RGE's for the
Yukawa coupling constants and gauge coupling constants do not have any
dependence on the soft breaking parameters up to two-loop level, 
the presence of these soft SUSY breaking terms does not affect the predictions for 
fermion masses and mixing angles. On the other
hand, since the evolutions of soft SUSY breaking parameters does depend on the
Yukawa coupling constants and gauge coupling constants, whether the EW
symmetry is broken (that is, the mass-square of the light Higgs doublet is
driven to be negative) may depend on the Yukawa sector, which can be used as a
test of the validity of the model. 

Even though SUSY breaking does not affect the running of the gauge coupling
constants and that of the Yukawa coupling constants, it could have
a large contribution to the threshold corrections, which is the subject of the 
next section.

\subsection{Threshold Corrections}

When the RGE analysis is performed, we usually consider $\beta$
function coefficient as a constant for each coupling constant between the two
relevant scales. This is done under the assumption that all the heavy modes
decouple at the same scale, the symmetry breaking scale. Threshold corrections
are the corrections due to the differences between the symmetry breaking scale
and the masses of the heavy particles decoupled from the spectrum after
symmetry breaking takes place. If all the heavy particles acquire masses
exactly the same as the symmetry breaking scale, there are no threshold
corrections. In practice, this is not the case. There are two possible sources
of threshold corrections:

\noindent{(i) GUT scale threshold corrections:} Due to the presence of many 
large Higgs representations in $SO(10)$, the GUT scale threshold corrections 
could be large. 

\begin{figure}
\centerline{
\psfig{file=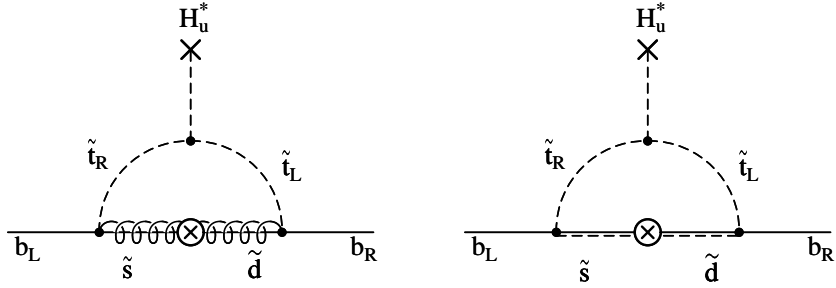,width=11cm}}
\caption{\label{threshold} Leading order contributions to 
the threshold corrections to the mass 
of the b-quark.}
\end{figure}

\noindent{(ii) SUSY threshold corrections:} 
Large threshold corrections to $m_{b}$, $\delta m_{b} / m_{b} \simeq -(0.15 
\sim 0.2)$, are needed in most $SO(10)$ models in order to have a prediction for $m_{b}$ 
consistent 
with experiment. 
The dominant contributions are from the diagrams shown in Fig.~\ref{threshold}.
They give a correction
\begin{equation}
\delta m_{b} \; / \; m_{b} \simeq (\tan\beta/50) I
\end{equation}
and $I$ is given by,\cite{Hall:1993gn,Rattazzi:1995gk,Babu:1999hn}  
\begin{equation}
I \simeq
\frac{50}{16\pi^{2}}\frac{\alpha_{G}}{\alpha_{2}}
\frac{\mu m_{\tilde{W}}}{m_{eff}^{2}}
[\frac{8}{3}\frac{\alpha_{3}}{\alpha_{G}}g_{3}^{2}
f(\frac{m_{\tilde{g}}^{2}}{m_{eff}^{2}})
-2\lambda_{t}^{2}f(\frac{\mu^{2}}{m_{eff}^{2}})]
\end{equation} 
where $f(x)=(1-x+x\ln x)/(1-x)^{2}$; $m_{\tilde{g}}$ and $\mu$ 
are the gluino mass and the $\mu$ term evaluated at the weak scale.; 
$m_{eff}^{2} \equiv \frac{1}{2}(m_{\tilde{b}}^{2}+ m_{\tilde{Q}}^{2})$ 
is the average of the squared masses of the $SU(2)$-singlet bottom squark and
the $SU(2)$-doublet third generation squarks. 
A large soft SUSY breaking parameter space can give rise to such a correction.
With the typical values $\alpha_{G}=0.75$ and 
$(\alpha_{3},\alpha_{2},y_{33}) \simeq (0.124,0.034,1)$ at $M_{weak}$,
$\delta_{b}$ can then be approximated as, with the assumption 
$2 m_{1/2} = 2 \frac{\alpha_{G}}{\alpha_{w}} m_{\tilde{W}}$,
\begin{equation}
I = (0.315) \; x \; t \; ( 0.69 f(t^2)  -  f(x^2) )
\end{equation}
where $t=(\mu/m_{0})$ and $x = m_{1/2}/m_{0}$ with $m_{1/2}$ and $m_{0}$
being the gluino mass and the common scalar mass respectively in the constrained MSSM (CMSSM).   
With $(t,x)=(5.2,2)$ which are typical values in CMSSM,\cite{Castano:1993ri}  
a correction
$(\delta m_{b} / m_{b} ) = -0.15$ for $\tan\beta=10$ is obtained.


\section{$SO(10)$ Models with Texture Assumptions}\label{modelt}

In what follows we discuss a few selected $SO(10)$ models combined with 
some texture ansatz for the mass matrices.

\subsection{Buchmuller and Wyler}

Buchmuller and Wyler\cite{Buchmuller:2001dc} assume symmetric mass textures 
for the up- and down-type quarks
\begin{equation}
M_{u,d} \sim \left(\begin{array}{ccc}
0 & \epsilon^{3}e^{i\phi} & 0\\
\epsilon^{3}e^{i\phi} & \rho \epsilon^{2} & \eta \epsilon^{2}\\
0 & \eta \epsilon^{2} & e^{i\psi}
\end{array}\right).
\end{equation}
Here $\rho = |\rho|e^{i\alpha}$ and $\eta = |\eta|e^{i\beta}$ are complex parameters 
of $\mathcal{O}(1)$. Using the $SO(10)$ relations, they have 
\begin{equation}
m_{\nu}^{Dirac} = m_{u}, \qquad m_{e} = m_{d}
\end{equation}
{\it assuming} the incorrect mass relations in the lighter two generations are lifted 
when higher dimensional Higgs representations are introduced. The right-handed 
neutrino Majorana mass matrix is generated by $\overline{126}_{H}$, and is assumed to 
have the following form
\begin{equation}
M_{\nu,RR} = \left(\begin{array}{ccc}
0 & M_{12} & 0\\
M_{12} & M_{22} & M_{23}\\
0 & M_{23} & M_{33}
\end{array}\right).
\end{equation}
With the relations
\begin{equation}
M_{12} : M_{22} : M_{33} = \epsilon^{5}:\epsilon^{4}:1, \qquad 
M_{23} \sim M_{22},
\end{equation}
the resulting effective neutrino mass matrix has the following form
\begin{equation}
M_{\nu}^{eff} = \left(\begin{array}{ccc}
0 & \epsilon e^{2i\phi} & 0\\
\epsilon e^{2i\phi} & -\sigma e^{2i\phi}+2\rho e^{i\phi} & \eta e^{i\phi}\\
0 & \eta e^{i\phi} & e^{2i\phi}
\end{array}\right) \cdot \frac{v_{1}^{2}}{M_{3}}
\end{equation}
where $M_{3}$ is the heaviest eigenvalue of $M_{\nu, RR}$ which is of 
the same order 
as $M_{33}$, and $\sigma$ is defined as $\sigma \epsilon^{4} = M_{22}/M_{3}$. 
The parameters $\eta, \; \rho$ and $\sigma$ are all of $\mathcal{O}(1)$, and 
$\eta$ and $\rho$ are determined using the quark masses. 
The effective neutrino masses form the following pattern,
\begin{equation}
m_{\nu_{1}} : m_{\nu_{2}} : m_{\nu_{3}} = \epsilon : \epsilon : 1
\end{equation}
with $\epsilon \sim 0.1$, we have 
$(m_{\nu_{2}}^{2}-m_{\nu_{1}}^{2})/(m_{\nu_{3}}^{2}-m_{\nu_{2}}^{2}) \sim 10^{-2}$, 
which is consistent with the LMA solution. 
Nevertheless, it is not clear whether the predicted solar angle is consistent with 
experiment or not. The prediction for $U_{e\nu_{3}}$ is of order 
$\mathcal{O}(\epsilon)$.

An interesting feature of this model is that, with $\epsilon \sim 0.1$, the scale 
of $M_{1}$ is about $10^{9} \; GeV$. The amount of baryonic asymmetry which 
is very sensitive to this scale is given by
\begin{equation}
Y_{B} \sim -\kappa \; sign(\sigma) \; \sin(\phi-\alpha) \times 10^{-9}.
\end{equation}
The result has the promise to be consistent with the observed value, 
once the parameters $\kappa$, $\sigma$, $\phi$ and $\alpha$ are fixed.

\subsection{Matsuda, Fukuyama and Nishiura}

Matsuda, Fukuyama and Nishiura\cite{Matsuda:1999yx} proposed Hermitian textures 
with four zeros in the context of $SO(10)$
\begin{eqnarray}\label{4zero}
M_{u,\nu} & = & \left(\begin{array}{ccc}
0 & A_{u,\nu} & 0\\
A_{u,\nu} & B_{u,\nu} & C_{u,\nu}\\
0 & C_{u,\nu} & D_{u,\nu}
\end{array}\right)
\\
M_{d,e} & = & \left(\begin{array}{ccc}
0 & A_{d,e}e^{i\alpha_{12}^{d,e}} & 0\\
A_{d,e}e^{-i\alpha_{12}^{d,e}} & B_{d,e} & C_{d,e}e^{i\alpha_{23}^{d,e}}\\
0 & C_{d,e} e^{-i\alpha_{23}^{d,e}} & D_{d,e}
\end{array}\right).
\end{eqnarray}
For a general matrix of the form
\begin{equation}
M = \left(\begin{array}{ccc}
0 & A & 0\\
A & B & C\\
0 & C & D
\end{array}\right),
\end{equation}
one can relate its three eigenvalues, $m_{1}$, $m_{2}$, and $m_{3}$, 
to the matrix elements by
\begin{eqnarray}
DA^{2} & = & -m_{1}m_{2}m_{3}\\
BD-A^{2}-C^{2} & = & m_{1}m_{2}+m_{2}m_{3}+m_{3}m_{1}\\
B+D & = & m_{1}+m_{2}+m_{3}.
\end{eqnarray}
Because of the observed fermion mass hierarchy, $|m_{3}| \gg |m_{2}| \gg |m_{1}|$, 
it is a good approximation to write 
$B=m_{2}$ and $D=m_{3}-m_{1}$ and express the mass matrix in terms of its three 
eigenvalues as
\begin{equation}
M \simeq \left(\begin{array}{ccc}
0 & \sqrt{-m_{1}m_{2}} & 0\\
\sqrt{-m_{1}m_{2}} & m_{2} & \sqrt{-m_{1}m_{3}}\\
0 & \sqrt{-m_{1}m_{3}} & m_{3}+m_{1}
\end{array}\right).
\end{equation}
To put this idea to work in the context of $SO(10)$, we first note that 
the set of equations given in Eq.(160)-(163) can be re-written as,
assuming $<120>$ is small and contribute to $M_{e}$ and $M_{d}$ only,
\begin{eqnarray}
M_{u} & = & Y^{10} <10> + \epsilon Y^{126} <\overline{126}>
\\
M_{d} & = & \alpha Y^{10} <10> + Y^{126} <\overline{126}> 
+ Y^{120} <120>
\nonumber\\
& = & \alpha M_{u} - (\alpha\epsilon-1) Y^{126} <\overline{126}> 
+ Y^{120} <120>
\\
r M_{e} & = & \alpha Y^{10} <10> -3 Y^{126} <\overline{126}> 
+ \delta Y^{120} <120>
\nonumber\\
& = & \alpha M_{u} - (\alpha\epsilon+3) Y^{126} <\overline{126}> 
+ \delta Y^{120} <120>
\\
r^{'} M_{\nu}^{Dirac} & = & Y^{10} <10> - 3 \epsilon Y^{126} <\overline{126}>
\\
s M_{\nu,LL} & = & \beta Y^{126} <\overline{126}>
\\
s^{'} M_{\nu,RR} & = & \gamma Y^{126} <\overline{126}>
\end{eqnarray}
where parameters $\alpha,\beta, \gamma$ are ratios of SM Higgs doublet VEVs 
from different $SO(10)$ representations.  
The symmetric (anti-symmetric) matrices $Y^{10,126}$ ($Y^{120}$) can be expressed 
in terms of the symmetric (anti-symmetric) part of the mass matrices 
$M_{u,d,e,\nu}$ as
\begin{eqnarray}
(1-\alpha \epsilon) Y^{10} <10> & = & (M_{u})_{s} - \epsilon (M_{d})_{s}\\
Y^{126} <\overline{126}> & = & \frac{1}{4} (M_{d})_{s} - \frac{1}{4} r (M_{e})_{s}\\
Y^{120} <120> & = & (M_{d})_{a}
\end{eqnarray}
where the subscripts $s$ and $a$ refer to symmetric part and anti-symmetric part, 
respectively. One can then solve for the elements in matrices $Y$'s in terms of 
quark masses. As the simple approximations, $B=m_{2}$ and $D=m_{3}+m_{1}$, 
work well for quark masses and CKM matrix elements, in order to obtain 
viable neutrino masses and mixing angles, a deviation must be made in the charged 
lepton mass matrix, $B_{e} = m_{\mu} (1+\xi)$ and 
$D_{e} =m_{\tau}+m_{e}-\xi m_{\mu}$. 
With $\xi \sim 0.01$, maximal $\nu_{\mu}-\nu_{\tau}$ mixing angle 
and the LMA solution can be accommodated. In this model, the allowed region for 
the leptonic CP violating Dirac phase can be obtained.

\subsection{Bando and Obara}

Bando and Obara\cite{Bando:2002tu,Bando:2003ei} pursue along the line of 
Matsuda, Fukuyama and Nishiura\cite{Matsuda:1999yx} to analyze mass matrices 
of the type given in Eq.(\ref{4zero}), and have a detailed analysis on all possible 
combinations of contributions from either $<10_{H}>$ or  
$<\overline{126}_{H}>$ for each non-vanishing entry. In other words, 
all possible ways the CG factor $(-3)$ due to $<\overline{126}_{H}>$ can appear 
in the neutrino Dirac mass matrix
\begin{equation}
\left(\begin{array}{ccc}
0 & \ast a_{\nu} & 0\\
\ast a_{\nu} & \ast b_{\nu} & \ast c_{\nu}\\
0 & \ast c_{\nu} & 1
\end{array}\right)
\end{equation}
where $\ast$ is either $1$ or $-3$ depending upon whether the coupling is due to 
$<10_{H}>$ or $<\overline{126}_{H}>$.
They found the following texture has best agreement with experiments
\begin{equation}
M_{u} = \left(\begin{array}{ccc}
0 & 126 & 0\\
126 & 10 & 10\\
0 & 10 & 126
\end{array}\right)
=\left(\begin{array}{ccc}
0 & a_{u} & 0\\
a_{u} & b_{u} & c_{u}\\
0 & c_{u} & 1
\end{array}\right).
\end{equation}
In this case, the Dirac neutrino mass matrix is
\begin{equation}
M_{\nu}^{Dirac} = \left(\begin{array}{ccc}
0 & -3a_{u} & 0\\
-3a_{u} & b_{u} & c_{u}\\
0 & c_{u} & -3
\end{array}\right) m_{t}
\end{equation}
and the right-handed neutrino Majorana mass matrix is generated by the 
coupling to $\overline{126}$ Higgs representation, and is of the form
\begin{equation}
M_{\nu,RR} = \left(
\begin{array}{ccc}
0 & r & 0\\
r & 0 & 0\\
0 & 0 & 1
\end{array}\right) m_{R}.
\end{equation}
The effective neutrino mass matrix is thus given by
\begin{equation}
M_{\nu}^{eff} = 
\left(\begin{array}{ccc}
0 & \frac{a^{2}}{r} & 0\\
\frac{a^{2}}{r} & \frac{2ab}{r}+c^{2} & c(\frac{a}{r}+1)\\
0 & c(\frac{a}{r}+1) & d^{2}
\end{array}\right)
\frac{m_{t}^{2}}{m_{R}}
\end{equation}
where $a=-3 a_{u}$, $b=b_{u}$, $c=-3c_{u}$, and $d=-3$.
The typical predictions for this type of mass matrices are
\begin{eqnarray}
\sin^{2} 2\theta_{23} & \sim & 0.98 - 1\\
\tan^{2} \theta_{12} & \sim & 0.29 - 0.46\\
|\theta_{13}| & \sim & 0.037 - 0.038\\
|m_{\nu_{3}}| & \sim & 0.053-0.059 \; eV\\
|m_{\nu_{2}}| & \sim & 0.003-0.008 \; eV\\
|m_{\nu_{1}}| & \sim & 0.0006-0.001 \; eV.
\end{eqnarray}

\section{$SO(10)$ Models with Family Symmetry in $4$-dimensions}\label{models}

A natural framework to accommodate small neutrino masses is a grand unified 
theory based on SO(10) in which a right-handed neutrino in each 
family is predicted and the see-saw mechanism can be implemented naturally. 
Many other models based on $SO(10)$ besides those mentioned in Section 5 
have been proposed to accommodate the observed 
fermion masses and mixing angles. Here we concentrate only on models which utilize 
family symmetry. We classify these models according to their intermediate 
symmetry below $SO(10)$ breaking scale and the family symmetry. 
Different symmetry breaking pattern of $SO(10)$ have different 
mass relations among the quark and lepton sectors, resulting in
different way to generate large leptonic mixing angles.

\subsection{Models with Symmetric Mass Textures}

Symmetric mass textures naturally arise if the $SO(10)$ is broken down 
to the SM gauge group through the left-right symmetry breaking chain. 
Due to its symmetric nature, this type of models tend to be more predictive 
compared to models with lop-sided/asymmetric mass texture.\\

{\noindent {\underline {\bf $SU(2)$ Family Symmetry}}}

\subsubsection{Chen and Mahanthappa}

The model proposed by Chen and Mahanthappa\cite{Chen:2000fp,Chen:2001pr,Chen:2002pa}
has $SU(2)$ family symmetry. Because $SO(10)$ breaks down 
through the left-right symmetry breaking chain, symmetric mass matrices arise. 
The field content of this model is given by, in terms of $SO(10) \times SU(2)$ 
quantum numbers,
\begin{eqnarray}
{\mbox Matter fields:} & 1(16,2), \; 1(16,1)
\nonumber\\
{\mbox Higgs fields:} & 5(10,1), \; 3(\overline{126},1)
\nonumber\\
{\mbox Flavon fields:} & 3(1,2), \; 3(1,3).
\nonumber
\end{eqnarray}
After the symmetry is broken, the following mass matrices are generated
\begin{equation}
M_{u,\nu_{LR}}  = 
\left( \begin{array}{ccc}
0 & 0 & \left<10_{2}^{+} \right> \epsilon'\\
0 & \left<10_{4}^{+} \right> \epsilon & \left<10_{3}^{+} \right> \epsilon \\
\left<10_{2}^{+} \right> \epsilon' & \left<10_{3}^{+} \right> \epsilon &
\left<10_{1}^{+} \right>
\end{array} \right)
 = 
\left( \begin{array}{ccc}
0 & 0 & r_{2} \epsilon'\\
0 & r_{4} \epsilon & \epsilon \\
r_{2} \epsilon' & \epsilon & 1
\end{array} \right) M_{U}
\end{equation}
\begin{equation}
M_{d,e} =  
\left(\begin{array}{ccc}
0 & \left<10_{5}^{-} \right> \epsilon' & 0 \\
\left<10_{5}^{-} \right> \epsilon' &  (1,-3)\left<\overline{126}^{-} \right>
\epsilon & 0\\ 0 & 0 & \left<10_{1}^{-} \right>
\end{array} \right)
 = \left(\begin{array}{ccc}
0 & \epsilon' & 0 \\
\epsilon' &  (1,-3) p \epsilon & 0\\
0 & 0 & 1
\end{array} \right) M_{D}
\end{equation}
where
\begin{equation}
\label{eq:higgsvev}
M_{U} \equiv \left<10_{1}^{+} \right>, 
\qquad
M_{D} \equiv \left<10_{1}^{-} \right>
\end{equation}
\begin{equation}
r_{2} \equiv \left<10_{2}^{+} \right> / \left<10_{1}^{+} \right>,
\quad
r_{4} \equiv \left<10_{4}^{+} \right> / \left<10_{1}^{+} \right>
\quad
p \equiv \left<\overline{126}^{-} \right> / \left<10_{1}^{-} \right>.
\end{equation}
$\epsilon M$ and $\epsilon^{'} M$ are the VEV's
accompanying the flavon fields. 
The mass hierarchy arises due to the Froggatt-Nielsen mechanism. 
The right-handed neutrino masses are generated due to $\overline{126}_{H}$, 
thus R-parity is preserved at all energy scales, and no additional proton 
decay modes are allowed, which is to be contrasted to the case when 
$\overline{16}_{H}$'s are implemented to generate right-handed neutrino 
Majorana masses (see the model of Babu, Pati and Wilczek discussed 
in the next section.) The right-handed neutrino mass matrix is given by
\begin{equation}
M_{\nu_{RR}} =   
\left( \begin{array}{ccc}
0 & 0 & \left<\overline{126}_{2}^{'0} \right> \delta_{1}\\
0 & \left<\overline{126}_{2}^{'0} \right> \delta_{2} 
& \left<\overline{126}_{2}^{'0} \right> \delta_{3} \\ 
\left<\overline{126}_{2}^{'0} \right> \delta_{1}
& \left<\overline{126}_{2}^{'0} \right> \delta_{3} &
\left<\overline{126}_{1}^{'0} \right> \end{array} \right)
 = 
\left( \begin{array}{ccc}
0 & 0 & \delta_{1}\\
0 & \delta_{2} & \delta_{3} \\ 
\delta_{1} & \delta_{3} & 1
\end{array} \right) M_{R}
\end{equation}
with $M_{R} \equiv \left<\overline{126}^{'0}_{1}\right>$.
The effective neutrino mass matrix is of the following form
\begin{equation}\label{mll}
M_{\nu_{LL}}^{eff} = \left(
\begin{array}{ccc}
0 & 0 & t\\
0 & 1 & 1+t^{3/2}\\
t & 1+t^{3/2} & 1
\end{array}
\right)\frac{d^{2}v_{u}^{2}}{M_{R}}
\end{equation}
where $t < 1$. This model can accommodate the LMA solution, in addition 
to LOW and ``Just SO'' VO solutions. Its prediction for $U_{e\nu3}$ is about $0.15$.
With $11$ input parameters, this model predicts $22(+9)$ masses, mixing angles, and 
CP violating phases for quarks and leptons (and right-handed neutrinos).

We note that in this model, the mass matrices, $M_{\nu_{LR}}$, $M_{\nu_{RR}}$ and 
$M_{\nu_{LL}^{eff}}$, have identical form. In other words, the texture considered 
in Eq.(\ref{cmtexture}) is invariant under the see-saw mechanism. The form 
invariance also occurs in a model of neutrino mixing\cite{Fritzsch:1999ee} which 
uses different texture.

\subsection{Models with Lop-sided/Asymmetric Mass Textures}

Models having $SU(5)$ as the intermediate symmetry have lop-sided Yukawa
matrices. This is due to the $SU(5)$ relation, $M_{d} = M_{e}^{T}$. This opens
up the possibility of large leptonic mixing angles due to the large mixing 
angle in the charged lepton mixing matrix.\\

{\noindent {\underline {\bf $U(1)$ family symmetry}}}

\subsubsection{Babu, Pati and Wilczek}

The model proposed by Babu, Pati and Wilczek\cite{Babu:1998wi}
utilizes the Abelian $U(1)$ as its family symmetry. 
Because it is based on dimension-$5$ operators to generate fermion masses, 
its field content, in terms of $SO(10)$ quantum numbers, is somewhat simple
\begin{eqnarray}
{\mbox Matter fields:} & 16_{1}, \; 16_{2}, \; 16_{3}
\nonumber\\
{\mbox Higgs fields:} & 1(10),\; 1 (16+\overline{16}), \; 45.
\nonumber
\end{eqnarray}
The mass matrices generated are given by
\begin{eqnarray}
M_{u} & = & \left(\begin{array}{ccc}
0 & \epsilon^{'} & 0\\
-\epsilon^{'} & 0 & \epsilon+\sigma\\
0 & -\epsilon+\sigma & 1
\end{array}\right) \cdot m_{u}
\\
M_{d} & = & \left(\begin{array}{ccc}
0 & \epsilon^{'} + \eta^{'} & 0\\
-\epsilon^{'}+\eta^{'} & 0 & \epsilon+\eta\\
0 & -\epsilon+\eta & 1
\end{array}\right) \cdot m_{d}
\\
M_{\nu_{LR}} & = & \left(\begin{array}{ccc}
0 & -3\epsilon^{'} & 0\\
3\epsilon^{'} & 0 & -3\epsilon+\sigma\\
0 & 3\epsilon+\sigma & 1
\end{array}\right) \cdot m_{u}
\\
M_{e} & = & \left(\begin{array}{ccc}
0 & -3\epsilon^{'} + \eta^{'} & 0\\
3\epsilon^{'}+\eta^{'} & 0 & -3\epsilon+\eta\\
0 & 3\epsilon+\eta & 1
\end{array}\right) \cdot m_{d}.
\end{eqnarray}
The right-handed neutrino Majorana mass matrix is 
generated by the effective operator, 
$\frac{1}{M}16_{i}16_{j}\overline{16}_{H}\overline{16}_{H}$ and is given by
\begin{equation}
M_{\nu_{RR}} = \left(\begin{array}{ccc}
x & 0 & z\\
0 & 0 & y\\
z & y & 1
\end{array}\right) \cdot M_{R}
\end{equation}
and the resulting effective neutrino mass matrix is given by
\begin{eqnarray}
& M_{\nu}^{eff} = \hspace{10cm}\nonumber\\
& \left(\begin{array}{ccc}
{\scriptstyle 9\epsilon^{'2}(x-z^{2})} \; &
{\scriptstyle 3\epsilon^{'}y(-3\epsilon^{'}z+(-3\epsilon+\sigma)x)} \; &
{\scriptstyle 3\epsilon(xy-(3\epsilon+\sigma)(x-z^{2})} 
\\
{\scriptstyle 3\epsilon^{'}y(-3\epsilon^{'}z+(-3\epsilon+\sigma)x)} \; &
{\scriptstyle -9\epsilon^{'2}y^{2}} \; &
{\scriptstyle (3\epsilon+\sigma)y(3\epsilon^{'}z-(-3\epsilon+\sigma)x)} 
\\
{\scriptstyle 3\epsilon(xy-(3\epsilon+\sigma)(x-z^{2})} \; &
{\scriptstyle (3\epsilon+\sigma)y(3\epsilon^{'}z-(-3\epsilon+\sigma)x)} \; &
{\scriptstyle (3\epsilon+\sigma)(-2xy+(3\epsilon+\sigma)(x-z^{2}))}
\end{array}\right) \cdot m_{\nu}^{\mbox{eff}}.\nonumber\\
\end{eqnarray}
The large atmospheric mixing comes from the effective neutrino mixing matrix, 
by choosing the value of parameter $y$. As $y$ also gives rise to small 
$V_{cb}$ value, the smallness of $V_{cb}$ and the maximality of the atmospheric 
mixing angle are thus related. This model can only accommodate SMA solution 
for the solar neutrinos. The LMA solution can be obtained if an intrinsic 
LH neutrino Majorana mass term arising from a dim-7 operator is included, 
assuming it dominates over the regular Type I seesaw 
term.\cite{Pati:2002ig,Pati:2002pe}  
A characteristic of this model is the presence of a new prominent 
dim-$5$ proton decay mode, $p \rightarrow \mu^{+}K^{0}$, in addition to the 
$p \rightarrow \overline{\nu}K^{+}$ mode. 
This is a consequence of utilizing the $16_{H}$ to generate 
neutrino masses.\cite{Babu:1997js} In this model, the $\theta_{13}$ angle 
is predicted to be about $5.5 \times 10^{-4}$. As all parameters are taken to
be real, CP is conserved in this model. With $11$ parameters, this model 
accommodates $18(+6)$ masses and mixing angles for quarks and leptons 
(and RH neutrinos), yielding $7$ predictions in accord with the data.

\subsubsection{Albright, Babu and Barr}

The model proposed by Albright, Babu and 
Barr\cite{Albright:1998vf,Albright:2000sz,Albright:2000dk,Albright:2001uh}
has $U(1) \times Z_{4} \times Z_{4}$ as family symmetry.
The model has the following particle content, in terms of $SO(10)$ representations,
\begin{eqnarray}
{\mbox Matter fields:} & 16_{1}, \; 16_{2}, \; 16_{3}, \; 2(16 \oplus \overline{16}), \;
2(10),\; 6(1) 
\nonumber\\
{\mbox Higgs fields:} & 
4 (10), \; 2 (16 \oplus \overline{16}), \; 1 (45), \; 5(1).  
\nonumber
\end{eqnarray}
The mass matrices in this model are generated by the Froggatt-Nielsen mechanism, 
and are given by
\begin{eqnarray}
M_{u} = & \left(\begin{array}{ccc}
\eta & 0 & 0\\
0 & 0 & \epsilon/3\\
0 & -\epsilon/3 & 1
\end{array}\right)\cdot m_{u}, \quad
M_{d} & =  \left(\begin{array}{ccc}
\eta & \delta & \delta^{'}e^{i\phi}\\
\delta & 0 & \sigma+\epsilon/3\\
\delta^{'}e^{i\phi} & -\epsilon/3 & 1
\end{array}\right)\cdot m_{d}
\nonumber\\
M_{\nu_{LR}} = & \left(\begin{array}{ccc}
\eta & 0 & 0\\
0 & 0 & \epsilon\\
0 & -\epsilon & 1
\end{array}\right)\cdot m_{u}, \quad
M_{e} & =  \left(\begin{array}{ccc}
\eta & \delta & \delta^{'}e^{i\phi}\\
\delta & 0 & -\epsilon\\
\delta^{'}e^{i\phi} & \sigma+\epsilon & 1
\end{array}\right)\cdot m_{d}.
\end{eqnarray}
The right-handed neutrino Majorana mass matrix is 
generated by the effective operator of the type, 
$\frac{1}{M}16_{i}16_{j}16_{H}16_{H}$, and is given by
\begin{equation}
M_{\nu_{RR}} = \left(\begin{array}{ccc}
c^{2}\eta^{2} & -b\epsilon\eta & a \eta\\
-b\epsilon\eta & \epsilon^{2} & -\epsilon\\
a \eta & -\epsilon & 1
\end{array}\right)\cdot \Lambda_{R}.
\end{equation}
And the effective neutrino mass matrix that accommodates the LMA solution is 
given by
\begin{eqnarray}
M_{\nu}^{eff} & \sim & \left(\begin{array}{ccc}
0 & \epsilon/(a-b) & 0\\
\epsilon/(a-b) & -\epsilon^{2}(c^{2}-b^{2})/(a-b)^{2} & -b\epsilon/(a-b)\\
0 & -b\epsilon/(a-b) & 1
\end{array}\right)
m_{u}^{2}/\lambda_{R}
\nonumber\\
& = & \left(\begin{array}{ccc}
0 & -\epsilon & 0\\
-\epsilon & 0 & 2\epsilon\\
0 & 2\epsilon & 1
\end{array}\right)
m_{u}^{2}/\lambda_{R}, \qquad ({\mbox choosing} \; a=1, \; b = c = 2).
\end{eqnarray}
This model can also accommodate the SMA, and ``Just So'' VO solutions.
An interesting property of this model is that the large mixing angle in 
atmospheric neutrinos is due to the lop-sided structure of $M_{e}$, in which 
$\epsilon \sim 0.1$ and $\sigma \sim 1$, giving rise to a large left-handed 
mixing angle in the $(2,3)$ block of $V_{e,L}$. The matrix $M_{e}$ 
is related to $M_{d}$ by the $SU(5)$ relation, $M_{e} = M_{d}^{T}$, thus such a 
lop-sided structure gives rise to a large mixing angle in the right-handed 
rotation matrix for the down-type quarks, $V_{d,R}$, making it un-observable.
The large solar mixing angle is due to the structure in the $(1,2)$ block of 
$M_{\nu}^{eff}$,
\begin{equation}
\left(\begin{array}{cc}
0 & -\epsilon\\
-\epsilon & 0
\end{array}\right).
\end{equation}
Because the large mixing angles in the atmospheric and solar neutrinos 
are due to different 
mass matrices, the prediction for $U_{e\nu_{3}}$ can be made to be extremely small. 
In this model, $|U_{e\nu_{3}}|$ is predicted to be $0.014$.
With $10$ parameters, this model accommodates all the $22$ masses, 
mixing angles, and CP violating phases at low energies.

\subsubsection{Maekawa}

Maekawa\cite{Maekawa:2001uk} has proposed a $SO(10)$ model combined with 
an anomalous $U(1)_{A}$ symmetry. This anomalous $U(1)$ symmetry is 
important for achieving DTS via 
Dimopoulos-Wilczek mechanism and for generating fermion mass hierarchy. 
The $U(1)_{A}$ charge assignments project out terms that destabilize 
the VEV of $45_{H}$, and thus guarantee that $<45_{H}>$ is along the 
$B-L$ direction, which otherwise can only be achieved with fine-tuning 
and introducing additional Higgs multiplets. These charge assignments 
also give rise to the fermion mass matrices of the following form:
\begin{equation}
M_{u} = \left(\begin{array}{ccc}
\lambda^{6} & \lambda^{5} & \lambda^{3}\\
\lambda^{5} & \lambda^{4} & \lambda^{2}\\
\lambda^{3} & \lambda^{2} & 1
\end{array}\right) <H_{u}>, \quad
M_{d} = \left(\begin{array}{ccc}
\lambda^{4} & \lambda^{7/2} & \lambda^{3}\\
\lambda^{3} & \lambda^{5/2} & \lambda^{2}\\
\lambda & \lambda^{1/2} & 1
\end{array}\right) <H_{d}>.
\end{equation}
The charged lepton mass matrix is $M_{e} = M_{d}^{T} \cdot \eta$, where 
$\eta$ characterizes the renormalization effects, and the Dirac neutrino 
mass matrix is given by
\begin{equation}
M_{\nu}^{Dirac} = \left(\begin{array}{ccc}
\lambda^{4} & \lambda^{3} & \lambda\\
\lambda^{7/2} & \lambda^{5/2} & \lambda^{1/2}\\
\lambda^{3} & \lambda^{2} & 1
\end{array}\right)\cdot \lambda^{2} \eta <H_{u}>.
\end{equation}
The right-handed neutrino Majorana masses are generated by couplings 
to $16$-dim Higgs representations
\begin{equation}
M_{\nu,RR} = \left(\begin{array}{ccc}
\lambda^{6} & \lambda^{5} & \lambda^{3}\\
\lambda^{5} & \lambda^{4} & \lambda^{2}\\
\lambda^{3} & \lambda^{2} & 1
\end{array}\right) \cdot \lambda^{9}
\end{equation}
and the resulting effective neutrino mass matrix is
\begin{equation}
M_{\nu}^{eff} = \left(\begin{array}{ccc}
\lambda^{2} & \lambda^{3/2} & \lambda\\
\lambda^{3/2} & \lambda & \lambda^{1/2}\\
\lambda & \lambda^{1/2} & 1
\end{array}\right) \cdot \lambda^{-5} \eta^{2} <H_{u}^{2}>.
\end{equation}
The LM matrix is then given by
\begin{equation}
U_{LM} = \left(\begin{array}{ccc}
1 & \lambda^{1/2} & \lambda\\
\lambda^{1/2} & 1 & \lambda^{1/2}\\
\lambda & \lambda^{1/2} & 1
\end{array}\right).
\end{equation}
The LMA solution can be accommodated in this model. The prediction 
for $U_{e\nu_{3}}$ is about $\lambda$ which is very close to the 
current bound from experiment. All order $\mathcal{O}(1)$ coefficients in the 
mass matrices are not specified in this model, thus the validity of this model 
is unclear.

\subsubsection{Shafi and Tavartkiladze}

Shafi and Tavartkiladze proposed a SUSY SO(10) model combined with an anomalous 
$U(1)_{H}$ symmetry\cite{Shafi:1999au} By extending matter content, 
the $U(1)_{H}$ can account for the observed mass hierarchy and mixing 
angles in the charged fermion sector. In this model, the three light lepton 
families are contained in the additional matter fields, $10_{i}$, rather than 
the three usual $16_{i}$ in which quarks reside. From this point of view, 
the quark-lepton unification is lost. The bi-large neutrino mixing 
is achieved by introducing three additional SO(10) singlet states. Due to the 
$U(1)_{H}$ charge assignment, only two of these singlets interact with the 
lepton doublets giving a $3 \times 2$ neutrino Dirac mass matrix, and the neutrino 
RH Majorana mass matrix is approximately $2 \times 2$. 
The LMA solution can be accommodated in this mdel.
\\

{\noindent {\underline {\bf $U(2)$ family Symmetry}}}

\subsubsection{Blazek, Raby and Tobe}

The model proposed by Blazek, Raby and Tobe\cite{Blazek:1999ue,Blazek:1999hz}
has $U(2) \times U(1)^{n}$ as its family symmetry. 
It has the following field content, in terms of $SO(10) \times U(2)$:
\begin{eqnarray}
{\mbox Matter fields:} & 1(16,2), \; 1(16,1), \; 1(1,2), \; 1(1,1)
\nonumber\\
{\mbox Higgs fields:} & 1 (10,1), \; 1(45,1)
\nonumber\\
{\mbox Flavon fields:} & 2(1,2), \; 1(1,3), \; 2(1,1_{A}).
\nonumber
\end{eqnarray}
The mass matrices of this model have the following form
\begin{eqnarray}
Y_{u} & = &
\left(\begin{array}{ccc}
\kappa_{1} \epsilon \rho &
(\epsilon^{'}+\kappa_{2}\epsilon)\rho &
0\\
-(\epsilon^{'}-\kappa_{2}\epsilon)\rho &
\epsilon\rho &
\epsilon r T_{\overline{u}}\\
0 &
\epsilon r T_{\overline{Q}} &
1
\end{array}\right) \lambda
\\
Y_{d} & = &
\left(\begin{array}{ccc}
\kappa_{1} \epsilon \rho &
\epsilon^{'}+\kappa_{2}\epsilon &
0\\
-(\epsilon^{'}-\kappa_{2}\epsilon) &
\epsilon\rho &
\epsilon r T_{\overline{d}}\\
0 &
\epsilon r T_{\overline{Q}} &
1
\end{array}\right) \lambda 
\\
Y_{\nu_{LR}} & = & 
\left(\begin{array}{ccc}
3 \kappa_{1} \epsilon \omega &
-(\epsilon^{'}-3\kappa_{2}\epsilon)\omega &
0\\
(\epsilon^{'}+3\kappa_{2}\epsilon)\omega &
3\epsilon\omega &
\frac{1}{2}\epsilon r T_{\overline{\nu}}\omega\\
0 &
\epsilon r T_{\overline{L}} \sigma &
1
\end{array}\right) \lambda
\\
Y_{e} & = & 
\left(\begin{array}{ccc}
3\kappa_{1} \epsilon &
-(\epsilon^{'}-3\kappa_{2}\epsilon) &
0\\
(\epsilon^{'}+3\kappa_{2}\epsilon) &
3\epsilon &
\epsilon r T_{\overline{e}} \sigma\\
0 &
\epsilon r T_{\overline{L}} \sigma &
1
\end{array}\right) \lambda
\end{eqnarray}
where $\omega = 2\sigma/(2\sigma-1)$ and $T_{f} =$ (Baryon number - Lepton number) 
of multiplet $f$. The right-handed neutrino Majorana mass matrix is 
generated by the effective operator of the type, 
$\frac{1}{M}16_{i}16_{j}\overline{16}_{H}\overline{16}_{H}$, and is given by
\begin{equation}
M_{\nu_{RR}} = \left(\begin{array}{ccc}
\kappa_{1} S & 
\kappa_{2} S &
0\\
\kappa_{2} S &
S & 
\phi\\
0 & \phi & 0
\end{array}\right).
\end{equation}
This model can accommodate LMA solution, in addition to SMA and 
``Just So'' VO solutions. In the case of LMA solution, its prediction 
for $U_{e\nu_{3}}$ is $0.049$. It has $16$ input parameters which 
yield $22$ masses and mixing angles.

\subsubsection{Raby}

Raby has proposed a $SO(10)$ model combined with $SU(2) \times U(1)$ family 
symmetry,\cite{Raby:2003ay} in which the ansatz proposed by Frampton, Glashow 
and Yanagida\cite{Frampton:2002qc} discussed in Sec. 3 naturally arises. 
The representations utilized in this model, in terms of $SO(10) \times SU(2)$, 
are given as follows:
\begin{eqnarray}
{\mbox Matter fields:} &  
1(16,2), \; 1(16,1), \; 3(1,1)
\nonumber\\
{\mbox Higgs fields:} & (10,1), \; (\overline{16},1), \; (45,1) 
\nonumber\\
{\mbox Flavon fields:} & (1,1), \; (1,1_{A}), \; (1,2), \; (1,\overline{2}), 
\; (1,3),
\nonumber
\end{eqnarray}
in addition to several vector-like Froggatt-Nielsen fields. For the 
purpose of discussion, we denote the three $SO(10)\times SU(2)$ singlets 
by $N_{i}$ ($i=1,2,3$) which play an important role in the neutrino sector, 
the $SU(2)$ (anti-)doublet flavon fields by 
($\tilde{\phi}$) $\phi$, and the $SU(2)$ singlet flavon fields by 
$\theta$ and $S_{i}$ ($i=1,2$).
In the charged fermion sector, the Yukawa matrices are given by
\begin{eqnarray}
Y_{u} & = &
\left(\begin{array}{ccc}
0 &
\epsilon^{'}\rho &
-\epsilon \xi\\
-\epsilon^{'}\rho &
\tilde{\epsilon}\rho &
-\epsilon\\
\epsilon\xi &
\epsilon &
1
\end{array}\right) \lambda
\\
Y_{d} & = &
\left(\begin{array}{ccc}
0 &
\epsilon^{'} &
-\epsilon\xi\sigma\\
-\epsilon^{'} &
\tilde{\epsilon} &
- \epsilon \sigma\\
\epsilon \xi &
\epsilon  &
1
\end{array}\right) \lambda 
\\
Y_{\nu_{LR}} & = & 
\left(\begin{array}{ccc}
0 &
-\epsilon^{'} \omega &
\frac{3}{2} \epsilon\xi\omega\\
\epsilon^{'} \omega &
3\tilde{\epsilon}\omega &
\frac{3}{2}\epsilon\omega\\
-3 \epsilon \xi \sigma &
-3 \epsilon \sigma &
1
\end{array}\right) \lambda
\\
Y_{e} & = & 
\left(\begin{array}{ccc}
0  &
-\epsilon^{'} &
3\epsilon \xi \\
\epsilon^{'} &
3 \tilde{\epsilon} &
3 \epsilon \\
-3 \epsilon \xi \sigma &
-3 \epsilon \sigma &
1
\end{array}\right) \lambda.
\end{eqnarray}
The ansatz of Frampton, Glashow and Yanagida is obtained by considering 
the following 
superpotential in which the three $SO(10) \times SU(2)$ singlets 
$N_{i}$ mix with the SM singlets in the three families, via the following 
superpotential
\begin{equation}\label{numix}
W_{\nu}=
\frac{\overline{16}}{\hat{M}}(N_{1}\tilde{\phi}^{a}16_{a} + N_{2}\phi^{a}16_{a} 
+ N_{3}\theta 16_{3}) + \frac{1}{2}(S_{1}N_{1}^{2}+S_{2}N_{2}^{2}).
\end{equation}
The complete neutrino mass terms are thus given by
\begin{equation}
\nu m_{\nu} \overline{\nu} + \overline{\nu} V N + \frac{1}{2}N M_{N} N
\end{equation}
where $m_{\nu}$ ($\propto Y_{\nu_{LR}}$) is the Dirac mass matrix due to the 
coupling between $\nu$ and $\overline{\nu}$, which is related to 
the up-quark mass matrix,  
and $V$ and $M_{N}$ are the Majorana mass matrices due to the coupling 
between $N_{i}$ and $\overline{\nu_{i}}$, and the self coupling of $N_{i}$ 
in Eq.(\ref{numix}),
\begin{equation}
V = \frac{v_{16}}{\hat{M}} \left(\begin{array}{ccc}
0 & \phi^{1} & 0\\
\tilde{\phi}^{2} & \phi^{2} & 0\\
0 & 0 & \theta
\end{array}\right), \qquad M_{N} = {\mbox diag}(M_{1},M_{2},0).
\end{equation}
After integrating out the heavy SM singlet neutrinos, $\overline{\nu}_{i}$ and 
$N_{i}$, the neutrino effective mass matrix is obtained,
\begin{equation}
M_{\nu}^{eff} = m_{\nu} (V^{T})^{-1}M_{N} V^{-1} m_{\nu}^{T}.
\end{equation} 
The key observation is that if one defines
\begin{equation}
D^{T} \equiv m_{\nu} (V^{T})^{-1} M_{N} \mathcal{P} = \left(
\begin{array}{cc}
a & 0\\
a^{'} & b\\
0 & b^{'}
\end{array}\right), \qquad 
{\mbox with} \quad 
\mathcal{P} = \left(\begin{array}{cc}
1 & 0 \\
0 & 1 \\
0 & 0
\end{array}\right),
\end{equation}
the effective neutrino mass matrix can be rewritten in the following form
\begin{equation}
D^{T} \hat{M}_{N}^{-1} D
\end{equation}
where
\begin{equation}
\hat{M}_{N} \equiv \left(\begin{array}{cc}
M_{1} & 0\\
0 & M_{2}
\end{array}\right)
\end{equation}
which is the ansatz proposed by Frampton {\it et al}. Note that in this model, 
because the mixing in the charged lepton sector is small, the LM matrix is 
approximately the diagonalization matrix of $M_{\nu}^{eff}$, and the prediction 
of a bi-large mixing pattern is not affected. Even though the bi-large mixing 
pattern can naturally arise in this model, the connection between the CP violation 
in neutrino oscillation and the sign of the baryogenesis, which exists in the 
model of Frampton, Glashow and Yanagida, is lost, due to additional CP phases 
and more complicated structure in this model.
\\

{\noindent {\underline {\bf $SU(3)$ Family Symmetry}}}

\subsubsection{Berezhiani and Rossi}

The model proposed by Berezhiani and Rossi\cite{Berezhiani:1998vn} 
has $SU(3)$ as family symmetry with some unspecified discrete 
symmetries imposed. It has the following field content, in terms of 
$SO(10) \times SU(3)$:
\begin{eqnarray}
{\mbox Higgs fields:} & 
2(10,1), \; 1 (16,1),\; (\overline{16},1), \; 2(45,1), \; 1 (54,1)
\nonumber\\
{\mbox Flavon fields:} &
1(1,\overline{6}), \; 3 (1,3),\; 1(1,8).
\nonumber 
\end{eqnarray}

The mass matrices of this model are given as follows:
\begin{eqnarray}
Y_{u} & = \left(\begin{array}{ccc}
y_{u} & 0 & 0\\
0 & y_{c} & 0\\
0 & 0 & y_{t}\\
\end{array}\right), \qquad
Y_{d} & = \left(\begin{array}{ccc}
y_{ut} D e^{i\xi} & Ae^{i\sigma} & C\\
Ae^{i\sigma} & y_{ct}D e^{i\xi} & B\\
\frac{1}{b}C & \frac{1}{b}B & D\\
\end{array}\right)
\\
Y_{\nu_{LR}} & = \left(\begin{array}{ccc}
y_{u} & 0 & 0\\
0 & y_{c} & 0\\
0 & 0 & \frac{1}{b}y_{t}\\
\end{array}\right), \qquad
Y_{e} & = \left(\begin{array}{ccc}
y_{ut} D e^{i\xi} & k_{3} Ae^{i\sigma} & \frac{1}{b}k_{2}C\\
k_{3}Ae^{i\sigma} & y_{ct}D e^{i\xi} & \frac{1}{b}k_{1}B\\
k_{2} C & k_{1} B & D\\
\end{array}\right).
\end{eqnarray}
The right-handed neutrino Majorana mass matrix is taken to be diagonal.
Because both $Y_{\nu_{LR}}$ and $M_{\nu_{RR}}$ are diagonal, the effective 
neutrino mass matrix is also diagonal. Thus the leptonic mixing matrix 
is proportional to $V_{e,L}$. Due to the lop-sided structure in the $(2,3)$ 
block of $Y_{e}$ arising from the $SU(5)$ breaking chain, 
the maximal mixing angle in atmospheric neutrinos is obtained. 
Nevertheless, the solar mixing angle in this model is very small, 
which is in the range of SMA solution. This model has $14$ input parameters 
in the Yukawa sector. The value of $\sin2\theta_{13}$ is predicted to 
be $\mathcal{O}(10^{-2})$.

\subsubsection{Kitano and Mimura}

Kitano and Mimura\cite{Kitano:2000xk} propose a $SO(10)$ model combined with 
$SU(3) \times U(1)_{H}$ family symmetry. The three families of matter fields 
form a triplet of $SU(3)$. The Yukawa couplings in this 
model have the form 
\begin{equation}
\sum_{i,j=1}^{3} (\frac{\Phi}{M_{pl}})^{x_{i}+x_{j}} 
\frac{\xi_{j} \xi_{j}}{M_{\ast}^{2}}.
\end{equation}
The field $\Phi$ is a singlet of $SU(3)$, but it has non-vanishing $U(1)_{H}$ 
charges. Thus its VEV $\lambda M_{pl}$ provides the mass hierarchy, if different 
generations, $i$, 
have different $U(1)_{H}$ charge, proportional to $-x_{i}$. The field 
$\xi_{i}$'s are $\overline{3}$ representation of $SU(3)$, whose VEV break 
the $SU(3)$ family symmetry
\begin{equation}
<\xi_{1}> \sim \left(\begin{array}{c}
0 \\ 0 \\ 1 \end{array}\right)M_{\ast}, \quad
<\xi_{2}> \sim \left(\begin{array}{c}
0 \\ 1 \\ 1 \end{array}\right)M_{\ast}, \quad
<\xi_{3}> \sim \left(\begin{array}{c}
1 \\ 1 \\ 1\end{array}\right)M_{\ast}
\end{equation}
where $M_{\ast}$ is the family symmetry breaking scale. With the $U(1)_H$ 
charge assignment, the mass matrices of the up- and down-type quarks are
\begin{eqnarray}
Y_{u} & \sim & \frac{1}{M_{\ast}^{2}}
(<\xi_{3}> \; <\xi_{2}> \; <\xi_{1}>) 
\; \left(\begin{array}{ccc}
\lambda^{6} & \lambda^{5} & \lambda^{3}\\
\lambda^{5} & \lambda^{4} & \lambda^{2}\\
\lambda^{3} & \lambda^{2} & 1
\end{array}\right)
\left(\begin{array}{c}
<\xi_{3}^{T}> \\ <\xi_{2}^{T}> \\ <\xi_{1}^{T}>
\end{array}\right)
\nonumber\\
& \sim &
\left(\begin{array}{ccc}
\lambda^{6} & \lambda^{5} & \lambda^{3}\\
\lambda^{5} & \lambda^{4} & \lambda^{2}\\
\lambda^{3} & \lambda^{2} & 1
\end{array}\right)
\\
Y_{d} & = & Y_{e}^{T} \sim \frac{1}{M_{\ast}^{2}}
(<\xi_{3}> \; <\xi_{2}> \; <\xi_{1}>) 
\; \left(\begin{array}{ccc}
\lambda^{5} & \lambda^{4} & \lambda^{4}\\
\lambda^{5} & \lambda^{4} & \lambda^{2}\\
\lambda^{3} & \lambda^{2} & 1
\end{array}\right)
\left(\begin{array}{c}
<\xi_{3}^{T}> \\ <\xi_{2}^{T}> \\ <\xi_{1}^{T}>
\end{array}\right)
\nonumber\\
& \sim & 
\left(\begin{array}{ccc}
\lambda^{5} & \lambda^{4} & \lambda^{4}\\
\lambda^{5} & \lambda^{4} & \lambda^{2}\\
\lambda^{3} & \lambda^{2} & 1
\end{array}\right).
\end{eqnarray} 
These Yukawa matrices give rise to the observed hierarchical masses
\begin{eqnarray}
m_{u} : m_{c} : m_{t} & \sim & \lambda^{6} : \lambda^{4} : 1
\\
m_{d} : m_{s} : m_{b} & = & m_{e} : m_{\mu} : m_{\tau} 
\sim \lambda^{4} : \lambda^{2} : 1
\end{eqnarray}
and the CKM matrix
\begin{equation}
V_{CKM} \sim \left(\begin{array}{ccc}
1 & \lambda & \lambda^{3}\\
-\lambda & 1 & \lambda^{2}\\
-\lambda^{3} & -\lambda^{2} & 1
\end{array}\right).
\end{equation}
The role of these $\xi$ fields is that it gives rise to the bi-maximal 
mixing pattern in the neutrino sector. With the $U(1)$ charge assignment, 
the neutrino mass matrix is
\begin{equation}
(<\xi_{3}> \; <\xi_{2}> \; <\xi_{1}>) 
\; \left(\begin{array}{ccc}
\lambda^{2} & \lambda & \lambda^{3}\\
\lambda & 1 & \lambda^{2}\\
\lambda^{3} & \lambda^{2} & \lambda^{4}
\end{array}\right)
\left(\begin{array}{c}
<\xi_{3}^{T}> \\ <\xi_{2}^{T}> \\ <\xi_{1}^{T}>
\end{array}\right)
\sim
\left(\begin{array}{ccc}
\lambda^{2} & \lambda & \lambda\\
\lambda & 1 & 1\\
\lambda & 1 & 1
\end{array}\right).
\end{equation}
To implement the above idea in $SO(10)$ is not trivial, because all 
the three families must have the same charge assignment under $U(1)_{H}$. 
This is overcome by introducing many Froggatt-Nielsen fields and diagrams 
so that the phenomenologically viable mass matrices discussed above are reproduced. 
The neutrino masses are generated by a complicated mechanism in which 
each SM singlet in the $16$-dim matter field mixes with a Froggatt-Nielsen 
field which transforms as singlet under $SO(10)$. Thus, instead of a $6 \times 6$ 
neutrino mass matrix, this model has a $9 \times 9$ neutrino mass matrix. The Dirac 
neutrino mass matrix is different from that of the up-type quark because they are due 
to different Froggatt-Nielsen diagrams. Large mixing angle in the atmospheric sector 
can be accommodated, and the LMA solution can also be accommodated.

\subsubsection{Ross and Velasco-Sevilla}

Ross and Velasco-Sevilla\cite{Ross:2002fb} 
utilize $SU(3)$ as the family symmetry in combination with $SO(10)$. 
Based on the formulation given in Sec. 2,\cite{King:2001uz} 
the up- and down-type quark mass matrices are of the form
\begin{equation}
M = \left(\begin{array}{ccc}
\lambda^{'} \epsilon^{8} & \lambda \epsilon^{3} & \lambda \epsilon^{3}\\
-\lambda \epsilon^{3} & \lambda^{''} \epsilon^{2} & \lambda^{''} \epsilon^{2}\\
-\lambda \epsilon^{3} & \lambda^{''} \epsilon^{2} & 1 + \lambda^{''} \epsilon^{2}
\end{array}\right) M_{33}.
\end{equation}
Note the expansion parameter $\epsilon$ is different for up- and down-quark sectors.
The Dirac neutrino mass matrix is of the form
\begin{equation}
M_{\nu}^{Dirac} = \left(\begin{array}{ccc} 
\mathcal{O}(\epsilon^{8}) 
& \epsilon^{3} (z + \epsilon (x+y) ) 
& \epsilon^{3} (x + \epsilon (x-y) )\\
\epsilon^{3} (z + \epsilon (x+y) ) 
& \epsilon^{2} (a w + \epsilon u) 
& \epsilon^{2} (a w - \epsilon u)\\
\epsilon^{3} (x + \epsilon (x-y) )
& \epsilon^{2} (a w - \epsilon u)
& 1
\end{array}\right) \cdot M_{\nu, \; 33}^{Dirac}
\end{equation}
where $M_{\nu, \; 33}^{Dirac}$ is the $(33)$ component of the Dirac neutrino 
mass matrix, and the real parameters $z, a, w$ and complex parameters $x, y, u$ 
are of $\mathcal{O}(1)$. Note that the higher order terms in symmetry 
breaking parameters have been included 
in $M_{\nu}^{Dirac}$ given above, as they are crucial for getting near 
bi-maximal mixing pattern.
The right-handed neutrino Majorana mass matrix is assumed to be diagonal
\begin{equation}
M_{\nu,RR} = \left(\begin{array}{ccc}
m_{1} & 0 & 0\\
0 & m_{2} & 0\\
0 & 0 & m_{3}
\end{array}\right).
\end{equation}
The mass eigenstates of the effective neutrino mass matrix are given by
\begin{eqnarray}
\nu_{1} & \simeq & \frac{|r|\nu_{e}-|z|e^{-i\xi}\nu_{b}}
{\sqrt{z^{2}+r^{2}}}
\\
\nu_{2} & \simeq & \frac{|r|e^{i\xi}\nu_{e}+|z|\nu_{b}}
{\sqrt{z^{2}+r^{2}}}
\\
\nu_{3} & \simeq & \nu_{a}
\end{eqnarray}
where $\xi = Arg(z) - Arg(r)$ and $r = \sqrt{2}(z u - a_{\nu} w y) / z$.
Here $\nu_{a}$ is the heaviest mass eigen states given by
\begin{equation}
\nu_{a} = 
\frac{(z+\epsilon x)(\nu_{\tau} + \nu_{\mu}) - \epsilon y (\nu_{\tau}-\nu_{\mu})}
{\sqrt{(z+\epsilon(x+y))^{2}+(z+\epsilon(x-y))^{2}}} 
\end{equation}
and $\nu_{b}$ is orthogonal to $\nu_{a}$. 
As one can see in $\nu_{3}$, $\nu_{\mu}$ contributes equally as $\nu_{\tau}$, thus 
the mixing angle between $\nu_{\mu}$ and $\nu_{\tau}$ is maximal. A large 
solar mixing angle is also obtained if $\tan^{2}\theta_{12} \simeq |z / r|^{2}$ 
is in the right parameter space. Solutions for these parameters that are 
consistent with other charged fermion masses and CKM matrix elements 
have been found; it can accommodate LMA, LOW and ``Just So'' VO solutions. 
The CHOOZ angel in this model is predicted to be about $U_{e\nu_{3}} \simeq 0.07$.

\subsection{Comparisons of Models and Other Issues}

\subsubsection{Distinguishing Models using CKM Unitarity triangle}

To distinguish various existing models, clearly we need more
precise results from the experiments. The most sensitive test are the three 
angles in the CKM unitarity triangle, as suggested in Ref. 141.
To illustrate this point, in Table \ref{compare},  
\begin{table}[b!]
\tbl{\label{compare}Predictions for the three angles of the CKM unitarity 
triangle from different models.}
{\begin{tabular}{lccc}\toprule
 & $\sin 2\alpha$ & $\sin2\beta$ & $\gamma$
\\
\hline\\
Albright-Barr\cite{Albright:2000dk}  &
$-0.2079$ & $0.6428$ & $64^{o}$
\\
& &&\\
Blazek-Raby-Tobe\cite{Blazek:1999hz} (LMA solution) &
$0.94$ & $0.39$ & $47^{o}$
\\
&&&\\
Berezhiani-Rossi\cite{Berezhiani:2000cg} (Ansatz B) &
$0.27$ & $0.75$ & $73^{o}$
\\
&&&\\
Chen-Mahanthappa\cite{Chen:2001pr} & 
$-0.8913$ & $0.7416$ & $34.55^{o}$
\\
& & &\\
Raby\cite{Raby:2003ay} &
$0.92$ & $0.50$ & $71.7^{o}$
\\
&&&\\ 
\hline
&&&\\
Experiments &
$-0.95 \sim 0.33$ & $0.79 \pm 0.12$ & $34^{o} \sim 82^{o}$
\\
&&&\\
\botrule
\end{tabular}}
\end{table}
predictions for the three angles of the CKM unitarity triangle from
various models are given. Clearly, the predictions from these models 
are very different. Many models will be ruled out as soon as one can 
pin down the values for these three angles more accurately. 

\subsubsection{Distinguish Models Using $\sin^{2}\theta_{13}$}

The two classes of models discussed in this section predict 
very different relations between
$U_{e\nu_{3}}$ and $\Delta m_{\odot}^{2} / \Delta m_{atm}^{2}$, as discussed in 
Ref. 56, 142, 143, 48.
The predictions for $\sin\theta_{13}$ of various $SO(10)$ models are 
summarized in Table.\ref{theta13}.
\begin{table}[b!]
\tbl{\label{theta13} Predictions for $\sin\theta_{13}$ of various models. 
The upper bound from CHOOZ experiment is $\sin\theta_{13} \lesssim 0.24$. 
First nine models use $SO(10)$. Last two models are not based on $SO(10)$.}
{\begin{tabular}{lccr}\toprule
Model &  family symmetry &  solar solution & $\sin\theta_{13}$\\
\hline\\
Albright-Barr\cite{Albright:2001uh} & $U(1)$ & LMA & 0.014\\
&&&\\
Babu-Pati-Wilczek\cite{Babu:1998wi} & $U(1)$ & SMA & $5.5 \times 10^{-4}$\\
&&&\\
Blazek-Raby-Tobe\cite{Blazek:1999hz} & $U(2) \times U(1)^{n}$ & LMA & 0.049\\
&&&\\
Berezhiani-Rossi\cite{} & $SU(3)$ & SMA & $\mathcal{O}(10^{-2})$ 
\\
&&&\\
Chen-Mahanthappa\cite{Chen:2002pa} & $SU(2)$ & LMA & 0.149\\
&&&\\
Kitano-Mimura\cite{Kitano:2000xk}& $SU(3) \times U(1)$ & LMA 
&  $\sim \lambda \sim 0.22$ \\
&&&\\
Maekawa\cite{Maekawa:2001uk} & $U(1)$ & LMA & $\sim \lambda \sim 0.22$ \\
&&&\\
Raby\cite{Raby:2003ay} & $3 \times 2$ seesaw with $SU(2)_{F}$ 
& LMA & $\sim m_{\nu_{2}}/2m_{\nu_{3}} \sim \mathcal{O}(0.1)$\\
&&&\\
Ross-Velasco-Sevilla\cite{Ross:2002fb} & $SU(3)$ & LMA & 0.07\\
\hline
\\
Frampton-Glashow\cite{Frampton:2002qc} & $3 \times 2$ seesaw &
LMA  & $\sim m_{\nu_{2}}/2m_{\nu_{3}} \sim \mathcal{O}(0.1)$ \\
$\quad$ -Yanagida &&&\\
&&&\\
Mohapatra-Parida\cite{Mohapatra:2003tw} & RG enhancement &
LMA & $0.08-0.10$\\
$\quad$ -Rajasekeran &&&\\
\botrule
\end{tabular}}
\end{table}
In the model of Chen and Mahanthappa,\cite{Chen:2002pa} the typical value 
for $\sin\theta_{13}$ is very close to the sensitivity of current experiments. 
In this model, the value of $U_{e\nu_{3}}=\sin\theta_{13}$ is related to the ratio  
$\Delta m_{\odot}^{2} / \Delta m_{atm}^{2}$ as 
\begin{equation}
U_{e\nu_{3}} \sim (\Delta m_{\odot}^{2} / \Delta m_{atm}^{2})^{1/3}.
\end{equation}
Thus as this ratio increases, the value of the angle $\theta_{13}$ increases. As the 
LMA solution is the most favored solution, the angle $\theta_{13}$ in this model 
is predicted to be very close to the current sensitivity of experiments. 
We note that $\theta_{13}$ of this order of magnitude leads to observable CP 
violation in neutrino oscillation.

In the model of Albright and Barr,\cite{Albright:2002if,Albright:2001xq}  
the relation between $\Delta m_{\odot}^{2} / \Delta m_{atm}^{2}$ 
and $\sin\theta_{13}$ is quite different: as the ratio   
$\Delta m_{\odot}^{2} / \Delta m_{atm}^{2}$ increases, the prediction for 
$|U_{e\nu_{3}}|$ decreases. This is due to the fact that the large mixing angles 
in the atmospheric sector and solar sector have different origins. The most favored 
LMA solution thus implies that the value of $U_{e\nu_{3}}$ is extremely small; 
a neutrino factory is needed in this case in order to pin down its value.

\subsubsection{$b-\tau$ unification}

In most models, the prediction for $m_{b}$ at the weak scale tends to be higher than 
the experimental observed value, and thus a threshold correction of the 
size $-(15-20) \; \%$ is needed to bring down $m_{b}$.  
Such a large threshold corrections for $m_{b}$ are expected 
due to loop diagrams of $SU(2)$-singlet bottom squark, the $SU(2)$-doublet 
third generation squark, gluinos and charginos, as discussed in Sec. 4.
Barr and Dorsner\cite{Barr:2002mw} suggested that, instead of these 
threshold corrections to $m_{b}$ being large and negative, 
$m_{b}/m_{\tau}$ may indeed be smaller than one at the GUT scale, 
and the deviation from the naive $b-\tau$ unification in $SU(5)$ is 
due to the large off-diagonal element of the charged lepton mixing 
matrix which also explain the large mixing in atmospheric neutrinos in models 
with lop-sided mass textures.

\subsubsection{CP Violation}

CP violation arises in these models from the complex phases in the VEV's 
of the scalar fields, and from the complex phases of the Yukawa coupling 
constants. Thus they are free parameters in the models. In the quark sector, 
the complex phase is constrained by the masses and the mixing angles, thus 
definite predictions for the three angles $(\alpha, \beta, \gamma)$ 
in the CKM unitarity triangle can be obtained. In the leptonic sector, 
on the contrary, it is not possible at this moment to obtain definite 
predictions for the three CP violating phases. The situation will 
be improved once the absolute scales of neutrino masses 
and the three mixing angles are known to much better precision.


\section{SUSY $SO(10)$ in Higher Dimensions}\label{modeled}

The idea of orbifold (SUSY) GUT's was first proposed by Kawamura to solve the 
doublet-triplet splitting 
problem,\cite{Kawamura:1999nj,Kawamura:2000ev,Kawamura:2000ir}  
and later developed by Altarelli and Ferugio\cite{Altarelli:2001qj} 
and Hall and Nomura.\cite{Hall:2001pg} 
The size of this type of extra dimensions are small, being  
inverse of the GUT scale, $R \sim 1/M_{GUT}$. 
To see how it works, let us consider the case with only one extra 
dimension, which is compactified on a $S^{1}/Z_{2}$ orbifold. 
The circle $S^{1}$ has radius $R$ and is defined by 
$y=y + 2\pi R$. Under $Z_{2}$, $y$ is mapped to $-y$. Thus the physical region 
can be taken as $0 \le y \le \pi R$. Various components transform under the $Z_{2}$ 
symmetry as follows
\begin{eqnarray}
A_{\mu}(x,y) & \rightarrow & A_{\mu}(x,-y) = P A_{\mu}(x,y) P^{-1}
\\
A_{5}(x,y) & \rightarrow & A_{5}(x,-y) = -P A_{5}(x,y) P^{-1}
\\
\Phi(x,y) & \rightarrow & \Phi(x,-y) = \pm P \Phi(x,y)
\end{eqnarray}
where $A_{\mu}(x,y)$ and $A_{5}(x,y)$ are the components of the gauge fields 
along the usual $4D$ and the $5th$-dimension, respectively; $\Phi(x,y)$ is a generic 
matter or Higgs field.   
The generators transform according to the following transformation rules,
\begin{equation}\label{sbo}
P T^{a} P^{-1} = T^{a}, \quad P T^{\hat{a}} P^{-1} = -T^{\hat{a}}.
\end{equation}
Here $T^{a}$ are generators of the residual symmetry group 
while $T^{\hat{a}}$ are the broken generators. 
The $5D$ bulk field can be decomposed into a infinite tower of $KK$ states
\begin{eqnarray}
\phi_{+}(x^{\mu},y) & = & \frac{1}{\sqrt{\pi R}} \sum_{n=0}^{\infty} \phi_{+}^{(n)}
\cos \frac{ny}{R}\\
\phi_{-}(x^{\mu},y) & = & \frac{1}{\sqrt{\pi R}} \sum_{n=1}^{\infty} \phi_{-}^{(n)}
\sin \frac{ny}{R}.
\end{eqnarray}
The mode $\phi_{+}$ is even under the $Z_{2}$ symmetry, 
\begin{equation}
P \phi_{+}(x,y) = \phi_{+}(x,-y) = + \phi_{+}(x,y).
\end{equation}
After compactification, it has a zero mode $\phi_{+}^{(0)}$.  
The $\phi_{-}$ is odd under $Z_{2}$ transformation, 
\begin{equation}
P \phi_{-} (x,y) = \phi_{-}(x,-y) = - \phi_{-}(x,y),
\end{equation}
thus it does not have a zero mode, and its $n$-th KK mode 
has a GUT scale mass, $(2n+1)/\pi R$. For the broken generators, 
the corresponding gauge bosons thus acquire GUT scale masses; 
only those corresponding to the un-broken generators have $4D$ 
zero modes, which are then identified as the gauge fields of the little group. 
From the conditions given in Eq.(\ref{sbo}), one sees that this symmetry 
breaking mechanism only works for non-Abelian symmetry, 
because the generators of an Abelian symmetry always commutes with the 
parity operator, $P$. In other words, it is not possible to reduce 
the rank of a group by Abelian 
orbifolding,\footnote{In very limited cases of orbifold breaking by outer 
automorphism, the rank reduction may be 
possible.\cite{Hebecker:2001jb,Quiros:2003gg}} and 
additional $U(1)$ symmetries survive, along with the SM gauge group, 
if the GUT symmetry has rank larger than $4$. Breaking these $U(1)$ symmetries 
can be achieved by the usual Higgs mechanism. In orbifold GUT models, 
because the GUT symmetry is broken by the orbifold boundary conditions, 
one does avoid the task of constructing 
symmetry breaking scalar potential which is usually non-trivial. 

One should also note that in higher space-time dimensions, supersymmetry 
is enlarged: In $5D$, $N=1$ SUSY has $8$ super charges; it corresponds 
to $N=2$ SUSY from the $4D$ point of view. By having different orbifold 
boundary conditions for different components in a ``super'' multiplet, one  
can reduce supersymmetry to $4D$ $N=1$, similar to the case of GUT breaking. 
To break both SUSY and the gauge symmetry by orbifolding, 
a larger discrete Abelian orbifold is thus needed when building a realistic model. 
A $N=2$ hypermultiplet can be decomposed into two $N=1$ chiral multiplets; 
and a $N=2$ vector multiplet can be decomposed into 
a $N=1$ vector multiplet (denoted by ``$V$'') and a $N=1$ chiral multiplet 
(denoted by ``$\Sigma$''). (See, for example, pp. 348-351 of Ref. 84.) 
To break $N=2$ SUSY down to $N=1$, we thus require $V$ to be even under 
the parity, and $\Sigma$ to be odd.

An immediate question one might ask is that how does the Georgi-Jarlskog $(-3)$ 
factor arise? Recall that in $4D$ GUT models, the GJ factor arises as the 
Clebsch-Gorden coefficients associated with the VEV's of scalar fields along 
certain symmetry breaking directions. In orbifold GUT scenario, because the GUT 
symmetry is broken by orbifold boundary conditions, the GJ factor must arise 
in some other way.

\subsection{Breaking SUSY $SO(10)$ by Orbifolding}

Several orbifoldings have been found to break $SO(10)$. 
The number of extra dimensions that has been considered is either one or two. 
Some orbifoldings break $SO(10)$ to only its maximal subgroups; 
others break $SO(10)$ fully down to $SU(3) \times SU(2)_{L} 
\times U(1)_{Y} \times U(1)^{'}$. Here we summarize 
various orbifolding that have been constructed.

\subsubsection{SUSY $SO(10)$ in $5D$}

Dermisek and Mafi\cite{Dermisek:2001hp} consider $SO(10)$ in $5D$ and the 
extra dimension is compactified on $S^{1}/(Z_{2}\times Z_{2}^{'})$ orbifold. 
The parities are chosen to be
\begin{eqnarray}
P & = & I_{5 \times 5} \otimes I_{2 \times 2}\\
P^{'} & = & diag(-1,-1,-1,1,1) \otimes I_{2 \times 2}
\end{eqnarray}
and the orbifold boundary conditions are given by
\begin{equation}
\begin{array}{lllllllll}
45_{{\mbox v}}: & V_{(15,1,1)}^{++} & V_{(1,3,1)}^{++} 
& V_{(1,1,3)}^{++} & V_{(6,2,2)}^{+-}
& 
\Sigma_{(15,1,1)}^{-+} & \Sigma_{(1,3,1)}^{-+} & \Sigma_{(1,1,3)}^{-+} 
& \Sigma_{(6,2,2)}^{--}
\\
&&&&&&&&
\\
10_{{\mbox H}}: & 
H_{(1,2,2)}^{++} & H_{(6,1,1)}^{+-} & 
H_{(1,2,2)}^{c \; --} & H_{(6,1,1)}^{c \; -+}.&&&&
\end{array}
\end{equation}
After compactification, the parity $Z_{2}$ reduces $N=2$ SUSY to $N=1$ SUSY 
in $4D$, and the parity $Z_{2}^{'}$ reduces $SO(10)$ to $G_{PS}$. 
The residual symmetry below the compactification scale is 
the Pati-Salam group $SU(4)_{c} \times SU(2)_{L} \times SU(2)_{R}$ on 
the 4D ``hidden'' brane at fixed point $y=\pi R/2$, which is then 
broken down to $G_{SM}$ by the usually Higgs mechanism,
and the symmetry on the ``visible'' brane at the fixed point $y=0$ is $SO(10)$.

Kim and Raby\cite{Kim:2002im} pursue along this line, and analyze 
the renormalization group evolution in this model: The $5D$ gauge coupling 
constant unification scale $M_{\ast}$ (to be contrasted with the $4D$ unification 
scale $M_{GUT}$) is found to be $\sim 3 \times 10^{17} \; GeV$ and the 
compactification scale is found to be $\sim 1.5 \times 10^{14} \; GeV$. 

Kyae and Shafi\cite{Kyae:2002ss,Kyae:2002hu} consider different parities, 
\begin{eqnarray}
P & = & diag(I_{3 \times 3}, I_{2 \times 2}, -I_{3 \times 3}, 
-I_{2 \times 2})\\
P^{'} & = & diag(-I_{3 \times 3}, I_{2 \times 2}, I_{3 \times 3}, 
-I_{2 \times 2})
\end{eqnarray}
and the $Z_{2} \times Z_{2}^{'}$ charge assignments for the components of 
the $SO(10)$ gauge field is 
\begin{equation}
\begin{array}{lllllllll}
45_{{\mbox v}}: & V_{(8,1,0)}^{++} & V_{(1,3,0)}^{++} 
& V_{(1,1,0)}^{++} & V_{(3,\overline{2},-5/6)}^{+-} & 
V_{(\overline{3},2,5/6)}^{+-} & V_{(\overline{3},1,-2/3)}^{--} 
& 2 V_{(3,2,1/6)}^{-+} 
& V_{(1,1,1)}^{--}
\end{array}.
\end{equation}
With these boundary conditions, they are able to break $SO(10)$ down to 
$SU(3) \times SU(2)_{L} \times U(1)_{Y} \times U(1)^{'}$.

\subsubsection{SUSY $SO(10)$ in $6D$}

Asaka, Buchmuller and Covi {\it et al}~\cite{Asaka:2001eh} consider 
$SO(10)$ in $6D$ and the extra 
dimensions are compactified on a $T^{2}/(Z_{2}\times Z_{2}^{GG} 
\times Z_{2}^{PS})$ orbifold. The idea is based on the observation 
that a simple extension of the SM gauge group to 
$SU(3) \times SU(2)_{L} \times U(1)_{Y} \times U(1)'$ (which has the same  
rank as $SO(10)$) is the common symmetry subgroup of the Pati-Salam gauge group, 
$SU(4) \times SU(2)_{L} \times SU(2)_{R}$ ($G_{PS}$) and 
the Georgi-Glashow gauge group, $SU(5) \times U(1)$, ($G_{GG})$. 
The first parity $P$ breaks supersymmetris down to $N=1$ in $4D$, 
upon compactification on $T^{2}/Z_{2}$. The other two parities break 
the $SO(10)$ gauge symmetry, and can be taken to be
\begin{equation}
P_{GG} =  \left(\begin{array}{ccccc}
\sigma_{2} &&&&\\
& \sigma_{2} &&&\\
&& \sigma_{2} &&\\
&&& \sigma_{2} &\\
&&&& \sigma_{2}
\end{array}\right), \quad 
P_{PS} = 
\left(\begin{array}{ccccc}
-\sigma_{0} &&&&\\
& -\sigma_{0} &&&\\
&& -\sigma_{0} &&\\
&&& -\sigma_{0} &\\
&&&& -\sigma_{0}
\end{array}\right)
\end{equation}
in the vector representation of $SO(10)$.
At the fixed point of $Z_{2}^{GG}$, SUSY $G_{GG}$ is respected; at the 
fixed point of $Z_{2}^{PS}$, SUSY $G_{PS}$ is respected. The charge assignments for 
the gauge fields are chosen such that component fields belonging to the symmetric 
subgroup have positive parity and those belonging to the coset space have negative 
parity. At the intersection of two $5D$ subspaces of the $6D$ bulk, 
in which $G_{PS}$ and $G_{GG}$ are un-broken, respectively, 
the extended SM gauge group, $SU(3) \times SU(2)_{L} \times [U(1)]^{2}$ 
is realized. At this intersection (which is also one of the fixed points 
of the orbifold transformations), the electroweak symmetry and the 
additional $U(1)^{'}$ are broken by the usual Higgs mechanism.

\subsection{Fermion Mass Hierarchy in SUSY $SO(10)$ Models in Higher Dimensions}

A few mechanisms have been proposed to solve the fermion mass hierarchy problem 
in orbifold GUTs. Some models address the realistic mass relations;  
the gauge symmetry breaking in this type of models is due to both orbifolding 
and the usual Higgs mechanism from which the GJ factor $-3$ needed to 
achieve the realistic mass relations arises as the CG coefficient associated with 
the VEVs of some Higgs multiplets. Other models make use of mechanisms that 
are purely higher dimensional, {\it e.g.} overlap wave function, to generate the 
mass hierarchy. No mechanisms have been found to generate the CG factor 
by compactification.

\subsubsection{Hall et al}

Hall {\it et al}~\cite{Hall:2001xr} proposed three $SO(10)$ models in $6D$, 
and the two extra dimensions are compactified on $T^{2}/Z_{2}$, 
$T^{2}/Z_{6}$, and $T^{2}/(Z_{2}\times Z_{2}^{'})$ tori, respectively, 
in each of these three models. In the first model there is $N=1$ SUSY in the bulk, 
and the other two models have $N=2$ SUSY. These models incorporate a mechanism 
proposed by Hall {\it et al}~\cite{Hall:2001pg,Hall:2001rz} to solve the fermion 
mass hierarchy problem in which the correct mass relations are generated by mixing 
the brane localized matter fields with additional matter fields that propagate 
along the fixed line. The generic Yukawa interactions for a brane-confined field, 
$\psi(x)$, and a bulk field, $\Phi(x,y)$, are given as follows:
\begin{equation}
\mathcal{L} \sim \int dy 
\{ \lambda_{0} \delta(y) \psi^{3} + \lambda_{1} \delta(y) \psi^{2}\Phi(y) 
+ \lambda_{2} \delta(y) \psi \Phi(y)^{2} + \lambda_{3} \delta(y)\Phi(y)^{3}
+ \lambda_{4} \Phi(y)^{3}\}.
\end{equation}
The effective Lagrangian below the compactification scale is
\begin{equation}
\lambda_{0} \psi^{3} + \frac{\lambda_{1}}{V^{1/2}} \psi^{2} \Phi_{(0)} 
+ \frac{\lambda_{2}}{V} \psi \Phi_{(0)}^{2}
+ \frac{\lambda_{3}}{V^{3/2}} \Phi_{(0)}^{3}
+ \frac{\lambda_{4}}{V^{1/2}} \Phi_{(0)}^{3}
\end{equation}
where $V=M_{string} \cdot R$ is the volume factor. Thus different Yukawa 
coupling constants in the $4D$ effective Lagrangian have 
different volume suppression factors, depending upon 
how many bulk fields are involved in the interactions. Therefore by having 
different matter multiplets locate at different locations in the bulk, the 
mass hierarchy can be generated.

\subsubsection{Albright and Barr}

Albright and Barr\cite{Albright:2002pt} 
consider $SO(10)$ in $5D$ and the fifth dimension is compactified 
on a $S^{1}/(Z_{2}\times Z_{2}^{'})$ orbifold. All quarks and leptons  
and most of the Higgs fields employed in Ref. 131 
are confined to the $SO(10)$ 3-brane; only the $SO(10)$ gauge 
fields and a $10$- and a $45$-dimensional Higgs fields are placed in the 
$5D$ bulk. As a consequence, most features of the $4D$ 
model\cite{Albright:2000sz} also exist here; the only exception is that 
because the DTS problem is solved by orbifolding, the Higgs superpotential 
in this case is simpler as the terms needed for DTS are absent. 
The parity $Z_{2}$ breaks $N=2$ SUSY (from the $4D$ point 
of view) down to $N=1$ SUSY; $Z_{2}^{'}$ breaks $SO(10)$ down to 
the Pati-Salam group. The orbifold boundary conditions under 
$Z_{2} \times Z_{2}^{'}$ are given by
\begin{equation}
\begin{array}{lllllllll}
45_{{\mbox v}}: & V_{(15,1,1)}^{++} & V_{(1,3,1)}^{++} 
& V_{(1,1,3)}^{++} & V_{(6,2,2)}^{+-}
& 
\Sigma_{(15,1,1)}^{-+} & \Sigma_{(1,3,1)}^{-+} & \Sigma_{(1,1,3)}^{-+} 
& \Sigma_{(6,2,2)}^{--}
\\
&&&&&&&&
\\
45_{{\mbox H}}: & 
H_{(15,1,1)}^{++} & H_{(1,3,1)}^{++} 
& H_{(1,1,3)}^{++} & H_{(6,2,2)}^{+-}
& 
H_{(15,1,1)}^{c \; --} & H_{(1,3,1)}^{c \; -+} 
& H_{(1,1,3)}^{c \; --} & H_{(6,2,2)}^{c \; -+}
\\
&&&&&&&&
\\
10_{{\mbox H}}: & 
H_{(1,2,2)}^{++} & H_{(6,1,1)}^{+-} & 
H_{(1,2,2)}^{c \; --} & H_{(6,1,1)}^{c \; -+}.&&&&
\end{array}
\end{equation}
The VEV $<45_{H}>$ arising from the complete Higgs superpotential 
on the visible $SO(10)$ brane is along the $(B-L)$ direction, 
thus the GJ factor of $-3$ remain in this model. 
Because all the matter fields are confined to the $4D$ brane, 
the Yukawa sector of this $5D$ model is essentially the same as that given in 
Ref. 131. 

\subsubsection{Kitano and Li}

Kitano and Li\cite{Kitano:2003cn} propose a supersymmetric 
$SO(10)$ model in $5D$; the extra dimension is compactified 
on $S_{1}/Z_{2}$ orbifold, which breaks $N=2$ SUSY down to $N=1$, from the 
$4D$ point of view. The gauge symmetry $SO(10)$ can be broken either 
by the orbifold boundary conditions or by the usual Higgs mechanism. 
All three families of matter fields along with the gauge fields 
propagate in the bulk; a $45_{H}$, a pair of $16 \oplus \overline{16}$ 
which are needed to break the rank of the symmetry, and a $10_{H}$ 
are confined to the visible brane.

In this model, the fermion mass hierarchy is accommodated 
utilizing the overlap between zero mode profiles along the fifth dimension, as
discussed in Sec.\ref{qmass}. In a $5D$ SUSY theory compactified on $S^{1}/Z_{2}$ orbifold, 
the zero mode wave function of a $5D$ bulk field with a bulk mass term $m$ is 
localized exponentially as
\begin{equation}
f_{0}(y) \sim e^{-my}.
\end{equation}
In the $SO(10)$ symmetric limit, all fields in one family must have the same bulk mass 
term, $m_{i}$, resulting in unrealistic mass spectrum. 
When the $U(1)_{X}$ subgroup of $SO(10)$ 
is broken by the Higgs mechanism, the VEV of the scalar field $<\phi>$ 
which triggers this breaking also contributes to the bulk mass terms of 
the matter fields. The resulting bulk mass terms are of the form
\begin{equation}
m_{i} \rightarrow m_{i} - \sqrt{2} g_{X} Q_{X}^{i} <\phi>.
\end{equation}
Because different $SU(5)$ components of $SO(10)$, $1$, $\overline{5}$ 
and $10$, have different $U(1)_{X}$ charges, $Q_{X}^{i}$, 
a realistic mass spectrum can be obtained
\begin{equation}
Y_{u} \sim \left(\begin{array}{ccc}
\lambda^{6} & \lambda^{5} & \lambda^{3}\\
\lambda^{5} & \lambda^{4} & \lambda^{2}\\
\lambda^{3} & \lambda^{2} & 1
\end{array}\right), \;
Y_{d} = Y_{e}^{T} \sim  \left(\begin{array}{ccc}
\lambda^{4} & \lambda^{3} & \lambda^{3}\\
\lambda^{3} & \lambda^{2} & \lambda^{2}\\
\lambda & 1 & 1
\end{array}\right), \;
m_{\nu}^{eff} \sim \left(\begin{array}{ccc}
\lambda^{2} & \lambda & \lambda\\
\lambda & 1 & 1\\
\lambda & 1 & 1
\end{array}\right).
\end{equation}
The lop-sidedness of $Y_{e}$ thus gives the maximal atmospheric mixing angle and 
the LMA solution to the solar neutrino problem is accommodated. 

\subsubsection{Shafi and Tavartkiladze}

Shafi and Tavartkiladze considered SO(10) in 5D compactified on 
a $S^{1}/(Z_{2} \times Z_{2}^{'})$ orbifold.\cite{Shafi:2003ie} 
The flavor structure of this model arises at the fixed point which 
has $G_{PS}$ symmetry, thus the mechanism which generates fermion mass hierarchy  
and mixing angles is purely $4$-dimensional. By extending the matter content 
of their model and by imposing $U(1)_{H}$ 
symmetry, the charged fermion masses and mixing angles can be accommodated. 
Bi-large neutrino mixing pattern is achieved by imposing ``flavor democracy'' 
in the neutrino sector, which has been made possible due to the extension of 
matter content. 

\section{Conclusion}\label{conclude}

SUSY GUT is one of the promising candidates for physics beyond the standard
model: the hierarchy problem is solved, charge quantization is
explained, gauge coupling constants unification is achieved; as a consequence, 
a prediction for the weak mixing angle $\sin^{2}\theta_{w}$ is obtained. 
It provides a natural framework for small neutrino masses to arise, and it has the 
promise for baryogenesis. We have seen in this review how the fermion mass 
hierarchy can arise from a very contrained framework of $SO(10)$; this is 
achieved by imposing family symmetries. As proton decay has not been observed, 
SUSY GUT's in $4D$ are under siege. This situation can be alleviated 
if the SUSY GUT model is constructed in higher dimensions. 
The presence of extra dimensions also provides new ways to understand 
fermion mass hierarchy. On the experimental side, one hopes that more precise 
measurements for the masses and CKM matrix elements will enable us to 
distinguish these models, thus pointing out the right direction for model building. 
On the theory side, one hopes to obtain an understanding of the complicated 
symmetry breaking patterns and charge assignments that are needed in many 
of these models, which will then shed some light on Physics beyond the Standard Model.

\section*{Acknowledgements}
M.-C.C. and K.T.M. are supported in part by the US Department 
of Energy Grant No. DE-AC02-98CH10886 and DE-FG03-95ER40894, respectively.

\end{document}